\pdfoutput=1

\documentclass{jfm}

\usepackage{amsmath}
\usepackage{amssymb}
\usepackage{amsfonts}
\usepackage{graphicx}
\usepackage{natbib}
\usepackage{epsfig} 
\usepackage{psfrag}
\usepackage{lscape}
\newcommand{\vv}{{\bf v}}
\newcommand{\vs}{{\bf v}_{s}}
\newcommand{\vn}{{\bf v}_{n}}
\newcommand{\vns}{{\bf v}_{ns}}
\newcommand{\pa}{\partial}
\newcommand{\om}{\boldsymbol{\omega}}
\newcommand{\Om}{\boldsymbol{\Omega}}
\newcommand{\Reys}{{\it Re_s}}

\def\lsim{\mathrel{\rlap{\lower4pt\hbox{\hskip1pt$\sim$}}
    \raise1pt\hbox{$<$}}}                
\def\gsim{\mathrel{\rlap{\lower4pt\hbox{\hskip1pt$\sim$}}
    \raise1pt\hbox{$>$}}}                
\newcommand{\degree}{\ensuremath{^\circ}}
\newcommand{\f}{\frac}
\newcommand{\be}{\begin{equation}}
\newcommand{\ee}{\end{equation}}
\newcommand{\Efe}{{\bf F}}

\ifCUPmtlplainloaded \else
  \checkfont{eurm10}
  \iffontfound
    \IfFileExists{upmath.sty}
      {\typeout{^^JFound AMS Euler Roman fonts on the system,
                   using the 'upmath' package.^^J}%
       \usepackage{upmath}}
      {\typeout{^^JFound AMS Euler Roman fonts on the system, but you
                   dont seem to have the}%
       \typeout{'upmath' package installed. JFM.cls can take advantage
                 of these fonts,^^Jif you use 'upmath' package.^^J}%
      }
  \else
  \fi
\fi

\ifCUPmtlplainloaded \else
  \checkfont{msam10}
  \iffontfound
    \IfFileExists{amssymb.sty}
      {\typeout{^^JFound AMS Symbol fonts on the system, using the
                'amssymb' package.^^J}%
       \usepackage{amssymb}%
         \let\leq=\leqslant
         \let\geq=\geqslant
      }{}
  \fi
\fi

\ifCUPmtlplainloaded \else
  \IfFileExists{amsbsy.sty}
    {\typeout{^^JFound the 'amsbsy' package on the system, using it.^^J}%
     \usepackage{amsbsy}}
    {\providecommand\boldsymbol[1]{\mbox{\boldmath $##1$}}}
\fi

\newcommand\Rey{\mbox{\textit{Re}}}  

\newsavebox{\astrutbox}
\sbox{\astrutbox}{\rule[-5pt]{0pt}{20pt}}

\title[Superfluid spherical Couette flow]
{Superfluid spherical Couette flow}

\author[C. Peralta, A. Melatos, M. Giacobello and A. Ooi]%
{C. \ns P\ls E\ls R\ls A\ls L\ls T\ls A$^{1,2}$%
  \thanks{E-mail: cperalta@aei.mpg.de}
,\ns
A. \ns M\ls E\ls L\ls A\ls T\ls O\ls S$^2$, \break
\ns M. \ns G\ls I\ls A\ls C\ls O\ls B\ls E\ls L\ls L\ls O$^{3}$
\and A. \ns O\ls O\ls I$^4$}
\affiliation{
$^1$ Max-Planck-Institut f\"ur
Gravitationsphysik, Albert-Einstein-Institut, Am M\"uhlenberg 1,
D-14476 Golm, Germany \\ [\affilskip]
$^2$ School of Physics, University of Melbourne,
Parkville, VIC 3010, Australia\\[\affilskip]
$^3$Air Vehicles Division, Defence Science \& Technology
Organisation, Melbourne, VIC 3207, Australia \\[\affilskip]
$^4$Department of Mechanical and Manufacturing
Engineering, University
of Melbourne, Parkville, VIC 3010, Australia
}

\pubyear{2008}
\volume{XXX}
\pagerange{XXX--XXX}
\date{?? and in revised form ??}

\begin{document}

\maketitle

\begin{abstract}
We solve numerically for the first time the two-fluid,
Hall--Vinen--Bekarevich--Khalatnikov (HVBK) equations
for a He-II-like superfluid contained in a differentially  
rotating, spherical shell, generalizing previous  
simulations of viscous spherical Couette flow (SCF) and
superfluid Taylor--Couette flow.
The simulations are conducted for
Reynolds numbers in the 
range $1 \times 10^2 \leq \Rey \leq 3 \times 10^4$,
rotational shear $0.1  \leq \Delta \Omega / \Omega \leq 0.3$, and
dimensionless gap widths $ 0.2 \leq \delta \leq 0.5$. 
The system tends towards a stationary but unsteady state,
where the torque oscillates persistently, with
amplitude and period determined by $\delta$ and $\Delta \Omega /\Omega$.
In axisymmetric superfluid SCF, the number of
meridional circulation cells multiplies
as $\Rey$ increases, and their shapes become
more complex, especially in the superfluid
component, with multiple secondary cells arising for
$\Rey > 10^3$.
The torque
exerted by the normal component is approximately
three times greater in a superfluid with 
anisotropic Hall--Vinen (HV) mutual friction than
in a classical viscous fluid or a superfluid with
isotropic Gorter-Mellink (GM) mutual friction. HV mutual
friction also tends to ``pinch" meridional
circulation cells more than GM mutual friction.
The boundary condition on the superfluid component,
whether no slip or perfect slip, does
not affect the large-scale structure of the flow
appreciably, but it does alter the cores of the circulation cells, 
especially at lower $\Rey$. As $\Rey$ increases,
and after initial transients die away, the mutual
friction force dominates the vortex tension, and
the streamlines of the superfluid and normal
fluid components increasingly resemble each other.
In nonaxisymmetric superfluid SCF, three-dimensional
vortex structures are classified according
to topological invariants. For misaligned spheres, the
flow is focal throughout most
of its volume, except for thread-like zones where it
is strain-dominated near the equator (inviscid component)
and poles (viscous component). A wedge-shaped isosurface
of vorticity rotates around the equator at roughly
the rotation period. For a freely precessing outer
sphere, the flow is equally strain- and vorticity-dominated
throughout its volume. Unstable focus/contracting points
are slightly more common than stable node/saddle/saddle points
in the viscous component but not in the inviscid component.
Isosurfaces of positive and negative vorticity 
form interlocking poloidal ribbons (viscous component)
or toroidal tongues (inviscid component) which attach
and detach at roughly the rotation period.
\end{abstract}

\section{Introduction}

A diverse family of flow states, collectively
termed spherical Couette flow (SCF), is observed when a
viscous fluid fills a differentially rotating, spherical shell.
The flow state at any instant is determined by the
Reynolds number
$\Rey$, dimensionless gap width $\delta$, relative
angular velocity $\Delta \Omega$, and, importantly, the
history of the flow. Some of the states are steady;
others (usually, but not always, those with higher
$\Rey$, $\delta$, or $\Delta \Omega$) are
unsteady. At low Reynolds numbers
($\Rey \lsim 10^3$), the basic flow ($0$-vortex state) is
steady and symmetric about the equator. Above
a critical Reynolds number, that for small
gaps ($\delta \lsim 0.1$)
can be approximated by $\Rey_c \approx 41 (1+\delta) \delta^{-3/2}$,
a Taylor vortex develops on each side
of the equator \citep{khlebutin68,je00}. The meridional
velocity increases with $\Rey$
and $\delta$ \citep{buhler90,er95}, scaling
as $v_{\theta} \propto \delta^2 \Rey \, \Delta \Omega$
for $\delta \lsim 0.1$ and $\Rey \lsim 10^3$
\citep{yb86}. For wide gaps ($\delta \gsim 0.3$), the
flow does not develop Taylor vortices except under
special conditions
[e.g., counterrotation; see \citet{lbewr96,lk04}]. It is
unstable with respect to nonaxisymmetric
perturbations \citep{bmy78,yb86}.
At high Reynolds numbers ($\Rey \gsim 10^5$), the flow
develops spiral vortices, shear waves, and
herringbone waves, before entering a fully developed
turbulent state as $\Rey$ increases further \citep{nt88,ntz02,nzt02}.

The problem of {\it superfluid SCF}, for
example in He II, has not yet been explored numerically
\citep{hb04} or experimentally. It is not known how the
flow states differ from viscous SCF, and what transitions
are allowed between them.
Even in cylindrical (Taylor--Couette)
geometry, only a limited amount of information exists regarding
state transitions in the superfluid problem, for the special
cases of very small gaps ($\delta \sim 0.02$)
and small Reynolds numbers ($\Rey \lsim 380$) \citep{hbj95,hb00}.
Taylor vortices are detected in He II at the critical Reynolds numbers predicted
by linear stability theory ($\Rey_c \sim 278$) \citep{bj88,b92}, but the
theoretical predictions are valid only at
temperatures $T \gsim 2.0$ {\rm K},  close to the
transition temperature $T_c$, where the normal
fluid component dominates ($\gsim 90$ \% of the total density). The circulation
cells are elongated in the axial direction, and anomalous modes
(cells rotating in the opposite sense
to those in a classical fluid) are observed \citep{hb00}.
The streamlines of the normal and superfluid components
are appreciably different for $\Rey \lsim 10^2$
but increasingly resemble each other as $\Rey$ increases \citep{hb95,pmgo06b}.

In this paper, we employ a numerical solver recently developed
to solve the  Hall--Vinen--Bekarevich--Khalatnikov (HVBK) equations 
for a rotating superfluid \citep{pmgo05a} to study the {\it unsteady}
behaviour of SCF in classical (Navier--Stokes) fluids and
superfluids, in two and three dimensions. First, we perform
a set of axisymmetric experiments with rotational shear
in the range $0.1 \leq \Delta \Omega/\Omega \leq 0.3$ in
medium and large gaps ($0.2 \leq \delta \leq 0.5$).
The flow is unsteady. 
The torque, which oscillates persistently and quasiperiodically
(near but not at the rotation period),
can be up
to three times greater in a superfluid than in a Navier--Stokes
fluid at the same Reynolds number.
We assemble a partial gallery of vortex states, in the same
spirit as for classical SCF \citep{mt87a,mt87b,tdumas,je00};
a complete classification lies beyond the scope of this paper.
Second, we take advantage of the three-dimensional
capabilities of our numerical solver to investigate two
systems that exhibit nonaxisymmetric flow:
(i) a spherical, differentially
rotating shell in which the rotation axes of the inner
and outer spheres are mutually inclined;
and (ii) a spherical, differentially rotating shell in which
the outer sphere precesses freely, while
the inner sphere rotates uniformly or is at rest.
These systems have never been studied before.
We use standard vortex identification methods, introduced
by \citet{cpc90} in viscous flows, to
fully characterize the three-dimensional vortex structures
we encounter.

The paper is divided into seven parts. In Section
\ref{sec:hydroheii}, we review the HVBK theory
of He II.  We describe the numerical method in Section \ref{sec:nummet}
and validate it in Section \ref{nummethod}.
In Section \ref{sec:unsteady2d}, we present results for
axisymmetric superfluid SCF, empahisizing its time-dependent
behaviour. We investigate the effects of
grid resolution, spectral filtering, superfluid fraction,
(ani)stropic mutual friction, and
no-slip/perfect-slip boundary
conditions. In Section
\ref{sec:nonaxi}, we present results for nonaxisymmetric
superfluid SCF for misaligned and precessing spheres, emphasizing
again the time-dependent behaviour and vortical topology.
Laboratory and astrophysical applications are discussed briefly
in Section \ref{sec:lab_astro}.

\section{HVBK theory}
\label{sec:hydroheii}
\citet{hv56a,hv56b} and \citet{bk61} first devised
a two-fluid hydrodynamic model to describe rotating He II in the presence
of a high density of vortex lines with quantized circulation.
The HVBK model was rederived by \citet{hr77} from first 
principles, within the framework of classical continuum mechanics.
It employs thermodynamic variables associated with the fluid
as a whole, which satisfy conservation equations
of mass, momentum, and energy, as in the
work of \citet{gn67} and \citet{hills72}
on the theory of mixtures.

In the full HVBK theory, the inertia of the vortex lines
is explicitly considered, with the superfluid
density regarded as an independent thermodynamic
variable, resulting in a three-fluid
set of equations. Vortex line inertia
is relevant when studying superfluid flow
near solid boundaries, as it explicitly
includes healing [where the superfluid
density decreases near a boundary; \citet{donnelly91}]
and relaxation (which prevents
the superfluid fraction from changing instantaneously
when the thermodynamic state is altered).
We do not consider these issues in this paper.
Instead, we use the equations of \citet{hr77} in
the HVBK limit where the vortex inertia is zero.

\subsection{HVBK equations of motion}
\label{subsec:hbvk_eqns}
The incompressible HVBK equations which describe the motion
of the superfluid (density $\rho_s$, velocity $\vs$) and
normal fluid (density $\rho_n$, velocity $\vn$) components
take the form \citep{hr77,bj88}
\begin{equation}
\label{eq:hvbk1} \frac{\pa \vn}{\pa t} + ( \vn \cdot \nabla) \vn = -\nabla \sigma_n
+ \nu_n \nabla^{2} \vn + \frac{\rho_s}{\rho} {\bf F},
\end{equation}
\begin{equation}
\label{eq:hvbk2} \frac{\pa \vs}{\pa t} + ( \vs \cdot \nabla) \vs = -\nabla \sigma_s
- \nu_s {\bf T} - \frac{\rho_n}{\rho} {\bf F},
\end{equation}
\begin{equation}
\label{eq:incompress}
\nabla \cdot \vn = \nabla \cdot \vs = 0,
\end{equation}
where $\sigma_s$ and $\sigma_n$ are defined as
\begin{equation}
\label{eq:sigmas}
\sigma_s = U - TS + \frac{p}{\rho} - \frac{\rho_n}{2 \rho} (\vn - \vs)^2
+ \frac{\rho_s \nu_s |\om_s|}{\rho},
\end{equation}

\begin{equation}
\label{eq:sigman}
\sigma_n = U + \frac{\rho_s}{\rho_n} TS + \frac{p}{\rho} + \frac{\rho_s}{2 \rho} (\vn - \vs)^2
+ \frac{\rho_s \nu_s |\om_s|}{\rho}.
\end{equation}
Here, $p$ is the pressure, $\rho = \rho_s + \rho_n$ is the
total density, $\nu_n$ is the kinematic viscosity,
$\nu_s$ is the stiffness parameter (defined in Section \ref{subsec:mut_fric}),
$\om_s = \nabla \times \vs$ is the macroscopic vorticity
(averaged over many vortex lines), and $U$ and $S$ are the internal
energy and entropy per unit mass, which we take to be uniform at a
given temperature $T$.
We define the mutual friction ${\bf F}$ and
vortex tension ${\bf T}$ in (\ref{eq:hvbk1}) and
(\ref{eq:hvbk2}) in the next section.
The incompressible limit corresponds
formally to infinite first and second sound speeds \citep{sonin87}.
\footnote{This is a good approximation in neutron stars, for
example, an important application where 
the flow is subsonic. Note that we model systems with $\rho_n \neq 0$, which
often sustain heat currents. However, as long as the flow is slower
than the speed of second sound, and no
external heat source is present, the fluid can
be treated as isothermal.}
Effective pressures
$p_{s}$ and $p_{n}$ are defined by $\nabla p_s = \nabla p -
\frac{1}{2} \rho_n \nabla (\vns^2)$ and $\nabla p_n = \nabla p +
\frac{1}{2} \rho_s \nabla (\vns^2)$, with $\vns = \vn-\vs$.
In the incompressible limit, only the first viscosity coefficient and
mutual friction
can be included as dissipative processes; other transport
coefficients involve compression 
of the normal and superfluid components
\citep{sonin87,ac06}. 
\subsection{Mutual friction and vortex tension}
\label{subsec:mut_fric}

Quantized vortex lines mediate an
interaction between the normal fluid
and the superfluid component known as mutual friction.
The major source of mutual friction in liquid helium
is roton-vortex scattering
in the experimentally relevant temperature range
$1 \, {\rm K} \, \leq T \leq \, 2.17 \, {\rm K}$.
For a rectilinear vortex array the mutual friction
is anisotropic.
\citet{hv56a,hv56b} showed experimentally that second
sound propagates at different speeds parallel and
perpendicular to the rotation axis and is damped
in the latter direction.
They postulated the following form for the mutual
friction force per unit mass due to a
rectilinear vortex array:
\begin{equation}
\label{eq:hvforce}
 {\bf F} = \frac{1}{2} B \hat{\mbox{\boldmath$\omega$}}_{s} \times (
\mbox{\boldmath$\omega$}_{s} \times \vns - {\bf T}) + \frac{1}{2}B^{\prime}
(\mbox{\boldmath$\omega$}_{s}
\times \vns - {\bf T}).
\end{equation}
In (\ref{eq:hvforce}), $B$ and $B'$ are temperature-dependent, dimensionless
coefficients \citep{bdv83}. The first and
second terms on the right-hand side give the force per unit mass
along and perpendicular to the second sound wave vector respectively.
The $B$ coefficient attenuates the second sound,
while $B^{\prime}$ shifts its frequency.
The term ${\bf T}$ was not included in the original
derivation of \citet{hv56a,hv56b}. It was proposed by \citet{am66}, to
take into account the curvature of the vortex lines.

The vortex tension ${\bf T}$ arises from the local circulation around
a vortex line. It was added to the HVBK equations of
motion by \citet{hr77} [cf. (\ref{eq:hvforce})].
Consider a vortex line which is slightly curved. The force per unit length,
${\bf f}$, which tends to straighten the vortex, points towards
its centre of curvature and has magnitude $e/r$, where $e$ is the
energy per unit length and $r$ is the radius of curvature.
In vector form, this can be written ${\bf f} = e (\hat{\om}_s \cdot \nabla) \hat{\om}_s$, with $\hat{\om}_s = \om_s/|\om_s|$.
When extending this argument to many vortex lines,
the local superfluid velocity around each vortex line is determined
by the quantization rule $\oint \vs \cdot d {\bf \rm l} = \kappa = {h}/{m}$, 
where the integral
is calculated around a path enclosing the vortex core,
 $m$ is the mass of the bosonic entity forming the condensate (the
helium atom in He II, or two neutrons in a neutron superfluid),
and $h$ is Planck's constant. The mean area density
of vortex lines is $\omega_s/\kappa$. Hence the average straightening
force per unit volume of superfluid is
$(e \omega_s/\kappa) (\hat{\om}_s \cdot \nabla) \hat{\om}_s
= \rho_s \nu_s {\om}_s \times (\nabla \times \hat{\om}_s)$, with
$\nu_s = e/(\rho_s \kappa)$ \citep{am66,k65}.
In order to evaluate this force, one needs
the energy per unit length of vortex line, which is
given classically by $e =  \rho_s \kappa^2 \ln (b_0/a_0)/(4\pi)$, where
$b_0$ is the intervortex spacing and $a_0$ is the core radius
of the vortex.
The stiffness parameter, $\nu_s$, in (\ref{eq:hvbk2}) is then given by
$\nu_{s} = (\kappa/4\pi) \ln(b_0/a_0)$,
and the vortex tension force per unit mass, ${\bf T}$, is
written as \citep{bj88,hbj95}
\begin{equation}
\label{eq:tforce} {\bf T} = \nu_s \om_{s} \times (\nabla \times \hat{\om}_{s}).
\end{equation}
Note that $\nu_s$ has the dimensions of a kinematic viscosity,
but its physical meaning is very different: it
controls the oscillation frequency of Kelvin waves excited on
vortex lines \citep{hbj95}. 

Quantized vortices are not always organized into a rectilinear
array. If the counterflow speed $v_{ns}$ exceeds a threshold,
growing Kelvin waves are excited along the vortex lines and
the rectilinear array is disrupted to form a self-sustaining,
reconnecting, ``turbulent" vortex tangle \citep{donnelly91}.
Experimentally, this is observed in narrow channels carrying
a heat flux, where second sound waves 
are attenuated preferentially along $\vns$ 
independently of frequency (and hence velocity gradients),
and the temperature gradient is proportional to the cube of the heat
flux \citep{gm49,vinen57a,vinen57b}. These data can be
explained by an isotropic mutual friction, called
the Gorter-Mellink (GM) force. Usually, the GM force 
per unit volume is written as
${\bf f} = A \rho_{n} \rho_s v_{ns}^2 \vns$,
where $A$ is a phenomenological constant which is
a function of temperature and has values
$23.1 \, {\rm g}^{-1} \, {\rm cm} \,{\rm s} \, \leq A \leq \, 3310
\, {\rm g}^{-1} \, {\rm cm} \,{\rm s}$ in the 
temperature range $1.20 \, {\rm K} \, \leq T \leq \, 2.16 \, {\rm K}$
in liquid helium.
Re-writing it as a force per unit mass, as in
(\ref{eq:hvbk1}) and (\ref{eq:hvbk2}), we have
\begin{equation}
\label{eq:gmforce1} {\bf F} = A^{\prime} \left( \frac{\rho_{n}
\rho_{s} v_{ns}^{2}}{ \kappa \rho^{2}}\right) \vns,
\end{equation}
where
$A^{\prime} = B^3 \rho_n^2 \pi^2 \chi_1^2/3\rho^2 \chi_2^2$
is a dimensionless, temperature-dependent coefficient, related
to the original GM constant by $A^{\prime} = A \rho \kappa$, and
$\chi_1$ and $\chi_2$ are dimensionless constants of order unity
\citep{vinen57c, pmgo05a}.

\section{Pseudospectral solver}
\label{sec:nummet}
In this section, we describe our numerical method.
We start from a three-dimensional, 
pseudospectral, Navier--Stokes solver, originally developed 
by S. Balachandar to study viscous
flows around circular and elliptical cylinders
\citep{mittal95,mb95}, prolate spheroids \citep{mittal99}, and rotating spheres
\citep{bb02,giacobello05}. The solver is
modified in two steps to solve the Navier--Stokes
equation in a spherical
Couette geometry with time-dependent boundary conditions: 
\begin{enumerate}
\item The absorption filter applied at the outer boundary to
enforce outflow is switched off
and replaced by a Dirichlet boundary condition
(see Section 2.3). The filter
smoothly attenuates the radial diffusive terms
in the Navier--Stokes equations, but it is inappropriate
in SCF, which takes place in an enclosed geometry.

\item A third-order Adams-Bashforth scheme is used
to evolve the fields in time (see Section \ref{subsec:algo}), upgrading
the second-order Adams-Bashforth scheme in the original solver.

\end{enumerate}

The solver is then extended to
handle the superfluid HVBK equations, which are mathematically similar to a Navier--Stokes
equation coupled to an Euler equation with a forcing term.
This extension is quite challenging, so we explain the
method in enough detail (in this Section and the Appendices)
for the reader to reproduce and verify the results
if desired.
An early attempt to solve the spherical Couette problem with
a pseudospectral code based on spherical harmonics \citep{hollerbach00}
was stymied by numerical
instabilities arising from the sensitivity to boundary conditions
[\citet{hb04}, R. Hollerbach 2004, private communication];
the basis functions are defined globally,
so instabilities at the boundaries
rapidly contaminate the whole computational domain.
Our approach, based on restricted Fourier expansions in the
angular coordinates and Chebyshev polynomials in
the radial coordinate, solves these difficulties by
combinating a low-pass spectral filter \citep{don94} with special
boundary conditions for the superfluid \citep{k65,hr77, pmgo05a}.

\subsection{Geometry}
\label{sec:grids}

We consider the motion of an isothermal,
incompressible, rotating
superfluid, described by equations (\ref{eq:hvbk1}) and (\ref{eq:hvbk2}), contained
between two concentric spheres.
Points in the domain are defined by the
spherical coordinates ($r$, $\theta$, $\phi$), with
\begin{equation}
\label{esfericas} R_{2} \leq r \leq R_{1}, \, \, \, \, 0 \leq \theta \leq \pi, \, \,
\, \, 0 \leq \phi \leq 2 \pi,
\end{equation}
where $R_1$ and $R_2$ are the radii of the inner and  outer spheres,
respectively.
The spheres are assumed to rotate rigidly,
with angular
velocities $\Omega_1 (t)$ and $\Omega_2(t)$ respectively. 
The inner sphere rotates about an axis
parallel to the $z$ axis; the outer
sphere rotates about an axis that can be inclined with respect to the
$z$ axis, by an angle $\theta_0$. 
The spheres can accelerate or decelerate, for example, in response 
to the back-reaction torque  exerted
by the fluid, or because the outer sphere precesses freely
(see Section \ref{sec:tbc}). All variables
are made nondimensional using $R_2$ as a unit of length, and $\Omega_1^{-1}$
as a unit of time, unless indicated otherwise.
The viscous Reynolds number is defined as $\Rey=\Omega_{1} R_{2}^2/\nu_n$
and a ``superfluid" Reynolds number is defined as $\Rey_s=\Omega_{1} R_{2}^2/\nu_s$. For cases where only the inner (outer) sphere rotates,
we define $\Rey_1=\Omega_{1} R_{1}^2/\nu_n$ and  $\Rey_2=\Omega_{2} R_{2}^2/\nu_n$.

\subsection{Algorithm}
\label{subsec:algo}

The radial coordinate ($r$) is discretized using
a Gauss-Lobatto collocation
scheme \citep{boyd02,canuto88}.
The angular directions $\theta$ and $\phi$ are discretized uniformly.
The number of collocation points in the three coordinates
is $(N_r,N_\theta,N_\phi)$; their detailed coordinate
locations are defined in appendix \ref{sec:appendixA}.
The collocation points are shifted from the poles
in order to avoid the coordinate singularities at $\theta=0$, $\pi$.
Note that this displacement is small; for a typical
grid with $N_\theta = 200$, the first grid point is located at $\theta_1 \approx 7.854 \times 10^{-3}$ {\rm rad}.

In spherical coordinates,
the Courant-Friedrichs-Lewy (CFL) stability condition
for a convective-dominated
equation with time step $\Delta t$
can be written as a limit on
$CFL = {\rm max} ( u_r/\Delta r + u_\theta /r\Delta\theta
+ u_\phi /r \sin \theta \Delta \phi)\Delta t$, where 
$(\Delta r,\Delta \theta,\Delta \phi)$ are the $(r,\theta,\phi)$
grid spacings; the
maximum is taken over the whole computational domain.
Trial and error suggests that the integrator is stable for
$CFL \lsim 0.6$, as for the Navier--Stokes solver developed by
\citet{mittal99}.
We usually
take $ 10^{-5} \leq \Delta t \leq 10^{-3}$ in dimensionless units.

The velocity fields are expanded in terms of Chebyshev
polynomials in $r$ and  Fourier
polynomials in $\theta$ and $\phi$.
The expansions must obey the pole parity conditions
and be infinitely differentiable to avoid slow
convergence of the numerical scheme and any emergence
of Gibbs phenomena (see Section \ref{subsec:poleparity}). The final forms
of the expansions (which are different for scalars and vectors)
are presented in Appendix \ref{appendix:expansions}.
Differentiation in $r$ and $\theta$
is performed in physical space, multiplying 
by a differentiation matrix.
Azimuthal derivatives are calculated in wavenumber space.
The explicit form of the $r$ and $\theta$ differentiation
matrices is presented in Appendix \ref{cap2:numdif}.

The equations of motion 
(\ref{eq:hvbk1}) and (\ref{eq:hvbk2}) are discretized in time
using a time-split algorithm \citep{chorin68,canuto88},
using an explicit, third-order, Adams-Bashforth method for the non-linear
terms and an implicit Crank-Nicolson method for the diffusive
terms. The final form of the discretized equations
is presented in Appendix \ref{cap2:tempdisc}, including the
pressure correction step. The time-split algorithm is presented
in Appendix \ref{ap:solalgo}, with an explanation of how the original
second-order solver is upgraded to third-order accuracy.

\subsection{Pole parity}
\label{subsec:poleparity}

A number of numerical issues arise when a discretised field is expanded
in a Fourier series on a sphere \citep{m73,o74,mittal95,bb02}. 
In particular, the $\theta$ expansion is restricted to
the range $0 \leq \theta \leq \pi$, so
a Fourier series (periodic in $0 \leq \theta \leq 2 \pi$) can only
be used with some symmetry restrictions \citep{bb02}.
In a spherical grid, lines of  latitude and longitude intersect at two points,
and the spherical components of a vector field are discontinuous 
at the pole even when its Cartesian components are continuous
\citep{swarz79}, so the $\theta$ expansion must obey certain 
boundary conditions
at the poles in order to be compatible with the $\phi$ expansion.
This is called the pole
parity problem \citep{m73,o74,yee81,mittal99}.
Parity conditions are chosen to ensure that the
series expansions are differentiable at the poles, avoiding convergence problems
and the Gibbs phenomenon \citep{o74}. For a scalar field $s(\theta,\phi)$,
only certain modes are permitted in the expansion \citep{fornberg98}.
For odd (even) azimuthal wave numbers, the expansion of $s(\theta,\phi)$
must have odd (even) parity. 
For a vector field ${\bf v}(\theta,\phi)$, the $r$ component
is continuous across the poles, but the $\theta$ and $\phi$
components change sign. The radial component of the vector field follows
the same parity rule as for a scalar.
For the tangential and azimuthal components,
expansions with odd (even) azimuthal wave numbers must have even (odd)
parity. 
The forms of the final expansions 
are given in Appendix \ref{appendix:expansions}.

\subsection{Spectral filter}
\label{subsec:polefilter}
\begin{figure}[htpb]
\begin{center}
\includegraphics[scale=0.55]{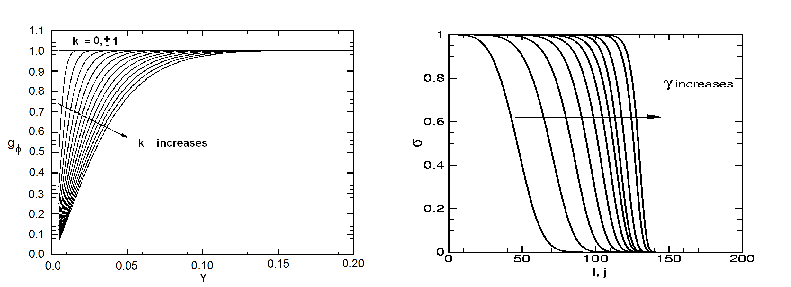}
\caption{(a) Pole filter $g_\phi$ as a function of cylindrical
radius $Y$ and azimuthal wavenumber $k$, with
$ 7 \leq k \leq 24$. The effects of the filter
are greatest in a small region near the poles, whose
cylindrical radius increases with increasing $k$.
(b) Spectral filter $\sigma$ as a function of radial wavenumber $l$
(or equivalently, latitudinal wavenumber $j$) and
filter order $\gamma$, with
$4 \leq \gamma \leq 22 $. The filtering is weaker as $\gamma$ increases.
The modes $k=0,\pm 1$ are not filtered, since they
are important to get the correct stability characteristics. 
For all three wavenumbers, $g_\phi = 1$ for all values of $Y$.}
\label{fig:filtros}
\end{center}
\end{figure}

The geometry of the sphere makes grid points cluster naturally
near the poles.
In order to deal with the
clustering, spectral methods use a filter to suppress
high-wavenumber modes near the poles 
in the $\phi$ expansion \citep{usr71,fm97,bb02,giacobello05}.
From previous studies on the stability of swirling flow past a sphere,
it is known
that the $k=\pm1$ modes ($k$ is the azimuthal wave number) are the most unstable \citep{na93}.
From the CFL condition, it can be
deduced that the time step is determined
by the $k=\pm 1$ modes
if $\beta_{ljk}$, $\gamma_{ljk}$, and $\delta_{ljk}$
decay faster than $k^{-2}$ \citep{bagchi02}. A
filter that fulfills these conditions was devised by \citet{bb02},
in which the coefficients of the $\phi$ expansion are
multiplied by
\begin{equation}
\label{eq:filconv} g_\phi (r,\theta,k) = 1 - \exp[-\xi(k)Y^{\phi(k)}].
\end{equation}
In (\ref{eq:filconv}), we define $Y = r \sin \theta$, and
$\xi (k)$ and $\phi(k)$ are functions subject to the boundary conditions
$g_\phi(r,\theta_1,k) = k^{-1}$ and $g_\phi(kr,\theta_1,k) = 0.9$.
The exponential
form of the filter ensures that its effects are limited to a small
region near the poles of the sphere. Figure \ref{fig:filtros}a illustrates
the behaviour of $g_\phi$ as a function of $Y$ and $k$.

Aliasing arises because we are restricted to a finite
range of wavenumbers \citep{boyd02}. 
As a remedy, we adopt Orszag's $2/3$ anti-aliasing 
(``padding") rule,  which filters out waves with wavelengths
twice and thrice the grid spacing. \citet{o71a} showed that one obtains 
an alias-free computation on a grid with $N$ points by
filtering out the high wavenumbers
and retaining only $2N/3$ modes \citep{boyd02,canuto88}.

Spectral methods are sensitive to boundary conditions.
Oscillations generated
by the Gibbs phenomenon \citep{boyd02} contaminate the
solution and grow unstably with time. In order
to mitigate these instabilities, we 
multiply the coefficients of the $r$
expansion by an expression of the form \citep{vgh84,don94}
\begin{equation}
\label{eq:specfil}
\sigma \left( l, N_r \right)  =\exp\left(-
\left| \frac{l-N_c}{N_{r}-N_c} \right|^{\gamma} \ln \epsilon \right), 
\end{equation}
and the coefficients of the $\theta$ expansion by a similar
expression $\sigma ( l, N_\theta)$,
where $0 \leq |l| \leq N_{r}$ is the radial wave number,
$\epsilon =  2.2 \times 10^{-16}$ is the machine zero,
$\gamma$ is the (integer) order of the filter, and
$N_c$ is the central wavenumber of the filter.
A small order ($\gamma \lsim 16$) indicates
strong filtering, while a high order ($\gamma \gsim 16$) indicates
gentle filtering.
Figure \ref{fig:filtros}b shows the behaviour of
the spectral filter  (\ref{eq:specfil}) as a function of wavenumber, for
$ 4 \leq \gamma \leq 22$.

\subsection{Initial and boundary conditions}
\label{sec:bcond}
\subsubsection{Initial conditions for $\vn$ and $\vs$}
The velocity fields must be divergence-free initially, in order to satisfy the incompressibility
constraint (\ref{eq:incompress}). The easiest choice is $\vn = \vs = 0$.
However, the superfluid
velocity field is used to calculate the vorticity unit vector,
$\mbox{\boldmath$\hat{\omega}$}_s$, which
in turn appears in (\ref{eq:hvforce}) and (\ref{eq:tforce}) 
and must remain well defined. Additionally, the HVBK equations
describe a rotating superfluid, implying $\mbox{\boldmath$\omega$}_s \neq 0$ 
in general.   
A simple initial condition that satisfies $\nabla \cdot \vn = \nabla \cdot \vs =0$
and $\mbox{\boldmath$\omega$}_s \neq 0$
is the Stokes solution \citep{landau_fluidos}
\begin{equation}
\label{eq:stokes} \vn = \vs = \frac{r R_1^3 R_2^3}{R_2^3 - R_1^3}
\left[\frac{\Omega_1 - \Omega_2}{r^3} + \frac{\Omega_2}{R_1^3}
- \frac{\Omega_1}{R_2^3} \right] \sin \theta \, {\bf e}_\phi.
\end{equation}
This ansatz is an exact solution of the spherical
Couette problem in the limit $\Rey \rightarrow 0$ and
a respectable approximation for $\Rey \lsim 10$, where meridional
circulation, which is absent from (\ref{eq:stokes}), carries only
$\sim 0.01$ \% of the total kinetic
energy \citep{tdumas}.

\subsubsection{Boundary conditions for $\vn$}
\label{ssubsec:rsbc}
The normal fluid satisfies a no-penetration condition, $(\vn)_r =0$,
at the inner and outer spheres. It also satisfies a no-slip condition;
its tangential velocity equals that of the surface, like for
a viscous, Navier-Stokes fluid. The angular velocity
vector $\Om_2$, which is tilted with
respect to the $z$ axis in the $x$-$z$ plane by an angle
$\theta_0$, can be written in spherical
polar coordinates as
\begin{eqnarray}
\Om_2 & = & \Omega_2 [(\cos \theta_{0} \cos \theta + \sin \theta_{0} \sin \theta \cos \phi) {\bf e}_r
-  (\cos \theta_{0} \sin \theta 
- \sin \theta_{0} \cos \theta \cos \phi) {\bf e}_\theta \\
\nonumber
& &
- \sin \theta_{0} \sin \phi {\bf e}_\phi],
\label{eq:omega_inclinada}
\end{eqnarray}
while $\Om_1$ remains fixed parallel to the $z$ 
axis. The no-slip and no-penetration boundary conditions then reduce to
\begin{eqnarray}
\label{eq:bcvn}
\vn(R_1,\theta,\phi) = R_1 \Om_1 \times {\bf e}_r, \\ 
\vn(R_2,\theta,\phi) = R_2 \Om_2 \times {\bf e}_r.
\end{eqnarray}

\subsubsection{Boundary conditions for $\vs$}
\label{sec:super_bc}
The distribution of quantized vorticity in the superfluid component
determines the boundary conditions for $\vs$. Quantized
vortices in a cylindrical
container are arranged in a rectilinear array
parallel to the rotation axis if the rotation
is slow [$\Rey \lsim 268$; \citet{bj88,barenghi95}]
or axisymmetric \citep{hbj95,hb00,hb04}. Under these
conditions, the numerical
evolution is stable if the
vortex lines are parallel to the curved wall
(i.e. perfect sliding, $\om_s \times {\bf n} = 0$)
and perpendicular to the end plates.

In more general situations, e.g. noncylindrical containers,
nonaxisymmetric flows, or fast rotation, there
is no general agreement
on what boundary conditions are suitable \citep{hb00}.
This is especially true when the rectilinear vortex array
is disrupted by the Donnelly--Glaberson instability
to form an isotropic, turbulent vortex tangle (Section \ref{subsec:mut_fric}).
The radial component of the superfluid satisfies no penetration:
\begin{equation}
\label{eq:nopen}
(\vs)_r (R_1,\theta,\phi) = 0 = (\vs)_r (R_2,\theta,\phi).
\end{equation}
It is less clear how to treat the $\theta$ and $\phi$
components. Numerical solutions of the HVBK equations in cylindrical Couette geometries
are stable only if there is
perfect sliding at the inner and outer
surfaces \citep{hbj95,hb95,hb04}; numerical instabilities
grow at rough surfaces \citep{hbj95}.
In spherical containers, however, the vortex lines are neither
perpendicular to the walls nor
parallel to the rotation axis everywhere.
Previous attempts to solve the HVBK equations in spherical geometries
foundered partly because of these issues
[\citet{hb04}; R. Hollerbach 2004, private communication].

\citet{k65}
suggested that vortex lines can either slide along, or pin to, the boundaries,
or behave somewhere between these two extremes.
If the boundary is not moving, the vortices terminate perpendicular to the surface \citep{k65}.
The tangential velocity $\vv_L$ of a vortex line relative to a
rough boundary moving with velocity $\bf u$ is given by \citep{k65,hr77,hb00}
\begin{equation}
\label{eq:vL} \vv_L - {\bf u} = c_1 \hat{\om}_s \times ({\bf n} \times \hat{\om}_s)
 + c_2 {\bf n} \times \hat{\om}_s,
\end{equation}
where ${\bf n}$ is the unit normal to the surface, and $c_1$ and
$c_2$ are coefficients describing the relative ease of
sliding. The form of (\ref{eq:vL}) follows
from calculating the energy dissipated as
vortices slip along the surface \citep{k65}.
Equation (\ref{eq:vL}) is difficult to
include in HVBK theory, where each fluid element is threaded by
many vortex lines, because $\vv_L$ is the velocity of a single vortex line;
it cannot be calculated from
$\vv_{n}$ and $\vv_{s}$, which are averaged
over regions containing many vortex lines. Additionally,
the slipping parameters $c_1$ and $c_2$ must be evaluated at each
point on the surface, yet there is no experimental or theoretical study available
in the literature on the precise form of these parameters.
However, we can consider two simple limits of equation (\ref{eq:vL}). For $c_1 = c_2 \rightarrow \infty$,
the vortex lines slide freely along the surface and one requires
\begin{equation} \label{eq:bcs21}
\om_s \times {\bf n}= 0 \end{equation}
in order that $\vv_L$ remains finite; that is, the vortex lines
are oriented perpendicular
to the surface. On the other hand, for $c_1=c_2=0$, we have
rough boundaries with $\vv_L = {\bf u}$. In spherical Couette
geometries, this implies no-slip, i.e.
\begin{eqnarray}
\label{eq:bcs22}
\vv_s(R_1,\theta,\phi) = R_1 \Om_1 \times {\bf e}_r, \\ 
\vv_s(R_2,\theta,\phi) = R_2 \Om_2 \times {\bf e}_r.
\end{eqnarray}
We find empirically
that conditions (\ref{eq:bcs22}) and (3.12) lead to stable numerical evolution
in most scenarios studied in this paper. 

The existence of a vortex-free ($\om_s=0$) region adjacent to the boundaries,
whose thickness approaches the
intervortex spacing, is theoretically
predicted by minimizing the free energy of a vortex
array in a container \citep{hall60,sf68,hr77,hbj95}. However,
it has not been detected conclusively in experiments \citep{nd70,mms80}.
It is unclear how to treat this
boundary layer numerically within HVBK theory, which assumes a high density
of vortices, so we do not consider it further in this paper.

\subsubsection{Accelerating spheres}
\label{sec:tbc}
The angular velocities of the outer and inner spheres, $\Om_2$
and $\Om_1$, can vary with
time, either in a prescribed way or in reaction to the torque
exerted by the fluid. 

One scenario considered in this paper is free precession of
the outer sphere.
This situation is relevant to astrophysical systems like
neutron stars \citep{ja02,s77,l03,s05}
and to laboratory systems like superfluid-filled gyroscopes \citep{reppy65}.
Let the outer sphere be biaxial, with symmetry axis ${\bf e}_3$ and
constant total angular momentum ${\bf J}$, and resolve
the angular velocity $\Om$ into components \citep{shaham86}
\begin{equation}
\label{eq:bcp1} \Om = \frac{{\bf J}_{\|}}{I_{\|}}
+ \frac{{\bf J}_{\bot}}{I_{\bot}}
\end{equation}
parallel and perpendicular to the symmetry axis, 
with ${\bf J} ={\bf J}_{\|} + {\bf J}_{\bot}$,
where
$I_{\|}$ and $I_{\bot}$ are the associated moments of inertia.
The precession frequency $\Omega_p$ is then defined by
\begin{equation}
\label{eq:bcp2} 
\Om_{p} = {\bf J}_{\|} \displaystyle \left(\frac{1}{I_{\bot}} - \frac{1}{I_{\|}}
\right) = \Omega_p {\bf e}_3,
\end{equation}
and the velocity of any point 
on the surface of the outer sphere in the inertial (lab) frame is
given by
\begin{equation}
\label{eq:bcp3} \frac{d {\bf x}}{dt} = R_2 \Om^\prime \times {\bf e}_r
- R_2 \Omega_p {\bf e}_3 \times {\bf e}_r,
\end{equation}
where $ \Om^\prime = \Omega^\prime {\bf J}/J$ is the inertial-frame precession frequency.
The back-reaction of the fluid on the container
needs to be included when solving the HVBK equations self-consistently. The
viscous torque accelerates (decelerates) the container. To this must
be added any external torques $N_{\rm ext}$. Examples of $N_{\rm ext}$
are the electromagnetic torque on the crust of a neutron star
\citep{og69,m97,spitkovski04} or
the friction between a rotating container and its spindle in laboratory
experiments \citep{tsatsa72,tsatsa80}.
In this situation, $\Om_1$
and $\Om_2$ evolve according to
\begin{equation}
\label{eq:backtorque} I_{ij} \frac{d\Omega_j^{1,2}}{dt} = ({N}_{\rm ext}^{1,2} + {N}_{\rm int}^{1,2})_i,
\end{equation}
where $I_{ij}$ is the moment-of-inertia tensor,
and $N_{\rm int}$ is
the instantaneous viscous torque
exerted by the normal fluid on the shell \citep{landau_fluidos},
\begin{equation}
\label{eq:torque_int}
N_{\rm int}^{1,2} = \frac{1}{\Rey}\int d\phi  \int d \theta \, r \sin \theta \left(
\frac{\partial v_{n \phi}}{\partial r} - \frac{v_{n\phi}}{r}\right)_{r=R_1,R_2}.
\end{equation}
Equation (\ref{eq:backtorque})
is solved explicitly at each time step using a third-order
Adams-Bashforth
algorithm to get $\Omega_i(t + \Delta t)$, after advancing the flow 
using $\Omega_i(t)$ (see Section \ref{subsec:algo} and Appendix \ref{cap2:numdif}).

\section{Validation}
\label{nummethod}

To the best of the authors knowledge, the problem
of superfluid SCF has never
been solved before for $\Rey \gg 1$, save for
an inconclusive pioneering 
attempt  by R. Hollerbach [private communication, 2004; see
also \citet{hb04}], who encountered numerical instabilities when implementing
the cylindrical boundary conditions used by \citet{hbj95}.
Consequently, we cannot verify our code directly against
previous superfluid SCF results, and we are forced into
a different validation strategy: in the limit
$T \rightarrow T_c$, the superfluid component vanishes,  
equation (\ref{eq:hvbk1})
reduces to the classical Navier-Stokes equation, and
we validate our numerical scheme against the
wealth of numerical and experimental studies
available for viscous SCF.

Our three-dimensional pseudospectral HVBK solver reduces to
a classical Navier--Stokes solver if all the coupling terms [HVBK friction,
HVBK tension, $\nabla (\vns^2)$ and $\nu_s \nabla \omega_s$] are
removed, and the $\vv_s$ arrays are disabled.
We validate our solver for three types of SCF in
this regime: (i) inner sphere rotating, outer sphere stationary
($\Omega_1 \neq 0$, $\Omega_2 = 0$); (ii) inner sphere stationary,
outer sphere rotating ($\Omega_1=0$, $\Omega_2 \neq 0$); and
(iii) both spheres rotating ($\Omega_1 \neq 0$, $\Omega_2 \neq 0$).
For the parameter range explored in this Section, viz.
$50 \leq \Rey_1, \Rey_2 \leq 1200$
and $0.18 \leq \delta \leq 0.5$, a grid with
$(N_r, N_{\theta}, N_{\phi}) = (81, 200, 4)$ and $\Delta t =10^{-3}$ is
sufficient to fully resolve the flow. A flow is regarded as fully
resolved if the spectral mode amplitudes
decrease quasi-monotonically
with polynomial index. 

The meridional streamlines
drawn in the figures below correspond to the final steady state.
A steady
state is deemed to have been reached when the difference on the
viscous torques between the inner and outer spheres satisfies 
$|N_{\rm ext}^{(2)} - N_{\rm int}^{(1)}| \leq 10^{-8}$.
Torques are expressed in units of $\rho R_1^5 \Omega_1^2$
or $\rho R_2^5 \Omega_2^2$, unless otherwise indicated. 

\subsection{Inner sphere rotating, outer sphere stationary}
Following \citet{mt87a}, we focus on a moderately sized
gap $\delta = 0.18$ with $\Omega_1 \neq 0$ and $\Omega_2=0$.
We look for time-dependent
transitions between axisymmetric steady states characterized by zero, one,
or two Taylor vortices
on either side of the equatorial plane, and refer
to them as $0$-, $1$-, and $2$-vortex states respectively.
We obtain these basic flow
states, together with an intermediate ``pinched" vortex state \citep{roque},
as illustrated in Figures $4$, $7$ and $9$ of \citet{mt87a}.

\begin{figure}[htpb]
\begin{center}
\includegraphics[scale=0.35]{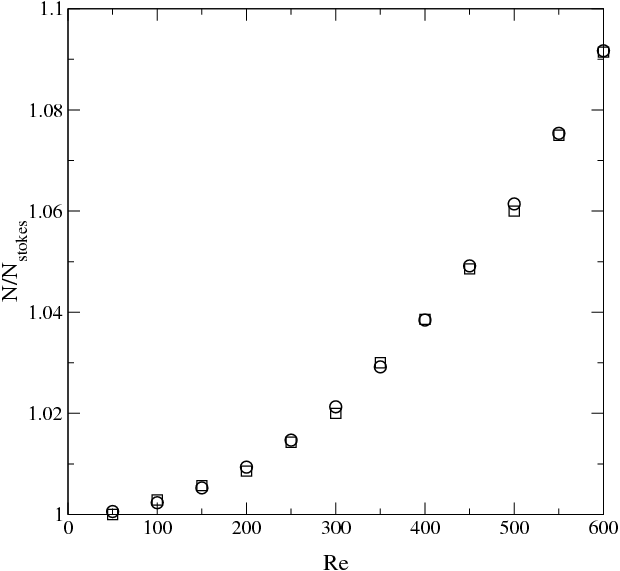}
\caption{Viscous inner torque on the inner sphere, in units of the 
Stokes torque $N_{\rm Stokes} = 16 \pi/[\Rey_1 (1-R_1^3/R_2^3)]$,
versus Reyolds number $\Rey$. The circles correspond to output
from our numerical solver. The squares denote data taken
from \citet{mt87a}.}
\label{fig:torque1}
\end{center}
\end{figure}

Figure \ref{fig:torque1} plots the steady-state viscous inner
torque (\ref{eq:torque_int}) as a function of Reynolds number for $50 \leq \Rey_1 \leq 600$, normalized to the torque exerted by
Stokes flow, $N_{\rm Stokes} = 16 \pi/[\Rey_1 (1-R_1^3/R_2^3)]$
\citep{mt87a}.
The square symbols record the values taken
from Figure $1$ of \citet{mt87a}, while the
circles are output from our numerical code. Each point
is obtained by starting from
an initially stationary fluid, not the Stokes solution (\ref{eq:stokes}),
and evolving in time until a steady state 
$|N_{\rm int}^{(2)} - N_{\rm ext}^{(1)}| \leq 10^{-8}$
is reached.
The points obtained
from our numerical simulations and the published results agree to three significant digits.

Once the basic $0$-vortex state is attained for $\Rey_1 = 600$, transitions
to pinched, $1$-vortex, and $2$-vortex
states can be induced by impulsively changing $\Rey_1$ (by reducing
the viscosity) to $650$,
$700$, and $900$ respectively. Below we check against \citet{mt87a,mt87b} 
whether our solver follows these transitions faithfully.

\subsubsection{$0 \rightarrow 1$ transition}
\label{sssec:0to1}
\begin{figure}[htpb]
\begin{center}
\includegraphics[scale=0.69]{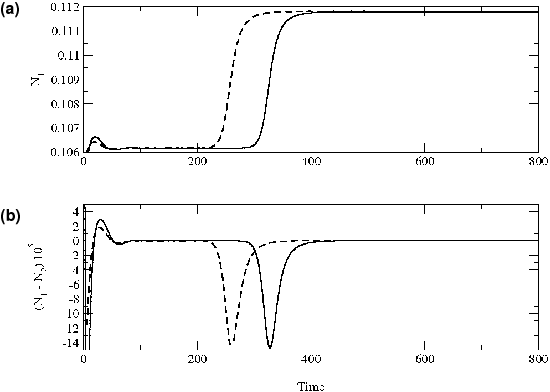}
\caption{Torque during the $0 \rightarrow 1$ transition. (a) Torque
on the inner sphere, $N_1$, as a function of time.
(b) Difference between the inner and outer torques as
a function of time. The dashed curves correspond
to numerical results from \citet{mt87b}, while the
solid curves are generated by our numerical code.}
\label{fig:torque0to1}
\end{center}
\end{figure}

We simulate the $0 \rightarrow 1$ transition by starting with a $0$-vortex
equilibrium at $Re = 650$ and then abruptly (over one time step) 
reducing the viscosity to give $Re = 700$, where the
equilibrum becomes unstable \citep{mt87b}. 
We obtain the intermediate states displayed
in Figure $10$ in \citet{mt87b}.
At the start of the sequence, the streamlines are not symmetric 
about the equator; the boundary between the counterrotating vortices at $t=230$
is displaced south of the equator. Then, at $t=315$, two wedges start to
form in the northen hemisphere, at $\approx 82 \, \deg$, and generate
a growing vortex at $t=320$, which evolves into a fully developed
Taylor vortex in the northen hemisphere at $t=400$.
The transitions 
occur at 
about the same time as Figures $10$a--$10$b in \citet{mt87b}, 
and about $68$ units of time
later than in Figures $10$c--$10$f in \citet{mt87b}.
Figure \ref{fig:torque0to1} shows how the torque evolves during 
the time interval covered by  the $0 \rightarrow 1$ transition. 
The transition, marked by a jump in the torque, occurs later 
than in the numerical experiments of \cite{mt87b} because it
is sensitive to the exact form of the initial perturbation, 
which in turn depends
on roundoff error, aliasing, and resolution \citep{mt87b}.
The sensitivity is exacerbated by the north-south asymmetry of the
$0 \rightarrow 1$ transition (cf. $0 \rightarrow 2$ below).
However, the shapes of the solid and dashed curves in Figure \ref{fig:torque0to1}
(especially the growth rate) agree to three significant digits if we slide
them on top of each other.
\vspace{1cm}

\subsubsection{$0 \rightarrow 2$ transition}
\label{sssec:0to2}
\begin{figure}[htpb]
\begin{center}
\includegraphics[scale=0.78]{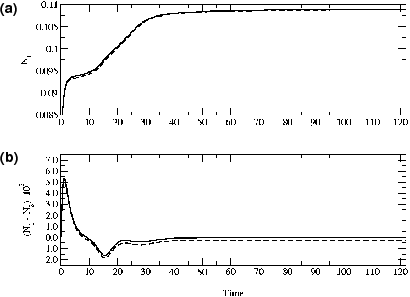}
\caption{Torque during the $0 \rightarrow 2$ transition. (a) Torque
on the inner sphere, $N_1$, as a function of time. (b) Difference 
between the inner and outer torque as
a function of time. The dashed curves correspond to
numerical results from \citet{mt87a}, while the solid curves
are generated by our numerical code.}
\label{fig:torque0to2}
\end{center}
\end{figure}

According to \cite{mt87b}, the $0 \rightarrow 2$ transition can be produced
by starting with a $0$-vortex equilibrium ($ 50 \lsim \Rey_1 \lsim 651$)
and impulsively increasing $\Rey_1$ above $775$,
where the vortex equilibrium is unstable. We therefore start with
the Stokes solution for $\Rey_1=650$ and suddenly increase
$\Rey_1$ to $800$. 
We obtain the $0 \rightarrow 2$ transitions of the meridional
flow as in Figure $4$ by \citet{mt87b}.
In Figure \ref{fig:torque0to2}, we plot the
torque on the inner sphere and the difference
between the inner and outer torques as functions of time (solid curve), 
together with
data taken from Figure $4$ of \citet{mt87b}
(dashed curve), showing an agreement to three significant
digits, after sliding the curves together.
In this case the transition is symmetric
with respect to the equator and occurs more quickly.
The $0 \rightarrow 2$ transition
is always symmetrical about the equator, as compared
to the $0 \rightarrow 1$ transition presented in
Section \ref{sssec:0to1}. In a bifurcation diagram (showing the relation
between torque and critical $\Rey_1$),
the $0$-vortex and $2$-vortex flows lie on the same
critical branch, while the $1$-vortex state
lies on a different, non intersecting branch \citep{mt87b}.

\subsection{Outer sphere rotating, inner sphere stationary}
We now allow the outer sphere to rotate ($\Omega_2 \neq 0$), while holding
the inner sphere fixed ($\Omega_1=0$). 
We reproduce the meridional streamlines for 
$\Rey_2 = 100$, $500$, $1000$, and $2000$, with $\delta=0.5$,
obtained by \citet{ds78} (Figures $3$, $4$, and $5$ of their paper), who used
a numerical method in which
the flow variables are expressed as a truncated series of $n$ orthogonal
Gegenbauer functions with variable coefficients, reducing
the Navier--Stokes equation to a set of ordinary differential
equations. At $\Rey_2=500$, the agreement is good, with
the streamlines showing the same distribution of
vortices in the northern hemisphere: one primary circulation
cell, slightly elongated in the direction of the rotation
axis, with its center located $\sim 40$ {\rm deg} over
the equator, and a small recirculation zone near the equator.
For $\Rey_2=1000$, the primary circulation
cell is more elongated, with most of the circulation
lying in a cylindrical sheath of radius approximately equal
to $R_1$, as predicted by the Taylor-Proudman theorem
\citep{p67}. For $\Rey_2=2000$, the agreement is
not as good. \citet{ds78} were unable to obtain a well defined flow pattern
for $\Rey_2 = 2000$, having been limited by
computational resources to only eight Gegenbauer polynomials,
which is not sufficient to follow
small vortex structures developing near the equator. The
higher-resolution results in
our simulations suggest that the small vortices
observed by \citet{ds78} near the
equator are probably low-resolution artifacts.

The steady-state dimensionless torque calculated by various
authors (including the present work)
is presented in Table \ref{table:omega2val}, together with bibliographic
information.

\begin{table}
\begin{center}
\begin{tabular}{ccc}
\hline
\hline $\Rey_2$
& $N_2$ ($\rho R_2^5 \Omega_2^2$)
& Reference
\\
\hline
$100$
&
$0.041745$
&
Present study
\\
&
$0.041750$
&
\citet{tdumas,dl94}
\\
&
$0.042450$
&
\citet{dq84}
\\
&
$0.041888$
&
\citet{ds78}
\\
&
$0.04160$
&
\citet{mj71a}
\\
\hline
$500$
&
$0.011979$
&
Present study
\\
&
$0.011985$
&
\citet{tdumas,dl94}
\\
&
$0.012282$
&
\citet{dq84}
\\
&
$0.011980$
&
\citet{ds78}
\\
\hline
$1000$
&
$7.2203 \times 10^{-3}$
&
Present study
\\
&
$7.2237 \times 10^{-3}$
&
\citet{tdumas,dl94}
\\
&
$7.7074 \times 10^{-3}$
&
\citet{dq84}
\\
&
$7.2382 \times 10^{-3}$
&
\citet{ds78}
\\ \hline
$2000$
&
$ 4.4483 \times 10^{-3}$
& Present study \\
&
$ 4.4478 \times 10^{-3}$
& \citet{ds78} \\
\hline
\end{tabular}
\end{center}
\caption{Comparison of numerical values obtained by various authors for the torque on the outer sphere, $N_2$, 
when the outer sphere is rotating and the inner sphere is stationary.}
\label{table:omega2val}
\end{table}

\subsection{Inner and outer spheres rotating}
\label{cap2:inner_outer_rotate}
The next step in the verification program is to consider
the rotation of both spheres. 
We follow \citet{p67} and \citet{mj71a},
who studied general axisymmetric flows between
rotating spheres, solving the Navier--Stokes equation
in terms of stream functions in a meridional plane.
\citet{p67} used a numerical
scheme based on finite differences (with typical
resolution of $40 \times 20$ mesh points), whereas
\citet{mj71a} used expansions in Legendre polynomials
(typically using up to $7$ terms).

Suppose the spheres counterrotate, with $\Omega_1=-\Omega_2$. 
We obtatin the meridional
streamlines and angular velocity profiles as shown
in Figures $4$-$5$ of \citet{mj71a}
for $\Rey_2 = 100$
and $500$ respectively, with $\delta = 0.5$. 
For $\Rey_2=100$, the agreement is good, with the
$0$-vortex state rotating clockwise in the
northern hemisphere, counterclockwise in the southern
hemisphere, and centered at $\approx 45$ {\rm deg}.
The angular velocity contours are nearly concentric
shells with values decreasing from $\Omega_2$ at the
outer shell to $-\Omega_2$ at the inner shell. At $\Rey_2=500$,
an additional counterclockwise vortex develops in the polar
region, near the inner sphere, because the
influence of the inner sphere strengthens
as the viscosity decreases. The locations
of this vortex and the main circulation cell (with
its center at $\theta \approx 30$ {\rm deg}) agree
with the results of \citet{mj71a}. 
The angular velocity profiles show a similar
pattern, forming
a cylindrical sheath parallel to the rotation
axes.

Now suppose that the spheres counterrotate, but
with $\Omega_1 = -2 \Omega_2$.
We do numerical simulations 
for
$\Rey = \Omega_{1} R_{2}^{2}/\nu = 100$ and $\Rey = 500$
respectively. We get good agreement with
the simulation results of \citet{mj71a} presented 
in Figures $6$ and $7$ of their paper.
The faster rotation of the inner sphere
produces an additional circulation cell
near its surface, both for $\Rey=100$
and $\Rey=500$. The center of
the secondary cell is slightly displaced towards
the equator in the latter case. The angular
velocity contours tend to
form a cylindrical sheath as the Reynolds
number increases \citep{p56}.

\begin{figure}[htpb]
\begin{center}
\centerline{\epsfxsize=12cm\epsfbox{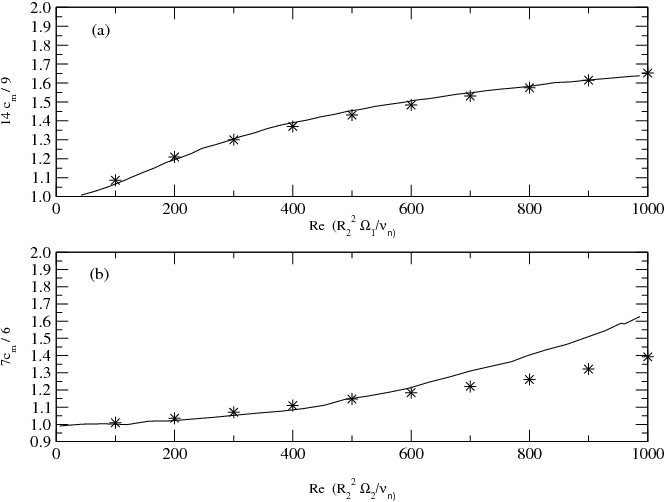}}
\caption{(a) Dimensionless torque on the inner sphere, $c_m = 7 \Rey N_1/16 \pi \rho R_2^5 \Omega_1^2$,
as a function of Reynolds number, $\Rey$. The spheres rotate in opposite directions,
with $\Omega_1 = -2 \Omega_2$. (b) Dimensionless torque on the inner sphere,
$c_m = 2 \Rey N_1/3 \pi \rho R_2^5 \Omega_2^2$
as a function of Reynolds number, $\Rey$. The spheres rotate in opposite directions,
with $\Omega_1 = - \Omega_2$.
The solid curve corresponds to data
taken from \citet{mj71a}. The asterisks are output from
the present study.}
\label{fig:torque8by9a}
\end{center}
\end{figure}

Figure \ref{fig:torque8by9a} plots the
dimensionless torque as a function of
the Reynolds number $\Rey_1$ or $\Rey_2$ (the definition used in each case is
indicated in the plots). When the inner and outer spheres rotate
in opposite directions, with $\Omega_1 = -2 \Omega_2$, the inner torque
is shown in Figure \ref{fig:torque8by9a}a; for $\Omega_1 = - \Omega_2$,
the inner torque is shown in Figure
\ref{fig:torque8by9a}b. 
When a steady state is reached, the difference between the inner and outer
torques approaches $\sim 10^{-8} \rho R_2^5 \Omega_{1,2}^2$. The solid curve
and asterisks represent the data obtained by \citet{mj71a}
and from our numerical simulations respectively.
The results coincide
to three significant digits for all the
Reynolds numbers considered when $\Omega_1 = -2 \Omega_2$. However,
the results diverge for $\Rey \geq 500$
when $\Omega_1 = -\Omega_2$.
This is a consequence of the low resolution in the expansions
used by \citet{mj71a}, who claimed to be unable
to reproduce small structures near the equator, of typical
size $\sim 0.3 \delta$, when comparing with
the study by \citet{p67}.
\citet{mj71a} used a maximum of $7$
modes in $r$ and $\theta$, whereas we use
$N_r=81$ and $N_\theta=200$ and can therefore
resolve vortical
structures as small as $10^{-4} \delta$. Note that
the torque is dominated by surface regions where the shear stresses
are stronger, e.g. where vortices cluster.

\begin{figure}[htpb]
\begin{center}
\includegraphics[scale=0.3]{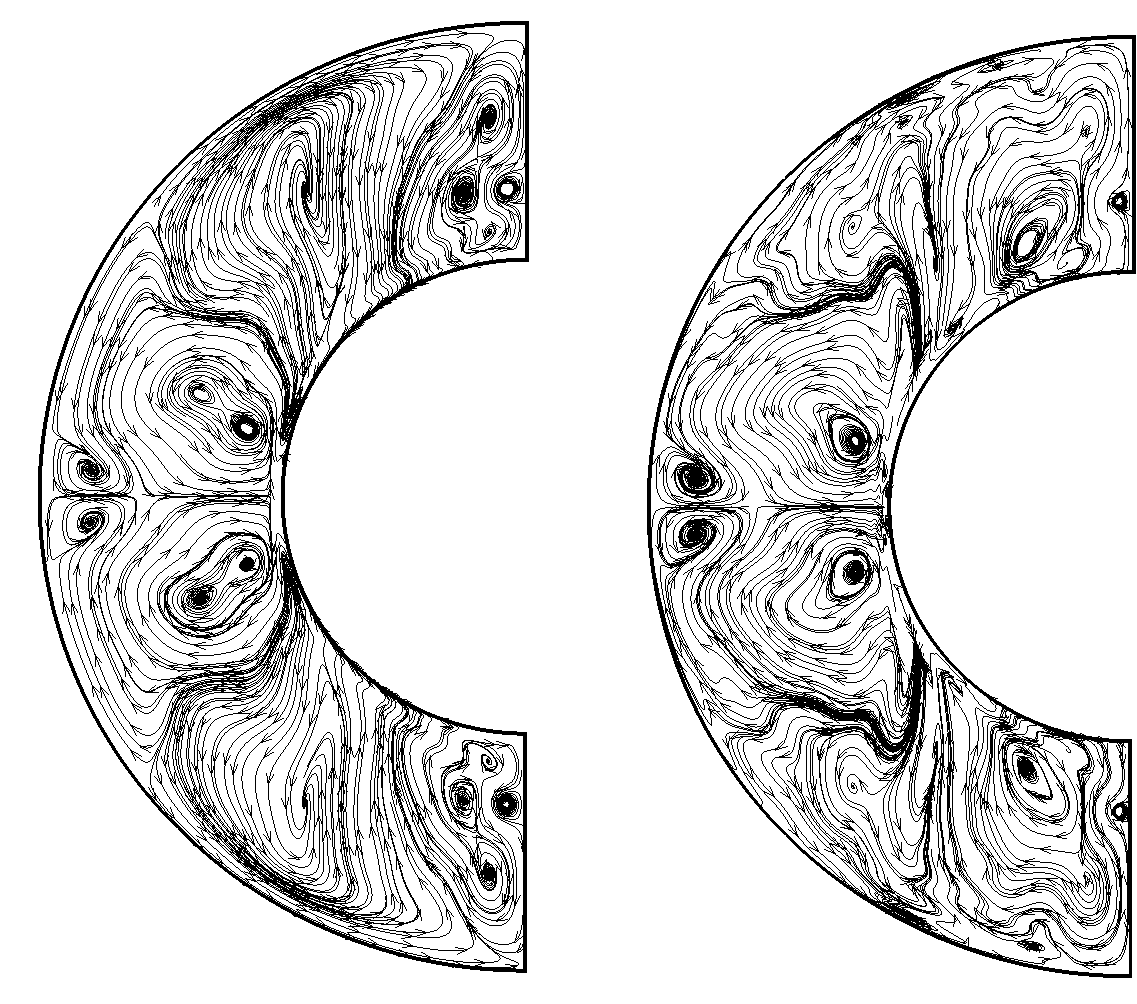}
\caption{Snapshots at $t=214$ of meridional streamlines for
the normal (left) and
superfluid (right) components in superfluid SCF, with
$\Rey=10^4$, $\delta=0.5$, and $\Delta \Omega=0.3$. Spectral
resolution:
$(N_r,N_\theta,N_\phi)=(150,400,4)$. Filter parameters: $(\gamma_r,\gamma_\theta)=(8,6)$.}
\label{fig:cap3_fig1}
\end{center}
\end{figure}

\section{Unsteady, superfluid SCF}
\label{sec:unsteady2d}
In this section, we investigate the {\it unsteady}
behaviour of SCF in classical (Navier--Stokes) fluids and
superfluids in two dimensions, by performing
a set of axisymmetric numerical experiments ($N_\phi=4$) with rotational
shear in the range $0.1 \leq \Delta \Omega \leq 0.3$, in
medium and large gaps ($0.2 \leq \delta \leq 0.5$).
For HV mutual friction, we use $B = 1.35$ and $B^{\prime} = 0.38$, 
the He II values at $T = 1.45$ {\rm K} \citep{bdv83,donnelly91,db98}.
\footnote{We consider adiabatic walls and divergence-free
$\vs$ and $\vn$. Although one expects the temperature to rise
continually in this scenario due to dissipation, we ignore
the influence of dissipation inhomogeneities in the superfluid flow.
This is equivalent to assuming $\rho_s v^2 \ll \rho_n c^2$,
where $v = v_n + v_s$ and $c$ is the second sound speed.
In all the simulations presented in this paper we have
$0.01 v^2 \ll 0.99 c^2$.
We can calculate
the rate of change of the internal energy of a unit mass of fluid
due to viscous heating from
$E = \Rey \sigma_{r \phi}^2 \Delta /2 \rho_n$,
where $\Delta$ is the total time of the simulation 
($\Delta \sim 10^2$) and  $\sigma_{r \phi}$ is the viscous stress tensor
\citet{landau_fluidos}.
For the parameters used in the simulations,
we have $10^{-8} \lsim E \lsim 10^{-4}$, which
is safe to ignore.}
For GM mutual friction, the parameter $A^{\prime} = 5.8 \times 10^{-3}$
(with $\chi_1/\chi_2 = 0.3$) at the same temperature
can be calculated from a fitting formula
derived by \cite{dthesis72}, which is consistent with previously
published experimental values \citep{vinen57c}. Stable
long-term evolution is difficult to achieve
for this value of $A^{\prime}$, so we take
$A^{\prime} = 5.8 \times 10^{-2}$ instead.
We compile a preliminary
gallery of vortex states, in the same
spirit as for classical SCF and
the validation experiments in Section
\ref{nummethod} \citep{mt87a,mt87b,tdumas,je00};
a complete classification lies outside the scope of this paper.
The torque
is observed to oscillate quasiperiodically, yet
persistently, accompanied by oscillations in the
vortical structure of the flow (Sections 
\ref{subsec:axi_example}, \ref{ssec:uns_tor}).
Resolution and filtering issues are discussed in Section
\ref{subsec:numissues}. The role played by the inviscid
superfluid, and the effect
of varying the strength and form of the mutual friction force,
are studied in Section \ref{subsec:visco_super} . It is observed that the
superfluid tends to destabilize the flow and increases
the torque. Boundary conditions are varied in Section \ref{subsec:bc_effect}.

\subsection{Meridional streamlines}
\label{subsec:axi_example}
Figure \ref{fig:cap3_fig1} depicts the meridional streamlines
of the normal (left) and superfluid (right) components in
superfluid SCF, for the special case $\Rey=10^4$, $\delta=0.5$, and
$\Delta \Omega=0.3$.
In the equatorial
zone ($60\degree \lsim \theta \lsim 120$\degree), we observe two large
circulation cells adjacent to the inner boundary. Each
large cell contains twin cores circulating in the same sense
(and therefore tending to repel).
Between the large cells and the outer boundary exist two
small vortices, occupying $\approx 20$ \% of the volume of
the large cells. The flow in each hemisphere is symmetric
about the equatorial plane. Away from the equator
($30 \degree \lsim |\theta - 90\degree| \lsim 90$\degree),
we observe a large cell (width $\approx 30 \degree$) at
mid latitudes, three vortices
in the normal component, and one small vortex in the superfluid
component.

The flow pattern described in the previous paragraph
is characteristic of moderately high Reynolds numbers
($\Rey \gsim 10^4$). The HV mutual friction
couples normal and
superfluid components strongly, so that their meridional
streamlines are similar.
At lower Reynolds numbers ($\Rey \lsim 10^3$),
the streamlines of the two components differ
markedly. The normal component behaves like a
viscous, Navier--Stokes fluid at low
$\Rey$, with a small
number ($\lsim 3$) of large circulation cells
on each side of the equatorial plane. The superfluid
is influenced less by the normal fluid, due
to the stiffness provided by the vortex tension force
\citep{hb95,swanson98}. Streamlines
of $\vs$ develop
multiple eddies and counter-eddies.
When GM mutual friction operates,
the normal and superfluid components
behave similarly, both at low
and high Reynolds numbers, but the superfluid displays
a richer variety of circulation cells, while
the normal component behaves like an uncoupled
Navier--Stokes fluid. The different effects of HV
and GM mutual friction are investigated
in Section \ref{subsec:visco_super}.

\begin{figure}[htpb]
\begin{center}
\includegraphics[scale=0.3]{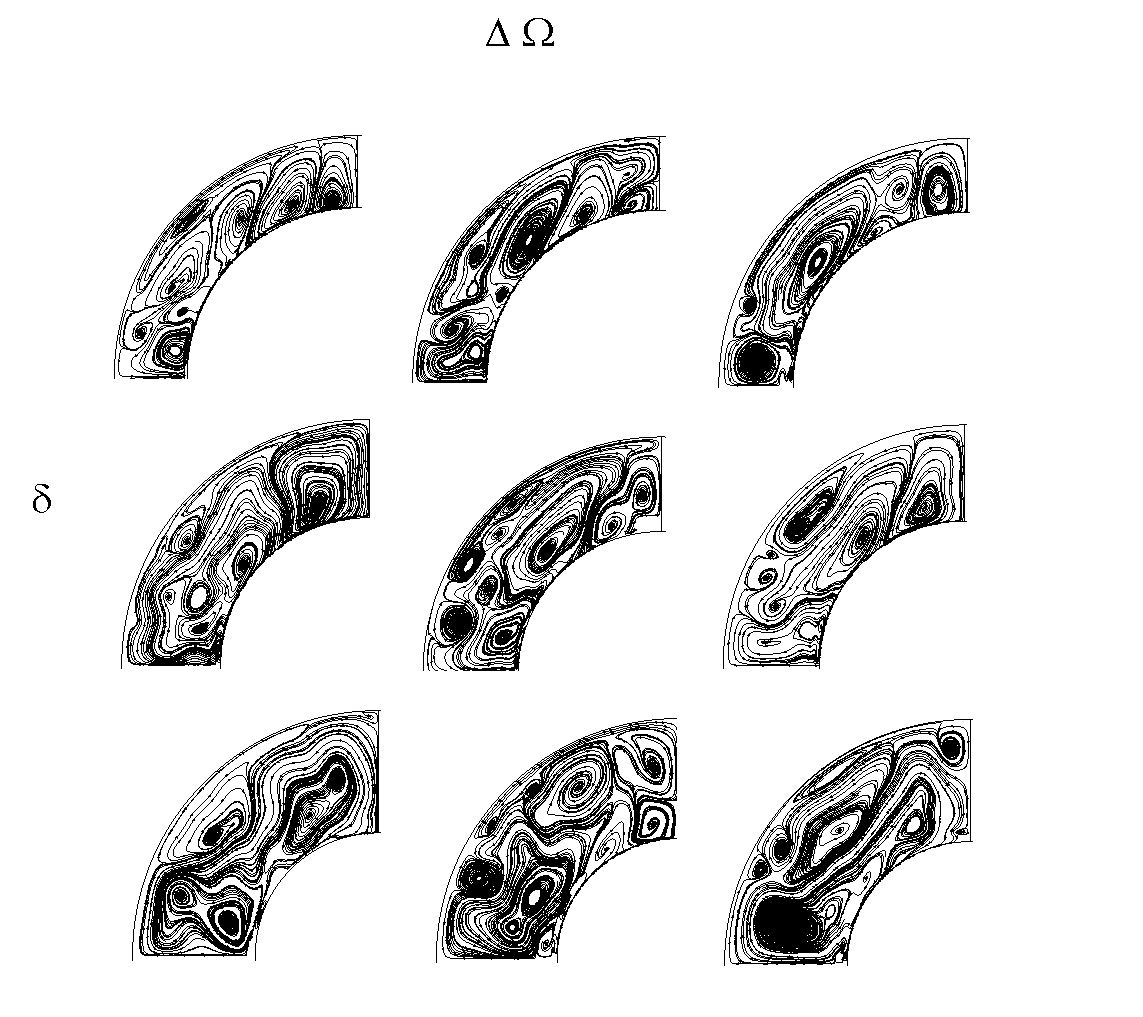}
\caption{Snapshots at $t=20$ of meridional streamlines for
the normal fluid component in superfluid SCF, with
$\Rey=3 \times 10^4$, $\delta=0.3$, $0.4$, $0.5$ (from top to bottom),
and $\Delta \Omega=0.1$, $0.2$, $0.3$ (from left to right).
The friction force if of GM form, with zero tension (${\bf T}=0$).
Spectral resolution:
$(N_r,N_\theta,N_\phi)=(120,250,4)$. Filter parameters: $(\gamma_r,\gamma_\theta)=(8,8)$.}
\label{fig:long_fig_normal}
\end{center}
\end{figure}

\begin{figure}[htpb]
\begin{center}
\includegraphics[scale=0.3]{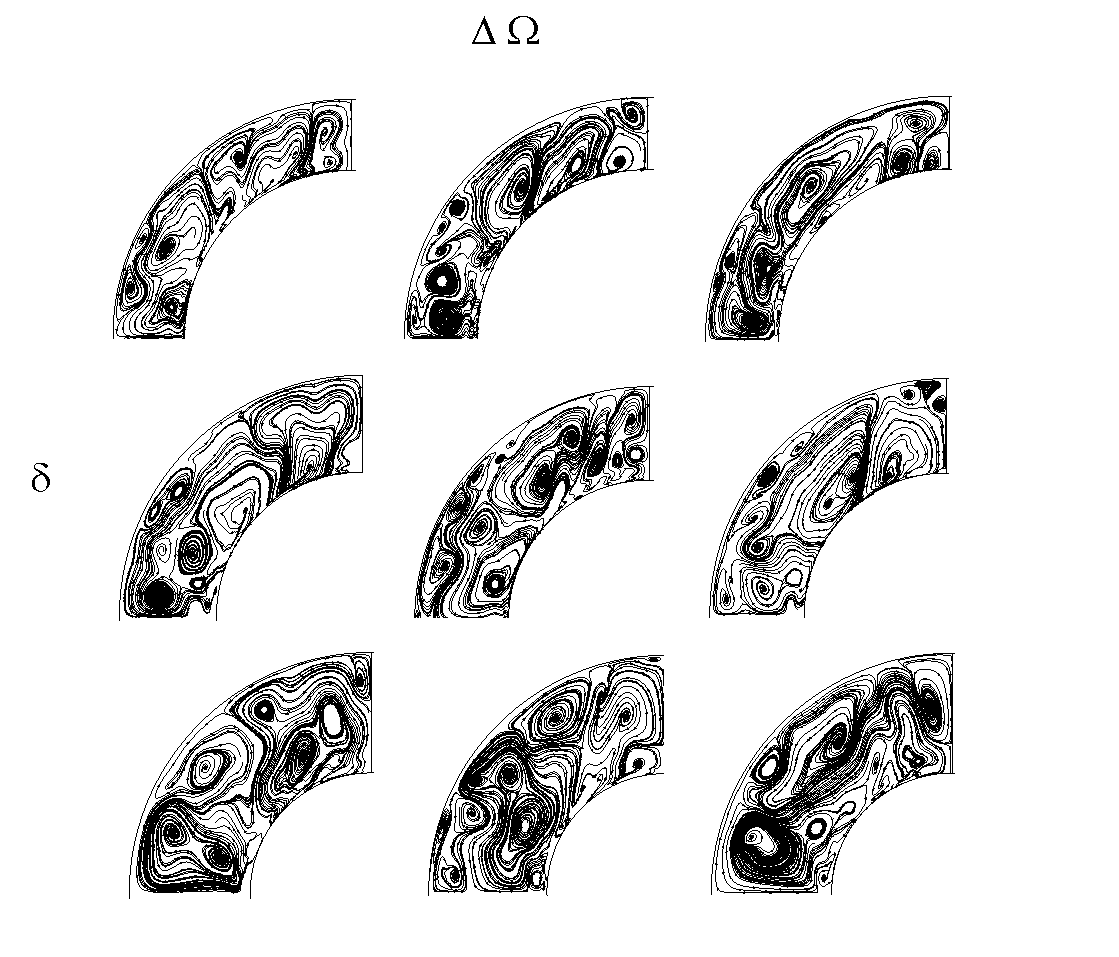}
\caption{Snapshots at $t=20$ of meridional streamlines for
the superfluid component in superfluid SCF, with 
$\Rey=3 \times 10^4$, $\delta=0.3$, $0.4$, $0.5$ (from top to bottom),
and $\Delta \Omega=0.1$, $0.2$, $0.3$ (from left to right).
The friction force if of GM form, with zero tension (${\bf T}=0$).
Spectral resolution:
$(N_r,N_\theta,N_\phi)=(120,250,4)$. Filter parameters: $(\gamma_r,\gamma_\theta)=(8,8)$.}
\label{fig:long_fig_super}
\end{center}
\end{figure}

Figures \ref{fig:long_fig_normal} and \ref{fig:long_fig_super}
show the meridional streamlines at $t=20$ and
$\Rey=3 \times 10^4$ for 
the normal and superfluid components,
with GM friction
and zero tension force. The rotational
shear $0.1 \leq \Delta \Omega \leq 0.3$ increases
from left to right; the dimensionless
gap width $ 0.3 \leq \delta \leq 0.5$ increases
from top to bottom. The flow is approaching, but has not reached,
a steady state at this time, with 
$|N_{\rm ext}^{(2)} - N_{\rm int}^{(1)}| \lsim 10^{-3}$.
The number
for equatorial and polar circulation cells remains
approximately constant as $\delta$ increases,
although the cells become progressively less
``stacked". By contrast, the flow becomes
less turbulent, and the number of cells decreases. Additionally, the
vortices are more {\emph stacked} with decreasing $\delta$.

\begin{figure}[htpb]
\begin{center}
\includegraphics[scale=0.3]{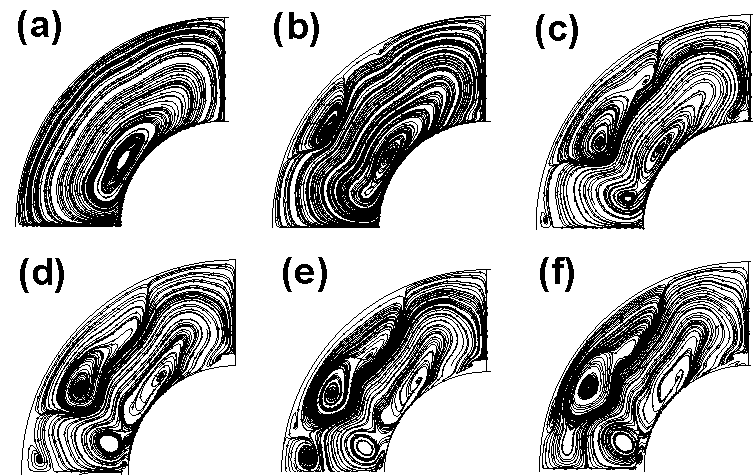}
\caption{Snapshots at $t=14$ of meridional streamlines for
the normal fluid component in superfluid SCF, with 
$\delta=0.5$, $\Delta \Omega=0.3$, and
$\Rey$ increasing from top left to bottom right. 
(a) $\Rey=100$, (b) $\Rey=300$, (c) $\Rey=1000$,
(d) $\Rey=3000$, (e)  $\Rey=10^4$, and (f)
$\Rey=3 \times 10^4$. The friction force is of
HV form; the coefficient of the tension force is $\Rey_s=10^{-5}$.
Spectral resolution:
$(N_r,N_\theta,N_\phi)=(120,250,4)$. Filter parameters: $(\gamma_r,\gamma_\theta)=(8,8)$.}
\label{fig:long_fig_normal2}
\end{center}
\end{figure}

\begin{figure}[htpb]
\begin{center}
\includegraphics[scale=0.3]{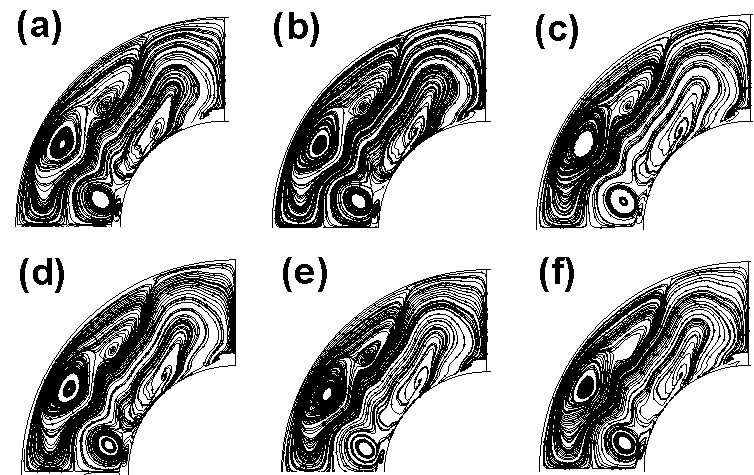}
\caption{Snapshots at $t=14$ of meridional streamlines for
the superfluid component in superfluid SCF, with 
$\delta=0.5$, $\Delta \Omega=0.3$, and
$\Rey$ increasing from top to left to bottom right. 
(a) $\Rey=100$, (b) $\Rey=300$, (c) $\Rey=1000$,
(d) $\Rey=3000$, (e)  $\Rey=10^4$, and (f)
$\Rey=3 \times 10^4$. The friction force is of
HV form; the coefficient of the tension force is $\Rey_s=10^{-5}$.
Spectral resolution:
$(N_r,N_\theta,N_\phi)=(120,250,4)$. Filter parameters: $(\gamma_r,\gamma_\theta)=(8,8)$.}
\label{fig:long_fig_super2}
\end{center}
\end{figure}

Figures \ref{fig:long_fig_normal2} and \ref{fig:long_fig_super2}
show  meridional streamlines at $t=14$ for 
the normal and superfluid components
as a function of Reynolds number, for fixed
$\delta$ and $\Delta \Omega$, HV friction,
and nonzero tension.
The streamlines of the normal fluid show
the $0$-vortex state at $\Rey=100$ 
(Figure \ref{fig:long_fig_normal2}a).
A secondary vortex develops near the outer shell
at $\Rey=300$
(Figure \ref{fig:long_fig_normal2}b), whose
size increases
with $\Rey$, elongating in the meridional
direction. Two additional vortices
form in the equatorial region at $\Rey=10^3$
(Figures \ref{fig:long_fig_normal2}c-f).
The streamlines of the superfluid component 
closely resemble the normal fluid component
for $\Rey \gsim 3 \times 10^3$ (see Figures \ref{fig:long_fig_super2}a-f).

\subsection{Unsteady torque}
\label{ssec:uns_tor}
\begin{figure}[htpb]
\begin{center}
\includegraphics[scale=0.55]{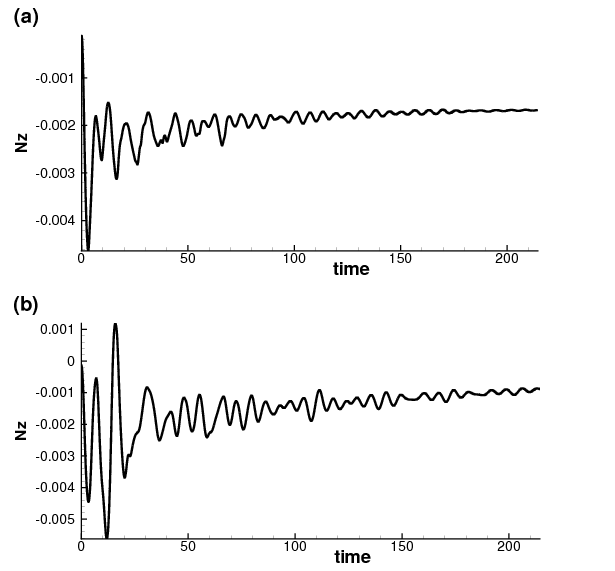}
\caption{Viscous torque exerted on the (a) inner and (b) outer
spheres as a function of time in superfluid SCF, with HV mutual friction,
$\delta=0.5$, $\Delta \Omega=0.3$, and
$\Rey=10^4$. Spectral resolution
and filter parameters are the same as in Figure \ref{fig:cap3_fig1}. Initially,
we set $\vn=\vs$ to the Stokes solution (\ref{eq:stokes}).}
\label{fig:cap3_torque1}
\end{center}
\end{figure}

The torque exerted by the normal fluid component on
the inner and outer spheres, calculated using
(\ref{eq:torque_int}), is plotted versus time in Figures
\ref{fig:cap3_torque1}a and \ref{fig:cap3_torque1}b. It oscillates,
with peak-to-peak amplitude $\sim 10^{-3}$ for
$t \leq 30$ and $\sim 10^{-5}$ for $t \geq 30$.
These oscillations, with period $\approx 2\pi /\Omega_1$,
persist as long as the differential rotation
is maintained, up to $t = 214$ in our longest 
simulation. They are observed at
all the Reynolds numbers considered in
this paper ($1 \times 10^2 \leq \Rey \leq 3 \times 10^4$).
The oscillation amplitude is greater for
HV friction; oscillations are still
observed for GM friction, but with peak-to-peak
amplitude $\sim 10^{-6}$.
In other words, superfluid SCF is intrinsically
unsteady and indeed quasiperiodic.

\subsection{Spectral resolution and filter}
\label{subsec:numissues}

\begin{figure}[htpb]
\begin{center}
\includegraphics[scale=0.5]{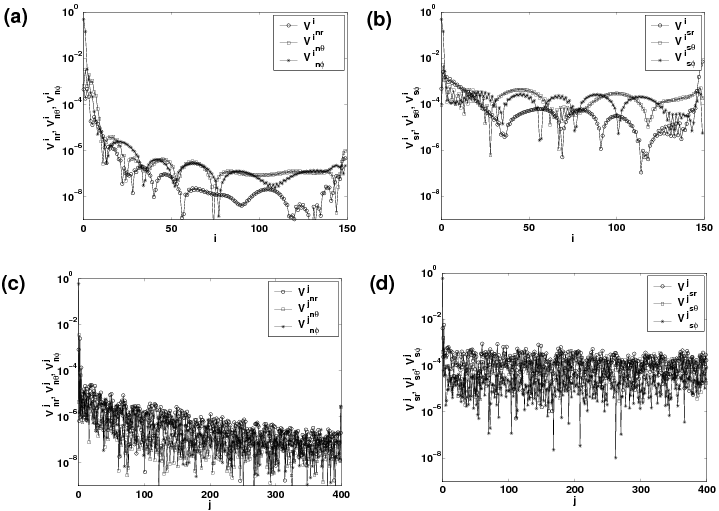}
\caption{Snapshot of spectral mode amplitudes at $t=0.3$,
as a function of the Chebyshev polynomial index
$i$ for (a) the normal velocity resolutes $(\vn)_{r}$, $(\vn)_\theta$, and
$(\vn)_\phi$, and (b) the superfluid velocity resolutes $(\vs)_r$, $(\vs)_\theta$,
and $(\vs)_\phi$; and 
as a function of the sine (cosine) polynomial index
$j$ for (c) the normal velocity resolutes $(\vn)_{r}$, $(\vn)_\theta$, and $(\vn)_\phi$, and (d) the superfluid velocity resolutes
$(\vs)_r$, $(\vs)_\theta$, and $(\vs)_\phi$.
The grid resolution is $(N_r,N_\theta,N_\phi)=
(150,400,4)$ and the exponential filters are switched off.
Simulation parameters: HV mutual friction, $\Rey=178$, $\delta=0.5$,
and $\Delta \Omega = 0.1$.}
\label{fig:cap3_unres1}
\end{center}
\end{figure}
\begin{figure}[htpb]
\begin{center}
\includegraphics[scale=0.5]{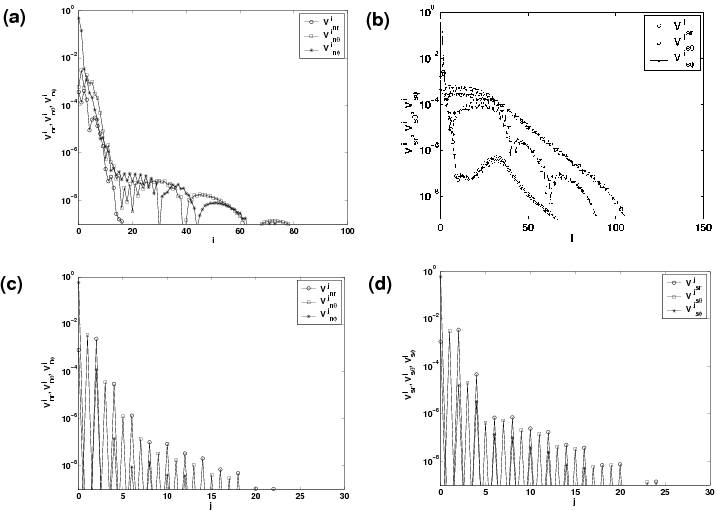}
\caption{Snapshot of mode amplitudes at $t=0.3$ as a function of the Chebyshev
polynomial index
$i$ for (a) $(\vn)_{r}$, $(\vn)_\theta$, and
$(\vn)_\phi$, and (b) $(\vs)_r$, $(\vs)_\theta$,
and $(\vs)_\phi$. Snapshot of mode
amplitudes at $t=0.3$ as a function of the sine (cosine) polynomial index
$j$ for (c)  $(\vn)_{r}$, $(\vn)_\theta$, and $(\vn)_\phi$, and (d)
$(\vs)_r$, $(\vs)_\theta$, and $(\vs)_\phi$.
The grid resolution is $(N_r,N_\theta,N_\phi)=
(150,400,4)$. The exponential filter has $(\gamma_r,\gamma_\theta)=(8,6)$.
Simulation parameters: HV mutual friction,
$\Rey=178$, $\delta=0.5$,
and $\Delta \Omega = 0.1$.}
\label{fig:cap3_unres3}
\end{center}
\end{figure}

In a well resolved simulation, the Chebyshev and Fourier
mode amplitudes decrease monotonically (on average) with
polynomial index. We prefer to maintain an amplitude
ratio of at least $10^4$ between the strongest and
weakest modes. \citet{giacobello05} found empirically
that this is sufficient to fully resolve vortical
structures in unsteady, three-dimensional transients
excited by the flow past a stationary and rotating sphere in
a classical, viscous Navier--Stokes fluid.
In this section, which is devoted
to axisymmetric flows, we are interested in the Chebyshev ($r$)
amplitudes $v^{i}_{nr,sr}$, $v^{i}_{n\theta,s\theta}$,
and $v^{i}_{n\phi,s\phi}$,
and Fourier ($\theta$) amplitudes
$v^{j}_{nr,sr}$, $v^{j}_{n\theta,s\theta}$, and $v^{j}_{n\phi,s\phi}$.
These coefficients do not correspond to $\beta_{ljk}$,
$\gamma_{ljk}$, and $\delta_{ljk}$ in equations
(\ref{eq:expansion_ur})--(\ref{eq:expansion_uphi});
$v^{i}_{nr,sr}$, $v^{i}_{n\theta,s\theta}$,
and $v^{i}_{n\phi,s\phi}$ are calculated
by transforming from
coordinate to wavenumber space in the $r$ direction,
and $v^{j}_{nr,sr}$, $v^{j}_{n\theta,s\theta}$, and $v^{j}_{n\phi,s\phi}$
are calculated by transforming from
coordinate to wavenumber space in the $\theta$ direction.

We start by comparing the mode amplitudes for a poorly
resolved and a well resolved simulation.
Figure \ref{fig:cap3_unres1} shows an example of
a poorly resolved simulation, with $\Rey=178$, $\delta=0.5$,
$\Delta \Omega = 0.1$, and $(N_r,N_\theta,N_\phi)=
(150,400,4)$.
The spectral filter is switched off. 
The mode amplitudes of the normal velocity components
decrease from $\sim 10^{-1}$ ($i=1$) to $\sim 10^{-6}$ ($i=150$),
which superficially suggests that the flow is well resolved. However,
the mode amplitudes of the superfluid velocity components
are roughly constant ($\sim 10^{-4}$) across all
Chebyshev ($1 \leq i \leq 150$) and Fourier ($1 \leq j \leq 400$) orders,
indicating that the run is poorly resolved.
Figure \ref{fig:cap3_unres3} shows a well resolved
simulation for the same parameters. The improvement
is achieved by switching on the spectral filters, and the
extent of the improvement depends on the strength of the filters,
with $\gamma_r=8$ and $\gamma_\theta=6$ in
Figure \ref{fig:cap3_unres3}. The Chebyshev
amplitudes of $\vn$ decrease by $\sim 8$ orders of magnitude
over the index range $1 \leq i \leq 25$. 
The Chebyshev amplitudes of $\vs$ decrease more gradually, by
$\sim 8$ orders of magnitude over the range
$1 \leq i \leq 100$. The Fourier amplitudes of $\vn$ and $\vs$
decay similarly (Figures  \ref{fig:cap3_unres3}c--d), with
only $N_\theta \approx 20$ required to resolve the flow
properly. For a weaker filter,
with $\gamma_r = \gamma_\theta=12$, the mode amplitudes
are unchanged to within $\sim 10$ \%, but $N_r \approx 20$
and $N_\theta \approx 100$ Chebyshev and Fourier modes are
required.

What happens in general when the exponential filter
is either switched off, as in Figure \ref{fig:cap3_unres1},
or maintained at a weak level ($\gamma_{r},\gamma_{\theta} \gsim 20$)?
For $\Rey \gsim 10^3$, we find that the evolution is stable for a short time
($t \lsim 10$), after which
$\vn$ and $\vs$ become unphysically large
and the numerical simulation breaks down. 
For $\Rey \sim 10^2$,
the evolution is stable for longer, provided
that perfect slip boundary conditions are applied to
$\vs$. Indeed, generally speaking, $\vs$ evolves less stably
for no-slip boundary conditions, which
promote the generation of superfluid vorticity.
Nevertheless, for a range
of SCF parameters, we observe that $\vn$ for a filtered
HVBK superfluid agrees well with $\vv$ for an unfiltered
Navier--Stokes fluid given identical boundary conditions
(see Section \ref{subsec:visco_super}), engendering
confidence that filtering does not cause
artificial behaviour.

\subsection{Effect of superfluid component}
\label{subsec:visco_super}
Laboratory experiments on the acceleration and deceleration
of He II in spherical vessels reveal a variety
of unsteady behaviour, e.g. sudden jumps and
quasiperiodic oscillations in angular velocity
\citep{tsatsa80}. It is not known what aspects
of this unsteady behaviour are caused by the nonlinear
hydrodynamics of the viscous normal component
of He II, or by the build-up of vorticity in the
inviscid superfluid component. We now explore this question.

In order to isolate how the superfluid component
influences the global dynamics of superfluid
SCF, we compare three particular cases: (i) a
one-component, viscous, Navier--Stokes fluid,
with Reynolds number $\Rey=10^4$; (ii) a one-component, inviscid,
HVBK superfluid, with ${\bf F} = {\bf T} = 0$
in (\ref{eq:hvbk1}) and (\ref{eq:hvbk2}); and
(iii) a two-component, HVBK superfluid, whose
normal component ($\Rey=10^4$) is coupled
to the superfluid component through HV
\citep{hv56a,hv56b} and GM \citep{gm49,donnelly91}
mutual friction (${\bf F} \neq 0$, ${\bf T} \neq 0$). To
make the comparison, we fix $\delta=0.5$ and $\Delta \Omega=0.3$.

\begin{figure}[htpb]
\begin{center}
\includegraphics[scale=0.3]{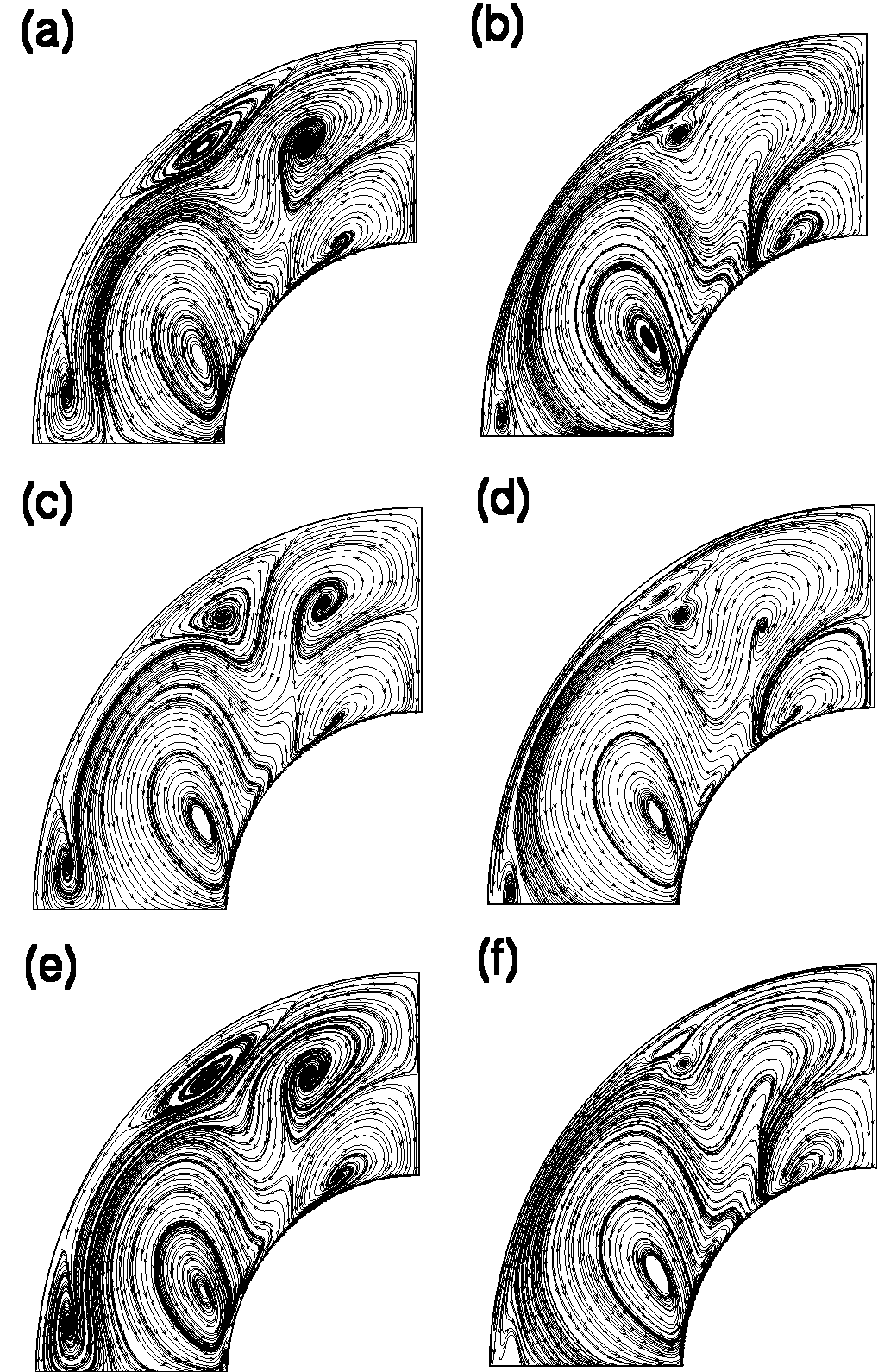}
\caption{Meridional streamlines in SCF for (a) viscous Navier--Stokes fluid
with $\Rey=10^4$, (b) pure one-component inviscid superfluid, (c) normal fluid component
of an HVBK superfluid with HV friction, (d) superfluid component
of an HVBK superfluid with HV friction,
(e) normal fluid component with GM friction,
and (f) superfluid component with GM friction.
All snapshots are taken at $t=6$. SCF parameters: $\delta=0.5$ and $\Delta \Omega=0.3$.}
\label{cap3_fig:comp_visc_sup2}
\end{center}
\end{figure}

Let us begin with case (i): a viscous, Navier--Stokes fluid.
Meridional streamlines, obtained by integrating the in-plane
components of the velocity field in the $x=0$ plane,
are displayed in
Figure \ref{cap3_fig:comp_visc_sup2}a, at time $t=6$.
The flow is
complicated, featuring three primary circulation cells:
one near the equator, and two near the poles 
(each about half the diameter of the equatorial 
cell). Two secondary, flattened vortices reside near
the outer boundary, one just above the equator and
the other centred at $r \approx 0.9$,
$\theta \approx 30$\degree. These structures develop from
a low-$\Rey$ Stokes flow at $t=0$.

Let us repeat the experiment with an HVBK superfluid
in which the normal and superfluid components
are completely {\it uncoupled}, i.e. the mutual friction
and tension are switched off (${\bf F} = {\bf T} = 0$).
This is case (ii).
Naturally, the normal component evolves exactly
like the classical Navier--Stokes fluid in
Figure \ref{cap3_fig:comp_visc_sup2}. As for
the superfluid, its meridional streamlines
at time $t=6$ are shown in Figure \ref{cap3_fig:comp_visc_sup2}b.
On large scales, the flow pattern resembles
Figure \ref{cap3_fig:comp_visc_sup2}a, with the
same number (three primary plus two secondary) of
circulation cells. However, the cells
have different shapes and diameters: the secondary
cells are smaller and less flattened than
in Figure \ref{cap3_fig:comp_visc_sup2}a, and
the volume of the largest (equatorial) primary
cell is three times greater than in
Figure \ref{cap3_fig:comp_visc_sup2}a.

Note that equations ($1.16$) and ($1.17$) evolve
independently when the coupling forces (${\bf T}$ and ${\bf F}$)
are switched off. The superfluid
equation of motion ($1.17$) reduces to the Euler equation for an ideal (inviscid)
fluid, so only
one boundary condition is required:
no penetration, $({\bf v}_s)_r=0$.
The same is true for HV and GM mutual friction:
the two forms of the friction force imply two
different orders of the system of equations
and hence two different sets of boundary conditions.
However, there are three reasons why we do not
treat HV and GM friction differently in our simulations:

\begin{enumerate}

\item The correct boundary conditions for the superfluid
are unknown. What ultimately determines the
boundary conditions for $\vs$ is
the distribution of quantized vorticity in the superfluid component.
In cylindrical containers, it is natural to assume
that the quantized vortices are arranged in a regular array
parallel to the rotation axis, if the rotation
is slow. Under these conditions, the numerical
evolution is stable if the vortex lines are parallel to the curved wall.
In spherical containers, the vortex lines are neither
perpendicular to the walls nor
parallel to the rotation axis everywhere.
In this paper, we test what boundary
conditions give the most stable evolution. Often,
no slip in the superfluid component works best,
even if it is not strictly mathematically correct for the GM force.
Physically, we justify this by assuming that vortices
pin to the boundaries, whereupon
they move at the same speed.
In other words, no slip corresponds
to pinning at the boundaries; since we ignore the presence
of the vortices in the fluid interior
(${\bf T}={\bf F}=0$ at  $R_1 < r < R_2$) but
not at the boundaries.

\item When we attempt to solve the the HVBK
equations numerically with ${\bf T} = 0$ in the HV friction force,
or with GM friction, using only
one boundary condition on the superfluid component (no penetration),
the results do not differ significantly from
the results presented in this paper.
For $\delta=0.5$, $\Delta \Omega = 0.3$, at $t=4$,
the outer torque differs by $10^{-6}$ between the two approaches.
One cannot see any significant difference in
the streamlines of both fluids, whether one uses the strictly
mathematically correct
set of boundary conditions for the superfluid (no penetration only)
or the simulations presented in this paper
(no slip in the superfluid component).

\item  The tension of the vortex lines, controlled
by the stiffness parameter $\nu_s$, can disrupt the flow and
destabilize its evolution. A high value
of $\nu_s$ ($\Rey_s \lsim 10^{4}$) stiffens the vortex array
against the drag of the
normal fluid.
Previous attempts to solve the HVBK equations in spherical geometries
foundered partly because of these issues.
By enforcing no-slip boundary conditions on the superfluid,
one ensures that the superfluid corotates with
the normal fluid at the boundaries,
reducing the friction force in that region.
As \citet{hbj95}  suggested (page 333, second last paragraph),
the appearance of a nonlinear boundary layer
may not be completely resolved by our code, or
indeed by any code to date.
Note also that $\nu_s$ is small ($\sim 10^{-5}$), so the influence
of the tension term in the simulations
presented in this paper is not strong.
We plan to study the effects of the tension force
in a future paper, when we are able to obtain
a better understanding of the nonlinear boundary
layer.
\end{enumerate}

When the mutual friction and tension forces are
switched on, as in case (iii),
the vortical
structures near the poles change. Consider first
what happens when ${\bf F}$ is of HV form.
For the normal
fluid component, displayed in
Figure \ref{cap3_fig:comp_visc_sup2}c, the flattened
vortex near the pole widens radially to about twice
the size of the same vortex with ${\bf F}=0$, while its latitudinal
extent remains unchanged.
The vortex near the equator halves in size,
compared to its counterpart in Figure \ref{cap3_fig:comp_visc_sup2}a.
The only appreciable changes in the superfluid
flow pattern, displayed in
Figure \ref{cap3_fig:comp_visc_sup2}d, are the
following: (i) the small vortex  near
the pole widens radially by $\sim 10$ \%; and
(ii) the larger circulation cell
near the pole comes to resemble the same cell
in the normal
component more closely. 

By contrast, when ${\bf F}$
is of GM form, the streamlines
of the normal fluid are very similar
to a classical viscous fluid at the same
Reynolds number, as we can appreciate
by comparing Figures \ref{cap3_fig:comp_visc_sup2}a
and \ref{cap3_fig:comp_visc_sup2}e. 
The superfluid component closely resembles
an uncoupled superfluid
(c.f. Figures \ref{cap3_fig:comp_visc_sup2}b
and \ref{cap3_fig:comp_visc_sup2}f). The
two components are loosely coupled because
GM friction is weaker than HV friction
[$|\Efe_{\rm HV} /\Efe_{\rm GM}| \sim 10^{5}$;
see \citet{pmgo05a,pmgo06b}].

\begin{figure}[htpb]
\begin{center}
\includegraphics[scale=0.5]{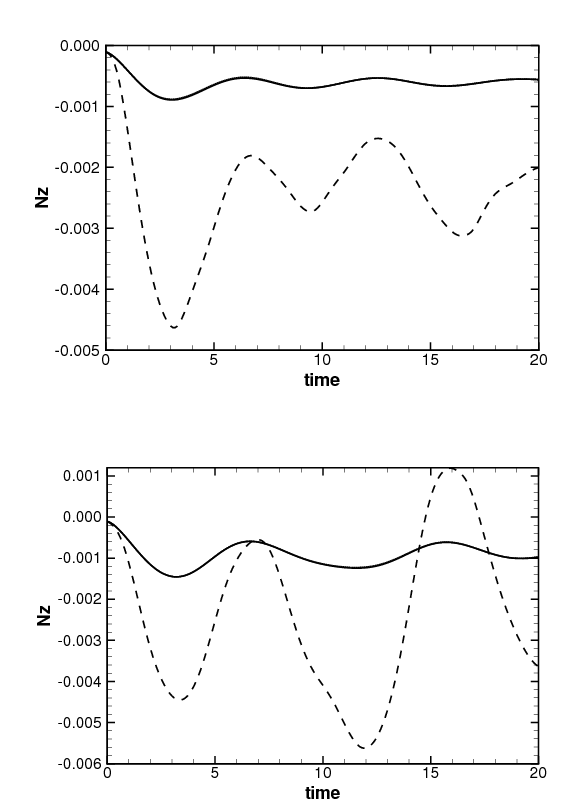}
\caption{Torque on the inner (upper plot)
and outer (lower plot) spheres, for a viscous fluid (solid curve),
a superfluid with HV mutual friction force (dashed curve), and
a superfluid with GM mutual friction (dotted curve). In all cases,
we take $\Rey=10^4$, $\delta=0.5$, and $\Delta \Omega=0.3$. The
solid and dotted curves differ by one part in $10^{4}$, and are
almost indistinguishable on the scale of the plot.}
\label{cap3_fig:comp_visc_sup1}
\end{center}
\end{figure}

The evolution of the torque on the inner and outer spheres
is plotted  in Figure \ref{cap3_fig:comp_visc_sup1} for
cases (i) and (iii). [The torque is zero in case (ii),
where the fluid is completely inviscid.] The solid
curve corresponds to a viscous, Navier--Stokes fluid,
while the dashed and dotted curves correspond to an HVBK
superfluid with HV and GM mutual friction respectively. Note
that, in an HVBK superfluid, the
torque is still exerted by viscous stresses in
the normal component. However, its magnitude is
modified by the presence of the superfluid component,
because the mutual friction modifies $\vn$
and hence $\partial (\vn)_i/\partial x_j$.

\begin{figure}[htpb]
\begin{center}
\includegraphics[scale=0.3]{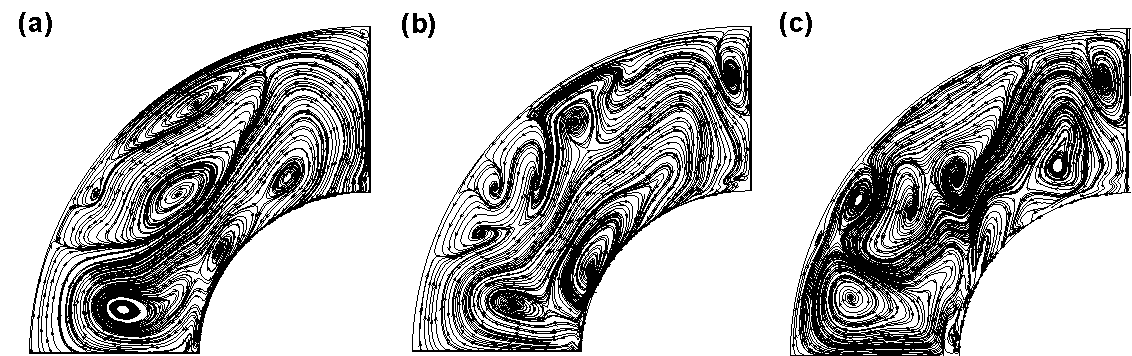}
\caption{Meridional streamlines in SCF for (a) viscous Navier--Stokes fluid,
(b) normal fluid component,
with HV mutual friction, and (c) superfluid component with HV mutual friction,
with $\Rey=10^4$. All snapshots are taken at $t=20$. SCF parameters:
$\delta=0.5$ and $\Delta \Omega = 0.3$.}
\label{cap3_fig:comp_visc_sup3}
\end{center}
\end{figure}

\begin{figure}[htpb]
\begin{center}
\includegraphics[scale=0.5]{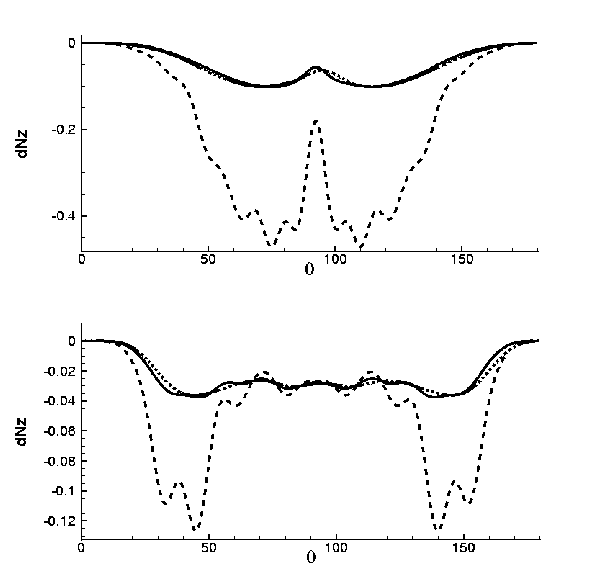}
\caption{Differential torque $dN_z/d(\cos \theta)$ as a function of colatitude
$\theta$ for the inner (upper plot)
and outer (lower plot) spheres, for a viscous Navier--Stokes fluid (solid curve),
a superfluid with HV mutual friction (dashed curve), and
a superfluid with GM mutual friction (dotted curve), at time $t=20$.
SCF parameters: $\Rey=10^4$, $\delta=0.5$, and $\Delta \Omega=0.3$.}
\label{cap3_fig:comp_visc_sup4}
\end{center}
\end{figure}

The torque exerted by the normal component increases
roughly thrice relative to case (i) when the superfluid component is included
with HV mutual friction. By
contrast, the torques exerted by a Navier--Stokes fluid
and a HVBK superfluid with GM
mutual friction differ by $6$ \%, which is barely
distinguishable on the scale of Figure \ref{cap3_fig:comp_visc_sup1}.
To understand this effect quantitatively,
consider the streamline snapshots of the Navier--Stokes
fluid, viscous HVBK  component, and inviscid HVBK
component at $t=20$, shown
in Figures \ref{cap3_fig:comp_visc_sup3}a--\ref{cap3_fig:comp_visc_sup3}c.
There are four circulation cells near the outer boundary
in the Navier--Stokes fluid and six in the normal
HVBK component. The magnitude of the torque
increases with the number of circulation cells,
because more circulation cells imply steeper
radial velocity gradients.
We observe this in the quantity $dN_{1,2}/d(\cos \theta)$,
which measures the differential contribution to the
torque as a function of colatitude
and is plotted in Figure \ref{cap3_fig:comp_visc_sup4}
for the inner and outer spheres. For example, we find
$|dN_1/d(\cos \theta)| <  0.1$ for the Navier--Stokes
fluid, but $|dN_1/d(\cos \theta)|$ is as large
as $0.5$ (at $\theta \approx 75 \degree$, $110 \degree$)
for the HVBK superfluid with HV mutual friction.
From equation (\ref{eq:torque_int}), we see that
$N_1$ and $N_2$
in the HVBK superfluid are greater due to larger
contributions from the first term in (\ref{eq:torque_int}), viz.
$\partial v_{n \phi}/\partial r$.
For example, at $r=R_2$ and $\theta \approx 45 \degree$,
we find $\partial v_{\phi}/{\partial r} \approx -0.73$
for the Navier--Stokes fluid and
$(\partial v_{n \phi}/{\partial r}) \approx -3.46$
for the normal component of the HVBK superfluid, 
whereas the second term in (\ref{eq:torque_int})
is similar in both systems, viz.
$v_{ n \phi}/r \approx v_\phi/r \approx 0.49$.

Behaviour of the sort just described was
predicted by \citet{hb95}, who solved the
HVBK equations numerically
inside infinitely long, differentially rotating cylinders.
They too observed that, as $\Rey$ increases,
the tension force diminishes, and the
friction force dominates.
Inside the circulation cells of the normal fluid, the
ratio $|{\bf T}|/|{\bf F}_{\rm HV}|$ decreases with
increasing $\Rey$.
\citet{hb95} suggested that,
at higher $\Rey$, the mutual friction
locks together $\vn$ and $\vs$, so that the
streamlines of the superfluid
and normal fluid are similar.

\begin{figure}[htpb]
\begin{center}
\includegraphics[scale=0.24]{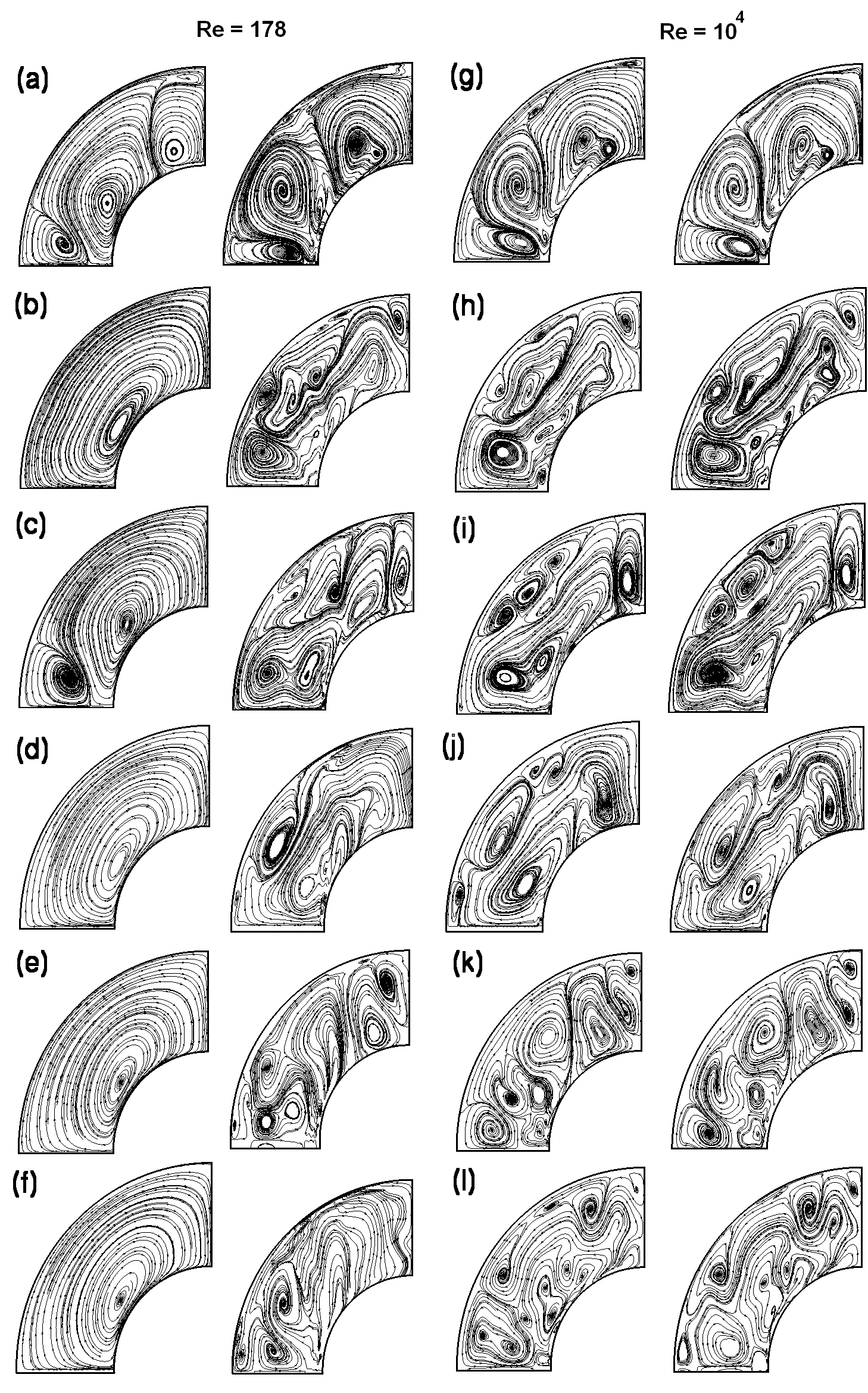}
\caption{Meridional streamlines in superfluid SCF
at low and high Reynolds numbers. Left half of figure:
$\Rey=178$,
at times  (a) $t=10$, (b) $t=20$, (c) $t=30$, (d) $t=40$, (e) $t=50$, and (f) $t=60$.
Right half of figure: $\Rey=10^4$, at times (g) $t=10$, (h) $t=20$, (i) $t=30$,
(j) $t=40$, (k) $t=50$, and (l) $t=60$. In each snapshot, the left-hand and right-hand
quadrants display the normal and superfluid components respectively.
SCF parameters: $\delta=0.5$, $\Delta \Omega=0.3$,
and $\Rey_s=10^5$.}
\label{cap3_fig:fig1}
\end{center}
\end{figure}

\begin{figure}[htpb]
\begin{center}
\includegraphics[scale=0.25]{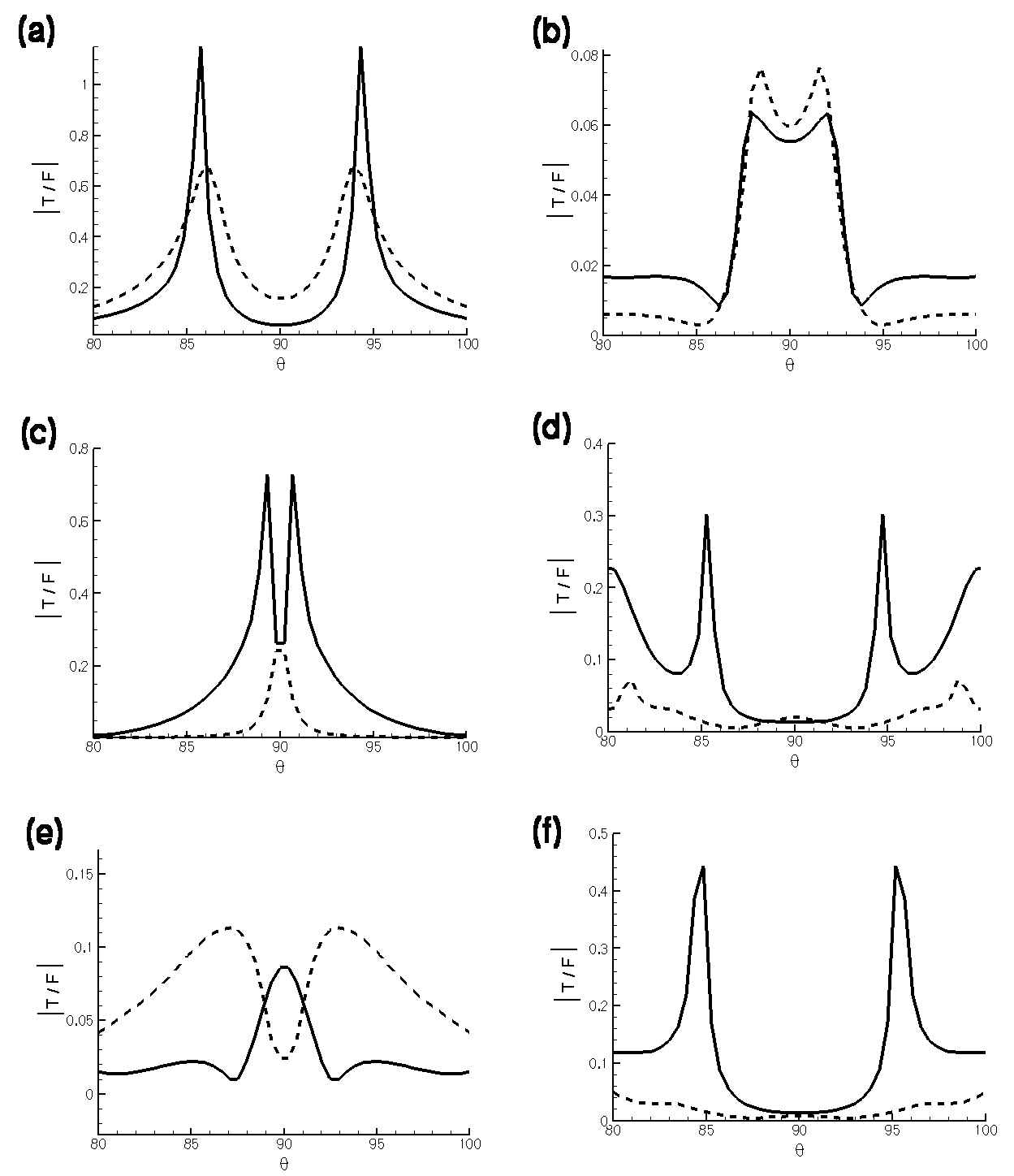}
\caption{Ratio of the tension force
 $({\bf T})$ to the HV mutual friction force $({\bf F}_{\rm HV})$ as
a function of
colatitude $\theta$ (in \degree), at the boundary
of the outer sphere, at times (a) $t=10$, (b) $t=20$, (c) $t=30$, (d) $t=40$, (e) $t=50$, and (f) $t=60$.
SCF parameters: $\delta=0.5$, $\Delta \Omega=0.3$,
$\Rey=178$ (solid curve), and $\Rey=10^4$ (dashed curve). Generally speaking,
$|{\bf T}|/|{\bf F}_{\rm HV}|$ decreases with $\Rey$ and $t$.}
\label{cap3_fig:tf_comp}
\end{center}
\end{figure}

In order to quantify how the
two-fluid coupling changes with $\Rey$,
consider a HVBK superfluid
with the same parameters as in
Figure \ref{cap3_fig:comp_visc_sup3}b
and \ref{cap3_fig:comp_visc_sup3}c, but with
$\Rey=178$ instead of $\Rey=10^4$. Figure \ref{cap3_fig:fig1}
compares a sequence of snapshots of the meridional
streamlines for $\Rey=178$
and $10^4$ at times $ 10 \leq t \leq 60$.
For $\Rey=178$, the normal component differs
markedly from the superfluid component, except
during the early stages of the evolution ($t \leq 10$).
Eventually, at $t \geq 30$, the normal fluid settles
down into a permanent $0$-vortex state, while
the superfluid develops $3$--$4$ vortices
near the equator. For $\Rey=10^4$, on the other hand,
the normal and superfluid components display
similar flow patterns, with the same number of vortices in
approximately the same locations. This occurs because
the HV mutual friction progressively dominates the
vortex tension as $\Rey$ increases and also as time passes.
Quantitative evidence is presented in
Figure \ref{cap3_fig:tf_comp},  where
$|{\bf T}|/|{\bf F}_{\rm HV}|$ is plotted as a function
of colatitude in the equatorial region
$80 \degree \leq \theta \leq 100 \degree$,
at the boundary of the outer sphere ($r=R_2$).
However, this initial transient soon disappears,
and the inequality reverses.
At $t=10$, we find $(|{\bf T}|/|{\bf F}_{\rm HV}|)_{\Rey=178}
< (|{\bf T}|/|{\bf F}_{\rm HV}|)_{\Rey=10^4}$ except
at $\theta \approx 85 \degree$ and $\theta \approx 95 \degree$
(see Figure \ref{cap3_fig:tf_comp}a).
At $t=20$, we find $(|{\bf T}|/|{\bf F}_{\rm HV}|)_{\Rey=178}
> (|{\bf T}|/|{\bf F}_{\rm HV}|)_{\Rey=10^4}$, except
in the narrow region
$87 \degree \lsim \theta \lsim 93 \degree$.
For $t \geq 30$, we find
$(|{\bf T}|/|{\bf F}_{\rm HV}|)_{\Rey=178} > (|{\bf T}|/|{\bf F}_{\rm HV}|)_{\Rey=10^4}$
throughout the equatorial region
$80 \degree \leq \theta \leq 100 \degree$
(except for a brief reversal at $t \approx 50$).
Thus, at low Reynolds
numbers ($\Rey \lsim 10^3$) and at early times, the tension force dominates
the mutual friction throughout most of the fluid,
while the opposite is true at high Reynolds numbers
and late times.
The stiffness of the superfluid
vortex array, encoded in the tension force, prevents
$\vs$ from following $\vn$, whereas, when
the mutual friction dominates, $\vs$
copies $\vn$ more closely.

As the tension is less
important at higher $\Rey$, one expects
the superfluid component to influence
the overall dynamics less as $\Rey$ increases.
This is reflected in the torque.
For a viscous fluid at $\Rey=178$, the torque
is half that for
a superfluid at the same Reynolds number. For
a superfluid with $\Rey=10^4$, the torque
doubles when compared with a
viscous fluid with the same Reynolds number.
In Section \ref{subsec:bc_effect},
we quantify how the boundary condition
on the superfluid component (indirectly) affects the torque.

\subsection{Effect of the boundary conditions}
\label{subsec:bc_effect}
The streamlines 
of the normal and superfluid components resemble each other ever
more closely as $\Rey$ increases, suggesting that the frictional
coupling is responsible, as argued in Section \ref{subsec:visco_super}.
Nevertheless, it is important to check how much of the similarity
arises from imposing the no-slip boundary condition on the
superfluid component at $r=R_1$, $R_2$, which in turn imposes
zero counterflow ($\vn = \vs$) at 
$r=R_1$, $R_2$. This matters, because it can be argued that the no-slip
condition is artificial. The physically correct boundary
conditions on $\vs$ are still uncertain, lying somewhere between
the following two extremes:
(i) quantized vortex lines slide freely along the surface, thereby
terminating perpendicular to it, as expressed by (\ref{eq:bcs22});
or (ii) quantized vortices are pinned
to the surface, so that $\vs$ does not slip relative to the surface,
as expressed by (\ref{eq:bcs21}) \citep{k65}.

To clarify these matters, we repeat two of the no-slip simulations
in Section \ref{subsec:visco_super} (with $\Rey=178$ and $\Rey=10^4$, as
well as $\delta=0.5$ and $\Delta \Omega = 0.3$) such that
the superfluid satisfies the perfect-slip
boundary condition (\ref{eq:bcs21}) instead of the no-slip
condition (\ref{eq:bcs22}). Perfect slip
implies that the vortex lines terminate perpendicular to the surfaces
of the inner and outer spheres.
The normal fluid satisfies the no-slip condition
(\ref{eq:bcvn}). Both $\vn$ and $\vs$ are initialized to
the Stokes solution (\ref{eq:stokes}).

\begin{figure}[htpb]
\begin{center}
\includegraphics[scale=0.24]{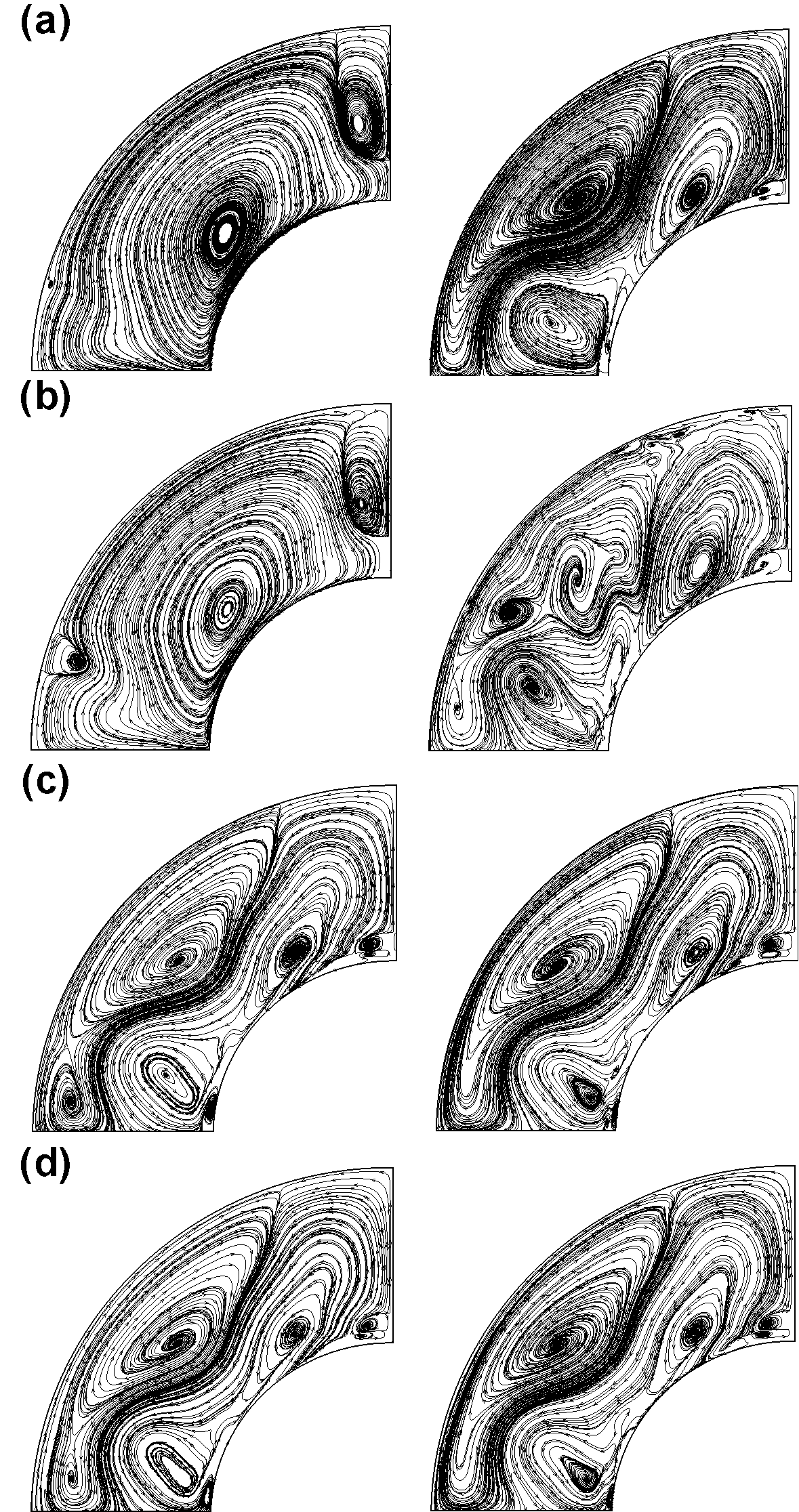}
\caption{Meridional streamlines in superfluid SCF for the normal (left panels)
and superfluid (right panels) components,
at time $t=18$, for  $\delta = 0.5$, $\Delta \Omega =0.3$, and $\Rey_s = 10^5$.
The panels illustrate the effect of boundary conditions.
(a) $\Rey=178$, perfect-slip boundary condition (\ref{eq:bcs21})
on $\vs$; (b) $\Rey=178$, no-slip boundary condition (\ref{eq:bcs22}) on $\vs$;
(c) $\Rey=10^4$, perfect-slip boundary condition (\ref{eq:bcs21}) ;
and (d)  $\Rey=10^4$, no-slip boundary condition.}
\label{cap3_fig:bc2}
\end{center}
\end{figure}

Figure \ref{cap3_fig:bc2} compares the results for
no slip and perfect slip at $t=18$. For both low
($\Rey=178$) and high ($\Rey=10^4$) Reynolds numbers, the
large-scale structure of the flow is the same under
both kinds of boundary conditions. For example,
in both Figures \ref{cap3_fig:bc2}a (perfect slip) and \ref{cap3_fig:bc2}b
(no slip), $\vs$ exhibits three large circulation cells: one near
the pole, centered near the inner
sphere, at $\theta \approx 30$ and $r\approx 0.6$;
one at mid-latitudes, centered near the outer sphere, at
$\theta \approx 45 \degree$ and $r \approx 0.8$; and
one at the equator, whose
diamater is half that of the polar cell. Similar structures
are also observed at $\Rey=10^4$ in Figures
\ref{cap3_fig:bc2}c (perfect slip) and
\ref{cap3_fig:bc2}d (no slip). On the other hand, the
detailed internal structure of the cells does depend
on the type of boundary condition employed, expecially
at lower $\Rey$. For example,
when there is no slip, the centers of the circulation
cells develop additional small
vortices, and
the streamlines at $r=R_1$, $R_2$ become jagged. 

In conclusion, therefore, the global resemblance of the
$\vn$ and $\vs$ streamlines at high $\Rey$ found in Section \ref{subsec:visco_super}
is not an artifact of imposing no slip on $\vs$ at the boundaries.
It is observed equally when perfect slip is allowed. Indeed, the
choice of boundary conditions affects the $\vs$ flow pattern
only as far as the small-scale structure in the cell cores
is concerned. (Of course, the $\vn$ streamlines are
almost independent of the boundary condition used
for the superfluid.) If HV mutual friction is replaced by
(weaker) GM mutual friction, the resemblance lessens, with $\vn$
tending to look like a Navier--Stokes flow (Figure \ref{cap3_fig:comp_visc_sup2}a)
and $\vs$ tending to look like an uncoupled
superfluid (Figure \ref{cap3_fig:comp_visc_sup2}b).

\begin{figure}[htpb]
\begin{center}
\includegraphics[scale=0.45]{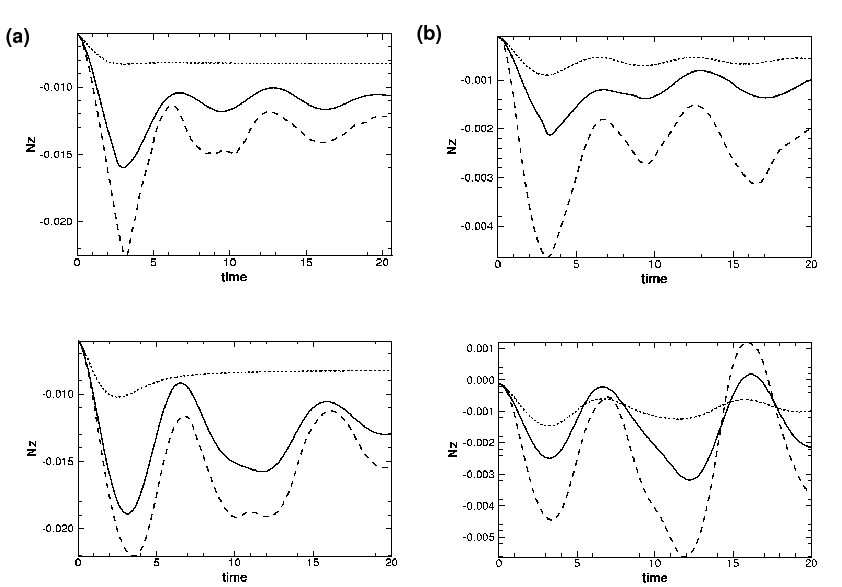}
\caption{Evolution of the inner torque (upper plots) and outer torque
(lower plots) for superfluid SCF with $\delta=0.5$, $\Delta \Omega=0.3$,
and Reynolds
number (a) $\Rey=178$ and (b) $\Rey=10^4$. The solid, dashed,
and dotted curves
correspond to an HVBK superfluid with perfect-slip boundary
condition on $\vs$, an HVBK superfluid with no-slip boundary condition on $\vs$, 
and a viscous Navier--Stokes fluid, respectively.}
\label{fig:cap3_compare_torques_slip_noslip}
\end{center}
\end{figure}

The torque on the outer and inner spheres is plotted versus time
in Figures \ref{fig:cap3_compare_torques_slip_noslip}a--\ref{fig:cap3_compare_torques_slip_noslip}b,
for perfect slip (solid curve) and no slip (dashed curve).
A viscous Navier--Stokes fluid is also plotted
for comparison (dotted curve). The perfect-slip
boundary condition on $\vs$ roughly halves
the amplitude of the torque compared to the no-slip boundary
condition.

\section{Nonaxisymmetric spherical Couette flow}
\label{sec:nonaxi}
We take advantage of the three-dimensional
capabilities of our numerical solver to investigate two
systems that exhibit nonaxisymmetric flow
(requiring spectral resolution $N_\phi \gsim 12$):
(i) a spherical, differentially
rotating shell in which the rotation axes of the inner
and outer spheres are mutually inclined;
and (ii) a spherical, differentially rotating shell in which
the outer sphere precesses freely, while
the inner sphere rotates uniformly or is at rest.
These systems have never been studied before.
We use standard vortex identification methods, introduced
by \citet{cpc90} in viscous flows, to
fully characterize the three-dimensional vortex structures
we encounter --- the first time this has been done for
an HVBK superfluid.
One incidental outcome of the work is to confirm that our
numerical method can resolve fine structures
in superfluid flow.

\subsection{Characterizing the flow topology}
\label{cap3:topology}
To understand the topology of a three-dimensional flow, one must classify
the vortices it contains. In its simplest guise, a vortex coincides
with a local pressure minimum, where the centrifugal acceleration balances
the radial pressure gradient.
However, this  criterion fails
in situations where unsteady strains or viscous effects
(at low Reynolds numbers) balance the centrifugal
force \citep{cpc90,jh95}. A second simplistic approach is
to use $|\om| = |\nabla \times \vv|$
to identify local concentrations of vorticity \citep{mmh85}. However, this
method can lead to confusion in wall-bounded flows, like ours, where
the vorticity produced by a wall-driven background shear can mask
the main vortical structures of the flow, and one
needs to know the location of the vortex core
beforehand \citep{r91}.

Several sophisticated alternatives to the simple pressure minimum and
vorticity magnitude tests have been developed. Most
are based on invariants of the
velocity gradient tensor \citep{hrm88}.
We describe two tests below:
the discriminant definition of
a vortex, which is insensitive to numerical error, especially
in Couette geometries \citep{fsg05}; and the
$\lambda_2$ definition, which is suited better to
open geometries, e.g.
the flow past a rotating sphere
\citep{giacobello05}.
\subsubsection{Discriminant definition of a vortex}
The velocity gradient tensor $A_{ij} = \partial v_i /\partial x_j$
measured by a nonrotating observer that moves locally with
the fluid can be decomposed into a symmetric ($S_{ij}$, or
rate-of-strain tensor) and an antisymmetric
($W_{ij}$, or rate-of-rotation tensor) part:
\begin{equation}
\label{eq:cap5_aijdec} A_{ij} = S_{ij} + W_{ij}
 = \frac{1}{2} \left( \frac{\partial v_i}{\partial x_j}
+ \frac{\partial v_j}{\partial x_i} \right)
+ \frac{1}{2} \left( \frac{\partial v_i}{\partial x_j}
- \frac{\partial v_j}{\partial x_i} \right).
\end{equation}
The eigenvalues $\lambda$ are roots of
the characteristic polynomial
$\lambda^3 + P_A \lambda^2 +Q_A \lambda +R_A = 0$,
whose coefficients are defined by
\begin{eqnarray}
P_A & = &  -{\rm tr} (A_{ij}) = - S_{ii}, \\
Q_A & = & \frac{1}{2}[P_A^2 - {\rm tr}(A_{ij}^2)]
    =\frac{1}{2} (P_A^2 - S_{ij} S_{ji} - W_{ij} W_{ji}),\\
R_A  & = & - {\rm det} (A_{ij})  = \frac{1}{2}(-P_A^3 + 3P_A Q_A - S_{ij} S_{jk} S_{ki}
- 3 W_{ij} W_{jk} W_{ki}).
\end{eqnarray}
These quantities are invariant 
under any  nonrotating coordinate transformation.
The quantity
$P_A$ is the trace of the velocity gradient tensor (that is, the continuity
equation); in an incompressible fluid, one has $P_A=0$.
The quantity $Q_A$ measures the excess of rotation over strain. The sign of
$R_A$ governs the stability
of the flow (see below). The scalar invariants
$R_A$ and $Q_A$ can be combined to form the discriminant
$D_A = Q_A^3 + 27 R_A^2/4$ \citep{sc94}.
\begin{itemize}
\item If $D_A > 0$ at some point in the flow, $A_{ij}$
at that point has one real and two complex conjugate eigenvalues. Following
the nomenclature of \citet{cpc90}, we say that such points
are focal in nature. Streamlines wrap around the axis of
the real eigenvector and describe a spiral in the plane
spanned by the two complex eigenvectors. The sense
of the spiral is determined by the sign of $R_A$.
For $R_A > 0$, the trajectories are attracted towards the axis
of the real eigenvector; the point is an unstable
focus/contracting (UF/C) \citep{cpc90}. For $R_A < 0$,
the trajectories are repelled away from
the eigenvector; the point is
a stable focus/stretching (SF/S).
\item If $D_A < 0$, all
the eigenvalues are real. We say that such points are
strain dominated \citep{sc94}. Streamlines either approach
or flee the point along the three independent,
intersecting, real eigenvectors. When projected onto
the three planes spanned by the eigenvectors, the trajectories
asymptotically approach the eigenvector axes along ``parabola-like"
or ``hyperbola-like" paths, depending on the sign of $R_A$.
If $R_A < 0$, we get a stable node/saddle/saddle
(SN/S/S); if  $R_A>0$, we get an unstable/node/saddle/saddle
(UN/S/S). The degenerate case $R_A=0$ corresponds to a two-dimensional flow.
\end{itemize}

In this paper, we plot topological isosurfaces
according to the
following colour scheme.
Regions with
$D_A>0$ which are SF/S and UF/C are coloured yellow
and light blue respectively.
Regions with $D_A<0$ which are SN/S/S and
UN/S/S are coloured orange and green respectively.

\subsubsection{$\lambda_2$ definition of a vortex}
\label{sec:l2_def}
\citet{jh95} realized that the existence of a pressure minimum
is not a sufficient and necessary condition for
a vortex core to be present. For example, an
unsteady flow
can create a pressure minimum even where there is no vortex. On
the other hand, centrifugal forces can cancel viscous forces
perfectly (e.g. K\'arm\'an's
viscous flow, or a Stokes flow at low $\Rey$), thereby
eliminating a pressure minimum even though a vortex
is present \citep{jh95}.

The $\lambda_2$ criterion seeks to locate
the pressure minimum in a plane perpendicular
to the vortex axis by
eliminating unsteady strains and viscous stresses
from the Navier--Stokes equations.
Decomposing $A_{ij}$ into
symmetric and anti-symmetric parts
and substituting into
the Navier--Stokes equation, we obtain,
for the symmetric part,
\begin{equation}
\label{eq:hessian}
\frac{\partial S_{ij}}{\partial t}+
v_{k} \frac{\partial S_{ij}}{\partial x_k}
- \frac{1}{\Rey}
\frac{\partial^2 S_{ij}}{\partial x_k \partial x_k}
+ S_{kj} S_{ik} + W_{kj} W_{ik}
= - \frac{\partial^2 p}{\partial x_i \partial x_j}.
\end{equation}
The Hessian of the pressure, $\partial^2 p/{\partial x_i \partial x_j}$,
contains information about pressure minima. The existence
of a pressure minimum requires the Hessian to have two positive
eigenvalues \citep{jh95}.
Ignoring the
unsteady strain [first term in (\ref{eq:hessian})]
and the viscous stress [third
term in (\ref{eq:hessian})],
a vortex is defined as a connected
region with two negative eigenvalues
of the tensor ${\bf S}^2 + {\bf W}^2$, which
is symmetric and has real eigenvalues only. Hence, if we
call the eigenvalues
$\lambda_1$, $\lambda_2$, and $\lambda_3$ and
order them such that
$\lambda_1 \geq \lambda_2 \geq \lambda_3$,
we define a vortex
to be a region where one has $\lambda_2 < 0$ \citep{jh95}.

\subsection{Misaligned spheres}
\label{sec:3D_incl}
In this section, we consider a spherical shell filled with
superfluid ($\delta=0.3$, $\Rey = 10^3$ or $10^4$),
whose inner and outer surfaces rotate at the same angular speed,
$\Omega_1 =\Omega_2 =1.0$, but about {\emph different
rotation axes}. The outer sphere rotates about an axis inclined
with respect to the $z$ axis, in the $x$-$z$ plane, 
by an angle $\theta_0 =3$\degree, while the inner sphere rotates
about the $z$ axis. We present a gallery
of some of the flows excited in this system and characterize
their topologies employing the methods in Section
\ref{cap3:topology}. Our results are merely a first attempt at this problem;
a lot remains to be learnt about 
the nonlinear physics behind such flows.

No-slip boundary conditions
(\ref{eq:bcvn}) are imposed on $\vn$, while the superfluid satisfies
either perfect slip 
(\ref{eq:bcs22}) or no slip (\ref{eq:bcs21}).
The mutual friction 
takes the GM form (\ref{eq:gmforce1}). We fail
to obtain stable evolution for HV mutual friction (\ref{eq:hvforce})
or large inclination angles ($\theta_0 \geq 3 \degree$);
the simulation becomes unresolved in $\phi$
unless $N_\phi \geq 20$ is used, straining
our computing budget.\footnote{We note here that post-processing
and visualization of the data is a very time-consuming task
when dealing with nonaxisymmetric flows.}

\subsubsection{Topology of the flow}
\label{cap3:topol_3dinc}
Investigations of the flow past a sphere
\citep{giacobello05} found that the $\lambda_2$ criterion is better
at identifying nonaxisymmetric vortex structures
than the discriminant criterion, validating the findings of
\citet{jh95}, who
noticed that vortical structures identified using $D_A$
tend to have noisy boundaries.
Recently, however, investigations of Taylor-G\"ortler vortices
in cylindrical containers by \citet{fsg05}
revealed that the vortex contours identified by
the $\lambda_2$ criterion can be severely perturbed
by numerical noise, unlike the discriminant
criterion. We find this too.
The discriminant criterion is better
suited to enclosed geometries than the $\lambda_2$
criterion, while the opposite is true for open
geometries. 

\begin{figure}[htpb]
\begin{center}
\includegraphics[scale=0.29]{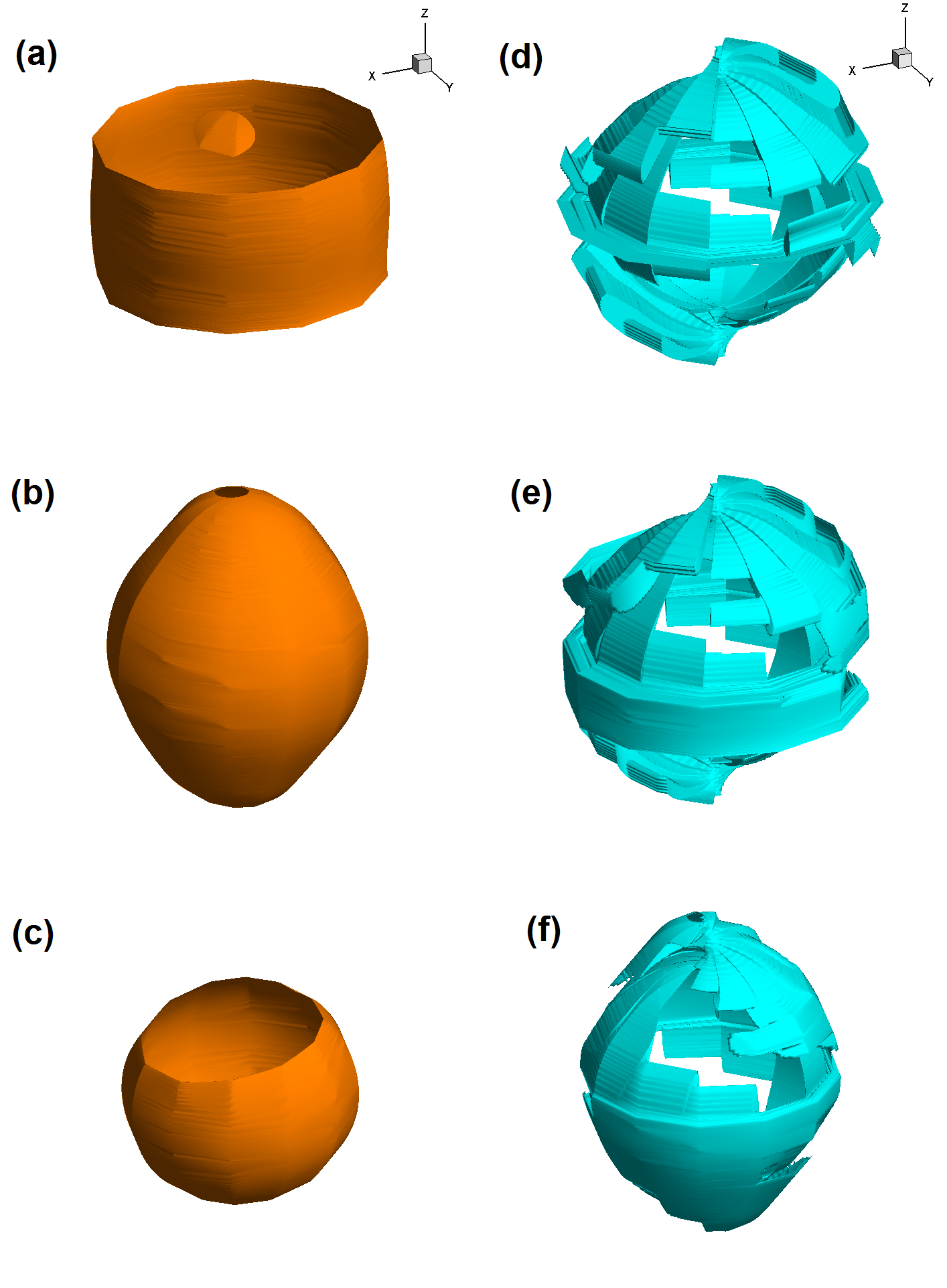}
\caption{Instantaneous flow topology for the normal component in superfluid SCF, with
$\Rey=10^3$, $\delta=0.3$, $\Delta \Omega=0$, and $\theta_0=3 \degree$.
Isosurfaces in orange for
(a) $\lambda_2 = -0.5$, (b) $\lambda_2 = -1$,
and $\lambda_2 = -3$, and in light blue for (d) $D_A = 10^{-4}$, (e) $D_A = 10^{-1}$,
and (f) $D_A = 1$. The $D_A$ criterion is to be preferred over the $\lambda_2$
criterion as it is less sensitive to the threshold.
Spectral
resolution and filter parameters:
$(N_r,N_\theta,N_\phi)=(100,200,12)$, $(\gamma_r,\gamma_\theta)=(10,10)$.}
\label{fig:cap3_3dfig1}
\end{center}
\end{figure}

In Figures \ref{fig:cap3_3dfig1}a--c,
we plot isosurfaces of
$\lambda_2$ ($\lambda_2=-0.5$, $-1$, $-3$)
for the normal component in superfluid SCF
with $\Rey=10^3$, at $t=50$.
It is hard to discern the regions of
the flow which are vorticity dominated. Moreover, when
varying the isosurface from
$\lambda_2 = -0.5$ to $\lambda_2 = -3$, we find that
its overall shape changes greatly, which is undesirable;
the visualization method
should be insensitive to the threshold \citep{giacobello05}.
By contrast, when we use the discriminant criterion,
as in Figures \ref{fig:cap3_3dfig1}d--f, focal regions in $\vn$
are cleary visible, and their shape is preserved despite
varying the threshold over four decades ($10^{-4} \leq D_A \leq 1$).

\begin{figure}[htpb]
\begin{center}
\includegraphics[scale=0.3]{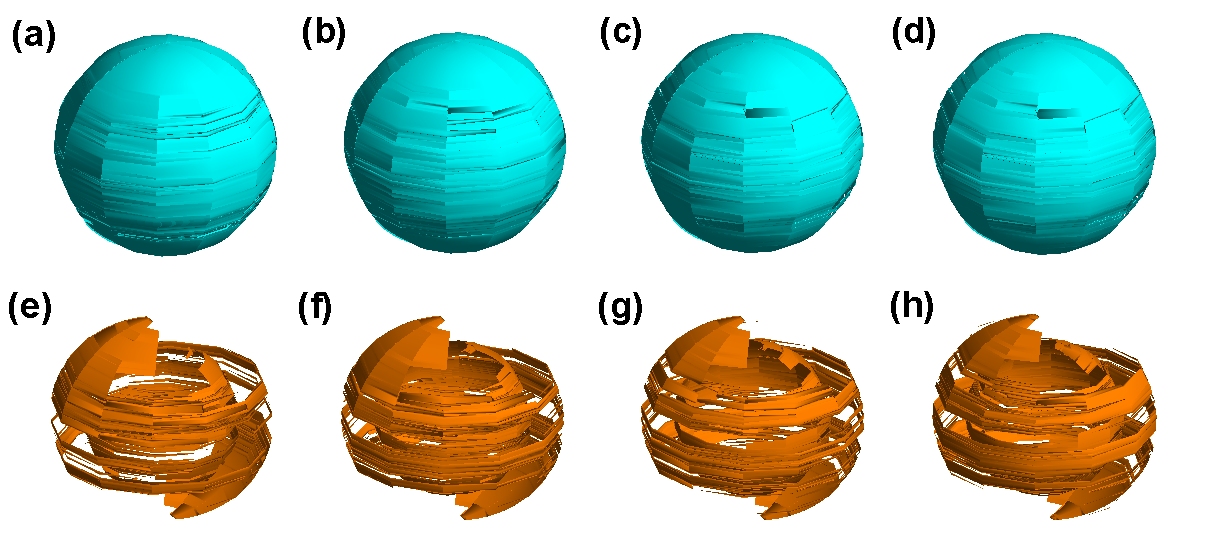}
\caption{Instantaneous flow topology of the superfluid component in superfluid
SCF with $\Rey=10^3$, $\delta=0.3$, $\Delta \Omega=0$, $\theta_0=3 \degree$,
and no-slip boundary conditions on $\vs$. Isosurfaces in light blue for
$D_A = 10^{-4}$ at (a) $t=20$, (b) $t=30$, (c) $t=40$,
and (d) $t=50$; and in orange for $D_A = -10^{-4}$ at (e) $t=20$, (f) $t=30$,
(g) $t=40$, and (h) $t=50$.
Spectral resolution and filter parameters:
$(N_r,N_\theta,N_\phi)=(100,200,12)$, $(\gamma_r,\gamma_\theta)=(10,10)$.}
\label{fig:cap3_3dfig10}
\end{center}
\end{figure}

\begin{figure}[htpb]
\begin{center}
\includegraphics[scale=0.3]{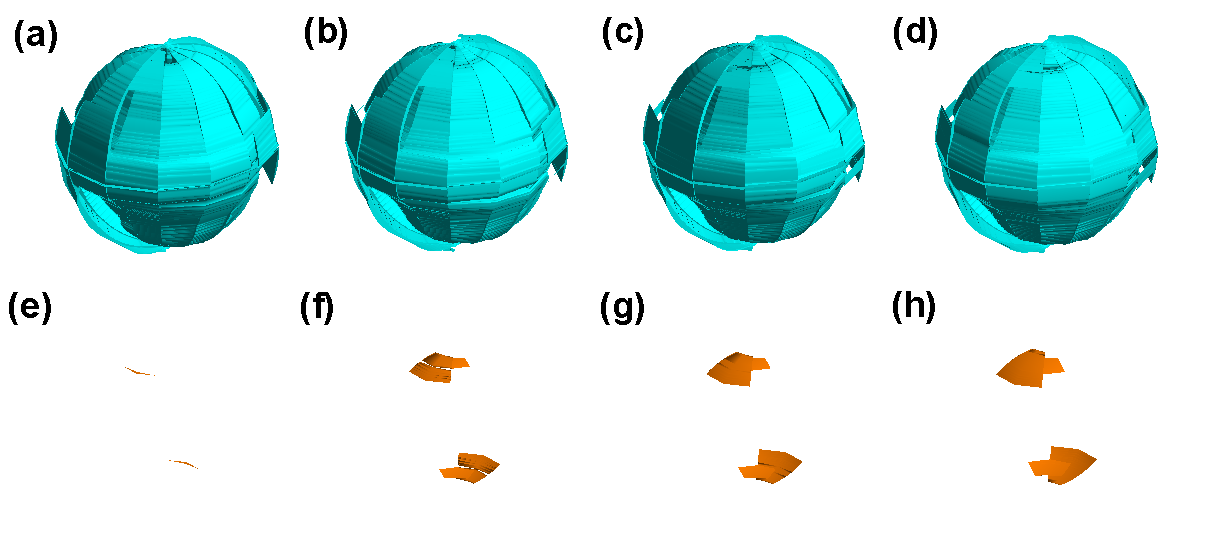}
\caption{Instantaneous flow topology of the normal fluid
component in superfluid SCF with
$\Rey=10^4$, $\delta=0.3$, $\Delta \Omega=0$, $\theta_0=3 \degree$,
and no-slip boundary conditions on $\vs$. Isosurfaces in light blue for
$D_A = 10^{-4}$ at (a) $t=20$, (b) $t=30$, (c) $t=40$,
and (d) $t=50$; and in orange for $D_A = -10^{-4}$ at (e) $t=20$, (f) $t=30$,
(g) $t=40$, and (h) $t=50$.
Spectral resolution and filter parameters:
$(N_r,N_\theta,N_\phi)=(100,200,12)$, $(\gamma_r,\gamma_\theta)=(10,10)$.}
\label{fig:cap3_3dfig11}
\end{center}
\end{figure}

The discriminant criterion allows us to
diagnose the topology of the inviscid superfluid component as well.
In Figure \ref{fig:cap3_3dfig10}, we present isosurfaces
of $D_A=10^{-4}$ (Figures \ref{fig:cap3_3dfig10}a--d)
and $D_A=-10^{-4}$ (Figures \ref{fig:cap3_3dfig10}e--h)
for $\vs$ in
superfluid SCF with $\Rey=10^3$ and
no-slip boundary conditions on $\vs$. Throughout
most of the volume, the flow is focal,
or vorticity-dominated.
Strain-dominated regions, shown
in orange, also exist, but are less widespread. They
have a threaded
structure (Figures \ref{fig:cap3_3dfig10}e--h),
which is maintained when
we increase the Reynolds number to
$\Rey=10^4$, because $\vs$
is weakly coupled to $\vn$ via
GM mutual friction (see Section \ref{subsec:visco_super}).
Perfect-slip boundary conditions
on $\vs$ do not alter the topology.
The normal fluid dynamics, on the other hand,
is almost completely dominated by vorticity, as
Figures \ref{fig:cap3_3dfig11}a--d show.
Strain-dominated regions are only
detected in small regions close
to the poles (see Figures \ref{fig:cap3_3dfig11}e--h).
This is natural:
the differential rotation is
small ($\Omega_2 \cos \theta_0 - \Omega_1 \approx 0.01$),
so the strain applied to the fluid is small.

\begin{landscape}
\begin{figure}[htpb]
\begin{center}
\includegraphics[scale=0.4]{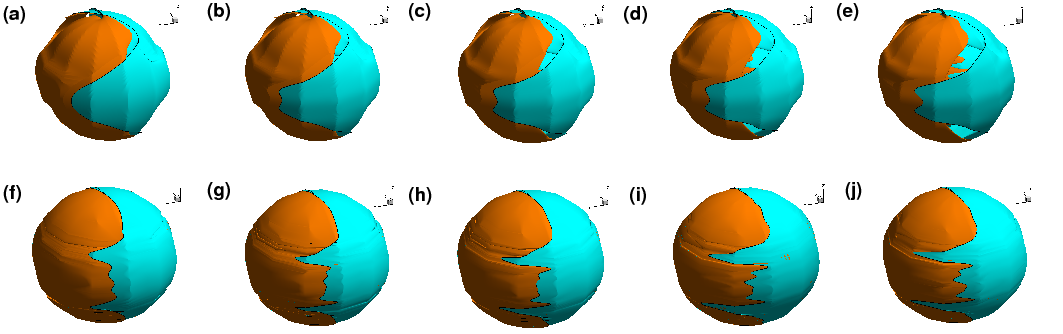}
\caption{Isosurfaces of vorticity for superfluid SCF with
$\Rey=10^4$, $\delta=0.3$, $\Delta \Omega=0$, $\theta_0=3 \degree$,
and no-slip boundary conditions on $\vs$.
Snapshots of the normal component $(\om_n)_\phi$ are shown at times (a) $t=10$, (b) $t=20$, (c) $t=30$,
(d) $t=40$, and (e) $t=50$. Snapshots of the superfluid component $(\om_s)_\phi$ 
are shown at times
(f) $t=10$, (g) $t=20$, (h) $t=30$,
(i) $t=40$, and (j) $t=50$. Positive vorticity $(\om_{n,s})_\phi=0.1$
is shown in light blue; negative vorticity $(\om_{n,s})_\phi=-0.1$
is shown in orange.
Spectral resolution and filter parameters:
$(N_r,N_\theta,N_\phi)=(100,200,12)$, $(\gamma_r,\gamma_\theta)=(10,10)$.}
\label{fig:cap3_3d_wphi1}
\end{center}
\end{figure}
\end{landscape}

The changes in the flow from one
snapshot to the next are hard to discern
in Figures \ref{fig:cap3_3dfig10} and \ref{fig:cap3_3dfig11}.
In order to reveal them more clearly,
we plot isosurfaces of $(\om_s)_\phi$ and  $(\om_n)_\phi$
at times $10 \leq t \leq 50$ in Figure \ref{fig:cap3_3d_wphi1}.
Positive values
[$(\om_{s,n})_\phi =0.1$] are
drawn in light blue; negative values [$(\om_{s,n})_\phi =-0.1$]
are drawn in orange.
Figures \ref{fig:cap3_3d_wphi1}a--e show how
the normal fluid
component evolves. The wedge-shaped isosurface $(\om_n)_\phi=-0.1$
develops a pointy extension at the equator that spreads
clockwise. Likewise, for the superfluid
component, the isosurface $(\om_s)_\phi=-0.1$
spreads clockwise in an equatorial band
located at $60 \degree \lsim \theta \lsim 150 \degree$.
Note that, although the
changes in the flow are easier to see, the
isosurfaces are threshold-sensitive.
For example,
the wavy contours in Figure \ref{fig:cap3_3d_wphi1}a--e
are not visible for $(\om_{s,n})_\phi = \pm 10^{-4}$.

\subsubsection{Unsteady torque}
\label{cap3:torque3d_incl}
\begin{figure}[htpb]
\begin{center}
\includegraphics[scale=0.45]{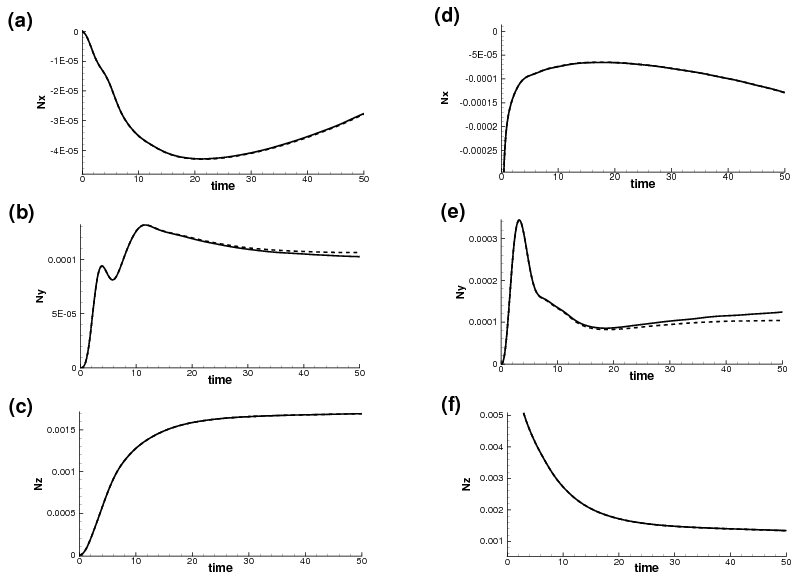}
\caption{Inner (left) and outer (right) torque versus time in
superfluid SCF with no-slip (solid curve) and perfect-slip (dashed curve) boundary
conditions on $\vs$, with
$\Rey=10^3$, $\delta=0.3$, $\Delta \Omega=0$, $\theta_0=3 \degree$,
and GM mutual friction. From
top to bottom, the plots show the (a) $x$, (b) $y$, and (c) $z$ components
of the inner torque, and the (d) $x$, (e) $y$, and (f) $z$ components
of the outer torque.
Spectral resolution: $(N_r,N_\theta,N_\phi)=(100,200,12)$. Filter parameters:
$(\gamma_r,\gamma_\theta)=(10,10)$.}
\label{fig:cap3_3dfig7}
\end{center}
\end{figure}

\begin{figure}[htpb]
\begin{center}
\includegraphics[scale=0.45]{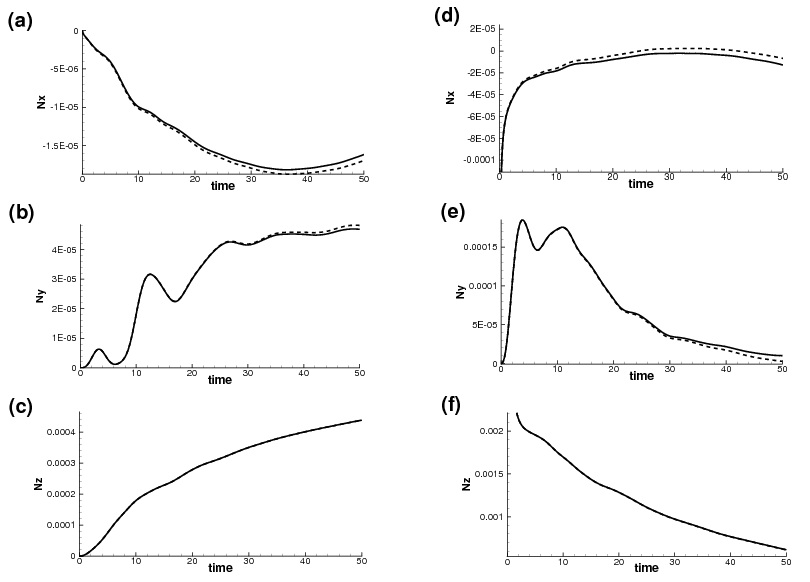}
\caption{Inner (left) and outer (right) torque  versus time
in superfluid SCF with no-slip (solid curve) and perfect-slip (dashed curve) boundary
conditions on $\vs$, with
$\Rey=10^4$, $\delta=0.3$, $\Delta \Omega=0$, $\theta_0=3 \degree$,
and GM mutual friction. From
top to bottom, the plots show the (a) $x$, (b) $y$, and (c) $z$ components
of the inner torque, and the (d) $x$, (e) $y$, and (f) $z$ components
of the outer torque.
Spectral resolution: $(N_r,N_\theta,N_\phi)=(100,200,12)$. Filter parameters:
$(\gamma_r,\gamma_\theta)=(10,10)$.}
\label{fig:cap3_3d_tor1e4}
\end{center}
\end{figure}

We present the evolution of the torque
on the outer sphere in Figures
\ref{fig:cap3_3dfig7} ($\Rey=10^3$) 
and \ref{fig:cap3_3d_tor1e4} ($\Rey=10^4$).
For $\Rey=10^3$, we have $N_z \sim  10 N_y$,
and $N_z \sim  40 N_x$, as expected
for small $\theta_0$. 
Moreover, $N_z$ tends to a constant value
$N_z \approx 1.7 \times 10^{-3}$ for
the outer torque and $N_z \approx 1.4 \times 10^{-3}$
for the inner torque at $t \geq 20$. The boundary condition
on $\vs$ has a negligible effect.
When it is changed from
no slip to perfect slip, $N_z$ decreases by
$\lsim 0.3$ \%, and 
$N_y$ decreases by $1$ \%  
(dashed curve in Figure \ref{fig:cap3_3dfig7}b).
For $\Rey=10^4$, the torque tends to
a constant value more gradually than for $\Rey=10^3$.
The differences between
no slip (dashed curves) and perfect slip (solid curves)
are slightly greater;
$N_x$ and $N_y$ 
(see Figures \ref{fig:cap3_3d_tor1e4}a--b
and \ref{fig:cap3_3d_tor1e4}d--e)
are $\sim 2$ \% larger for no slip.
Again, the dominant torque component
is $N_z$ ($ > N_y > N_x$).

The differences in the torque components arise from
asymmetries in the flow.
In an axisymmetric flow,
the greatest contribution to the torque comes
from regions containing a larger number of tightly packed
circulation cells.
The same is true in a nonaxisymmetric flow.
We calculate the torque from ${\bf N} = \int dS (\tau_{r \theta} {\bf e}_\theta +
\tau_{r \phi} {\bf e}_\phi )$, where $dS$
denotes the area element on the sphere and
$\tau_{ij}$ are the shear stresses.
We have $\tau_{r \theta} \neq 0$
and $\tau_{r \phi} \neq 0$ 
in a nonaxisymmetric flow, giving
$N_x \neq 0$ and $N_y \neq 0$. Note that
$N_x \neq N_y$ when avergared over time, since
the rotation axis is in the $x$-$z$ plane.

\subsection{Free precession}
\label{sec:3D_precess}
In this section, we consider a spherical rotating
shell filled with superfluid, where the outer sphere
precesses freely, while
the inner sphere rotates uniformly. We exaggerate
the biaxiality of the outer shell, taking
$\Omega_p=1.0$ for the body-frame precessional frequency
(defined in Section \ref{sec:tbc})
and $\Omega^\prime=2.0$ for the inertial-frame precession frequency
\citep{landau_mecanica,ja01}. This allows us to investigate
all the time-scales comprising the precession dynamics
using simulations of reasonable duration, something
that would be impossible for $\Omega_p \ll \Omega^\prime$.
The fixed angular momentum vector of the outer sphere
points in the $z$ direction
in the inertial frame of the inner sphere (which
rotates with $\Omega_1=1.0$). An expression
for the velocity of every point on the outer sphere
is given in Section \ref{sec:tbc}.
We consider a relatively low Reynolds number, $\Rey=10^3$, with
$\delta=0.3$ and no-slip boundary conditions on $\vs$.

\subsubsection{Topology of the flow}
\label{cap3:3D_precess_topol}
\begin{figure}[htpb]
\begin{center}
\includegraphics[scale=0.3]{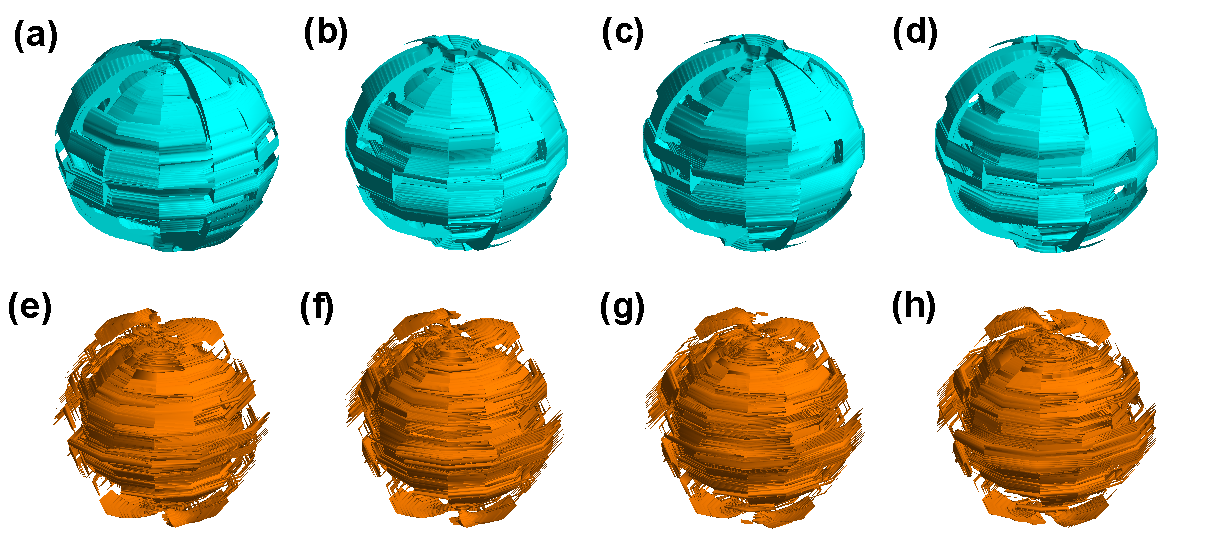}
\caption{Instantaneous flow topology for a rotating superfluid
contained within a freely precessing outer sphere and uniformly
rotating inner sphere. Discriminant isosurfaces
are shown
for the normal fluid component, for $D_A = 10^{-4}$ (light blue) at (a) $t=20$, (b) $t=30$, (c) $t=40$,
and (d) $t=50$, and for $D_A = -10^{-4}$ (orange), at (e) $t=20$, (f) $t=30$,
(g) $t=40$, and (h) $t=50$. The mutual friction is of GM form.
Simulation parameters: $\Rey=10^3$, $\delta=0.3$, $\Omega_1=1.0$,
$\Omega^\prime=2.0$, $\Omega_p=2.0$, $\theta_p=3 \degree$, and
no-slip boundary conditions on $\vs$.
Spectral resolution and filter parameters:
$(N_r,N_\theta,N_\phi)=(100,200,12)$, $(\gamma_r,\gamma_\theta)=(8,8)$.}
\label{fig:cap3_3dfig13}
\end{center}
\end{figure}

\begin{figure}[htpb]
\begin{center}
\includegraphics[scale=0.3]{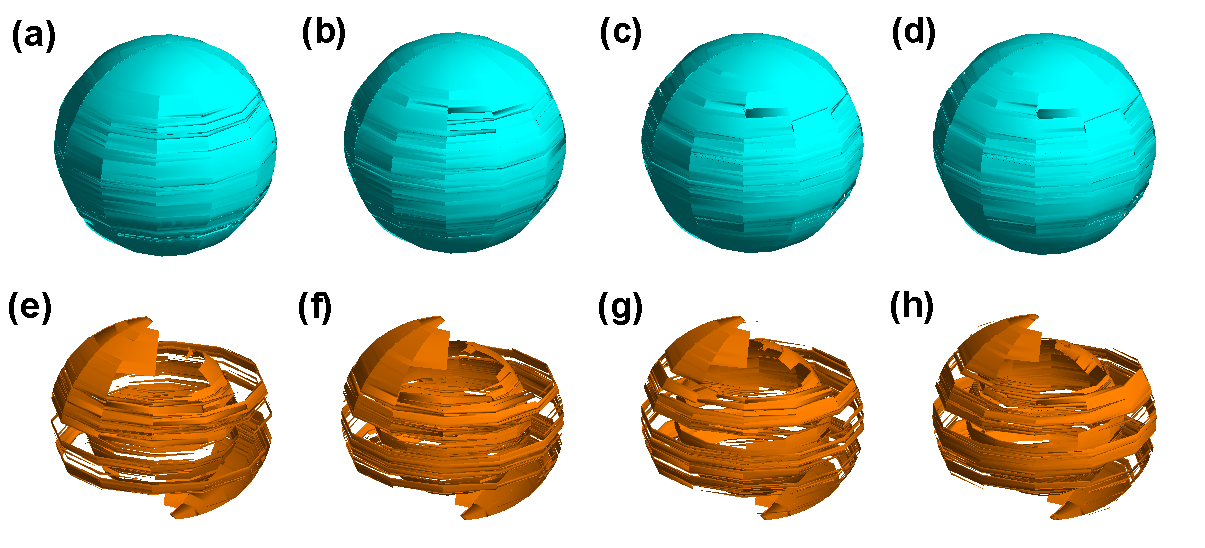}
\caption{Instantaneous flow topology for a rotating superfluid
contained within a freely precessing outer sphere and uniformly
rotating inner sphere. Discriminant isosurfaces
are shown for the superfluid component, for
$D_A = 10^{-4}$ (light blue), at (a) $t=20$, (b) $t=30$, (c) $t=40$,
and (d) $t=50$, and for $D_A = -10^{-4}$ (orange) at (e) $t=20$, (f) $t=30$,
(g) $t=40$, and (h) $t=50$.
The mutual friction is of GM form.
Simulation parameters: $\Rey=10^3$, $\delta=0.3$, $\Omega_1=1.0$
$\Omega^\prime=2.0$, $\Omega_p=1.0$, $\theta_p=3 \degree$, and
no-slip boundary conditions on $\vs$.
Spectral resolution and filter parameters:
$(N_r,N_\theta,N_\phi)=(100,200,12)$, $(\gamma_r,\gamma_\theta)=(8,8)$.}
\label{fig:cap3_3dfig14}
\end{center}
\end{figure}

The topology of the flow is illustrated in Figure \ref{fig:cap3_3dfig13}
for the normal fluid component and in Figure \ref{fig:cap3_3dfig14}
for the superfluid component. Unlike the misaligned
spheres in Section \ref{sec:3D_incl}, this flow is
influenced equally by strain and vorticity.
The UF/C topology is slightly more prevalent
(see the light blue isosurfaces in Figures \ref{fig:cap3_3dfig13}a--d)
than the SN/S/S topology in the normal fluid component.
In the superfluid component, the UF/C and
SN/S/S topologies are equally prevalent.
The UF/C regions (in light blue in
Figures \ref{fig:cap3_3dfig13}a--d) exhibit a complicated brick-like
structure (for $N_\phi=12$), while the SN/S/S regions are more
filamentary. The superfluid component is similar
to the normal fluid component but has smoother isosurfaces
(see Figure \ref{fig:cap3_3dfig14}), so
it is doubly difficult to distinguish transients in the flow.

\begin{landscape}
\begin{figure}[htpb]
\begin{center}
\includegraphics[scale=0.4]{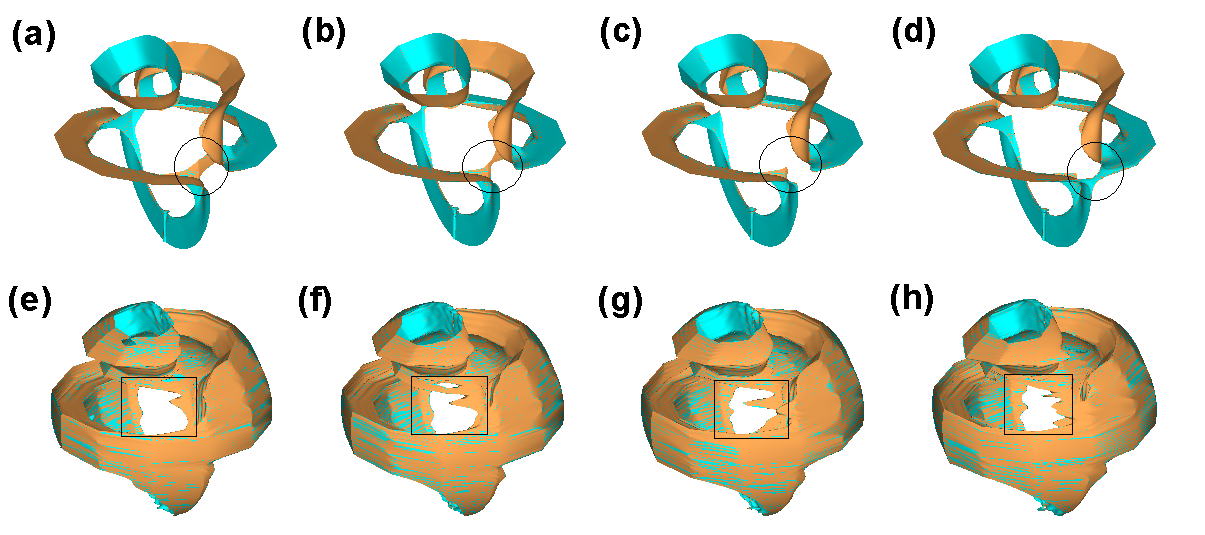}
\caption{Isosurfaces of vorticity with the Stokes solution
subtracted, projected along the instantaneous principal axis of
inertia of the outer sphere.
Snapshots of $\Delta \om_n \cdot {\bf e}_3$ are shown at
times (a) $t=20$, (b) $t=30$, (c) $t=40$, and
(d) $t=50$, and of $\Delta \om_s \cdot {\bf e}_3$ at times
(e) $t=20$, (f) $t=30$, (g) $t=40$, and (h) $t=50$. Positive vorticity
($\Delta \om_{n,s} \cdot {\bf e}_3 =0.1$)
is shown in light blue; negative vorticity ($\Delta \om_{n,s} \cdot {\bf e}_3=-0.1$)
is shown in orange.
Simulation parameters: $\Rey=10^3$, $\delta=0.3$, $\Omega_1=1.0$,
$\Omega^\prime=2.0$, $\Omega_p=1.0$, $\theta_p=3 \degree$, and
no-slip boundary conditions on $\vs$.
Spectral resolution and filter parameters:
$(N_r,N_\theta,N_\phi)=(100,200,12)$, $(\gamma_r,\gamma_\theta)=(8,8)$.}
\label{fig:cap3_3dfig15}
\end{center}
\end{figure}
\end{landscape}

When we plot isosurfaces of vorticity, in the
same manner as in Figure \ref{fig:cap3_3d_wphi1},
the results are unsatisfactory. The positive
and negative isosurfaces
are tightly interleaved and it is hard to make
out the underlying topology. However, the results improve
dramatically when we subtract
the vorticity of the Stokes solution from the
total vorticity and project $\Delta \om_{n,s}$
along the instantaneous principal axis of inertia, ${\bf e}_3(t)$,
of the outer sphere
(defined in Section \ref{sec:tbc}).
We present isosurfaces for $\Delta \om_n \cdot {\bf e}_3=\pm 0.1$
in Figures \ref{fig:cap3_3dfig15}a--d; as before, positive (negative)
isosurfaces are coloured light blue (orange).
Similarly, we present isosurfaces for
$\Delta \om_s \cdot {\bf e}_3=\pm 0.1$ in Figures \ref{fig:cap3_3dfig15}e--h.
We observe that isosurfaces of $\Delta \om_n \cdot {\bf e}_3$
form two interlocking ribbons of opposite sign which attach
($t=20$), detach ($t=40$), and attach again ($t=50$) at two
equatorial points (one of which is framed by a black circle). In
contrast, isosurfaces of $\Delta \om_s \cdot {\bf e}_3$ exhibit
a tongue-like structure in the equatorial plane (framed
by a black square), which grows clockwise from $t=20$
to $t=40$ and finally develops sawteeth at $t=50$
(see Figures \ref{fig:cap3_3dfig15}e--h).
We suspect that these three-dimensional structures
are not completely developed by $t=50$, i.e. we are
observing transients, but computational limitations
prevented us from extending the runs at the time
of writing.

\subsubsection{Unsteady torque}
\begin{figure}[htpb]
\begin{center}
\includegraphics[scale=0.5]{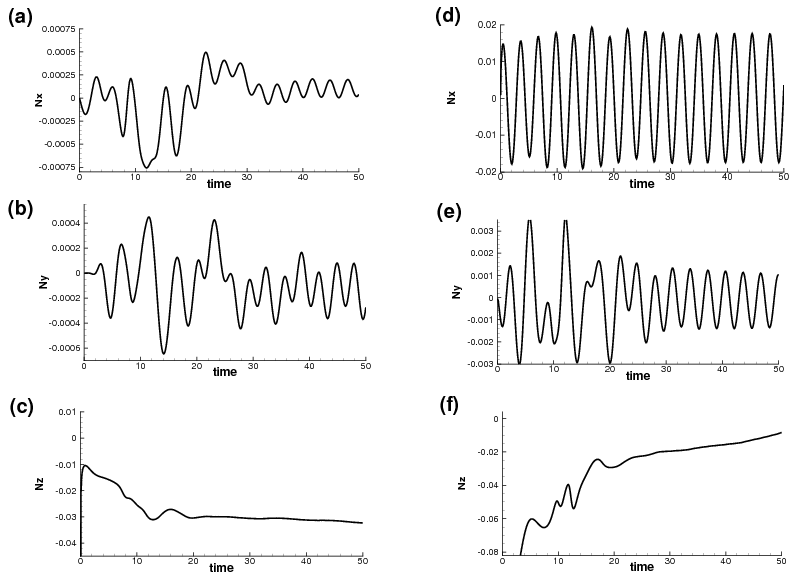}
\caption{Evolution of the viscous torque on the inner (left) and outer (right)
spheres for a rotating superfluid contained within a freely
precessing outer sphere and uniformly rotating inner sphere.
From top to bottom, the plots show the (a) $x$, (b) $y$, and (c) $z$ components
of the inner torque and the (d) $x$, (e) $y$, and (f) $z$ components
of the outer torque.
Simulation parameters: $\Rey=10^3$, $\delta=0.3$,
$\Omega^\prime=2.0$, $\Omega_p=1.0$, $\Omega_1=1.0$, $\theta_p=3 \degree$,
no-slip boundary conditions on $\vs$, and GM mutual friction.
Spectral resolution and filter parameters:
$(N_r,N_\theta,N_\phi)=(100,200,12)$, $(\gamma_r,\gamma_\theta)=(8,8)$.}
\label{fig:cap3_3dfig17}
\end{center}
\end{figure}

In Figure \ref{fig:cap3_3dfig17}, we plot the viscous
torque exerted by the fluid on the inner (Figures \ref{fig:cap3_3dfig17}a--c)
and outer (Figures \ref{fig:cap3_3dfig17}d--f) spheres
for $\Rey=10^3$. On the inner sphere, we find $N_z \sim 10^2 N_x \sim 10^2 N_y$.
On the outer sphere, we find $N_z \sim N_x \sim 10 N_y$.
The outer torque increases linearly up to
$t \approx 20$, when it reaches  
$|N_z| \approx 0.03$, while the inner torque decreases
to $|N_z| \approx 0.01$ over the same interval and
oscillates persistently ($\Delta |N_z| \approx 10^{-3}$,
period $\approx 3$). The other torque components oscillate
persistently for $t \geq 2$. For example, $({\bf N}_2)_x$ has
constant amplitude ($\approx 4 \times 10^{-2}$)
and period ($\approx 3$), $({\bf N}_1)_x$ has a smaller amplitude
($\approx 2 \times 10^{-4}$) but 
the same period, $({\bf N}_1)_y$
has period $\approx 3$ and
amplitude ranging from $2 \times 10^{-4}$ to
$4 \times 10^{-4}$, and $({\bf N}_2)_y$
has amplitude ranging from
$2 \times 10^{-3}$ to $5 \times 10^{-4}$.

\section{Conclusion}
\label{sec:lab_astro}

Superfluid SCF, like its classical (Navier--Stokes) counterpart,
is controlled by three global parameters: the
dimensionless gap width $\delta$, the Reynolds number $\Rey$,
and the rotational shear $\Delta \Omega/\Omega$.
In addition, it is a function of the form (isotropic
versus anisotropic) and dimensionless amplitude
of the mutual friction and vortex tension forces.
In this paper, we solve numerically the HVBK equations describing
superfluid SCF for a range of $\delta$, $\Rey$, and $\Delta \Omega/\Omega$
and study the time-dependent behaviour of the resulting flow. The numerical
solver is based upon
a pseudospectral collocation method.
Special attention is paid to the pole parity problem 
and to controlling the growth of global oscillations
(due to the Gibbs phenomenon) by filtering
out high spatial frequencies spectrally.
The solver accurately resolves flows covering the parameter range
$10 \leq \Rey \leq 10^5$, $0.2 \leq \delta \leq 0.9$,
and $0 \leq \Delta \Omega/\Omega \leq 0.3$. Grids with
resolution
$(N_r,N_\theta,N_\phi) = (150,400,4)$ and
$(100,200,12)$ are adopted
for the most challenging problems we attempt in two
and three dimensions respectively.

In two dimensional superfluid SCF,
persistent quasiperiodic oscillations are always observed in the
torque during steady differential rotation (after initial
transients die away), with typical period $\sim \Omega_1^{-1}$ and
fractional amplitude $\sim 10^{-2}$. The oscillation amplitude
increases as $\Rey$ increases.
The viscous torques exerted by a Navier--Stokes fluid
and an HVBK superfluid with GM
mutual friction differ by $6$ \%.
However, the torque roughly triples
for HV mutual friction.
The meridional streamlines are more complicated for HV friction, with
more small circulation cells near the outer sphere, and
the amplitude of the torque oscillations is greater.

In three dimensional superfluid
SCF, nonaxisymmetric vortex structures 
are classified according to topological
invariants. The discriminant criterion
is more instructive than the $\lambda_2$
criterion. For misaligned spheres, the
flow is focal (vorticity-dominated) throughout most
of its volume, except for thread-like zones where it
is strain-dominated near the equator (inviscid component)
and poles (viscous component). A wedge-shaped isosurface
of vorticity rotates around the equator at roughly
the rotation period. For a freely precessing outer
sphere, the flow is equally strain- and vorticity-dominated
throughout its volume. Unstable focus/contracting points
are slightly more common than stable node/saddle/saddle points
in the viscous component but not in the inviscid component.
Isosurfaces of positive and negative vorticity 
form interlocking poloidal ribbons (viscous component)
or toroidal tongues (inviscid component) which attach
and detach at roughly the rotation period.
Persistent torque oscillations
are observed in all the
three dimensional flows considered, with
period $\sim 6 \Omega_1^{-1}$.

A detailed knowledge of the global superfluid hydrodynamics
inside a neutron star is needed to understand the origin
of the timing irregularities --- glitches and timing noise --- observed in
over $100$ radio pulsars to date \citep{dmhd95,sl96,lss00,hthesis02,sfw03,hobbs04,pmgo05a,pmgo06a,pmgo06b,mpw07}. Glitches are 
characterized by a sudden increase in the angular velocity
of the pulsar, in the range 
$10^{-11} \lsim \Delta \Omega/\Omega \lsim 10^{-4}$ \citep{lg06,per06}. 
Pulsars also exhibit non-Gaussian
fluctuations in pulse arrival times over many years,
known as timing noise.
The long relaxation time following a glitch, and the
temperatures in a neutron star, independently imply that
the interior of the star is a {\emph neutron superfluid}
\citep{handbook04,lg06}.

Recently, it was shown that the {\emph global} pattern of superfluid circulation
in a neutron star exerts a dramatic influence on its rotation
and may play a central role in explaining the phenomena
of glitches and timing noise \citep{pmgo05a,pmgo06a,pmgo06b}.
For this reason, it is important to understand more
fully the axisymmetric and nonaxisymmetric dynamics
of superfluid SCF. The results in this paper represent
a first step along this path. One particular consequence
is that, if the meridional circulation is fast enough,
a vortex tangle ({\it superfluid turbulence}) is
alternatively created and destroyed in the
outer core of the star (and indeed any spherical
container). For example, before a glitch, differential
rotation in the outer core drives a nonzero, poloidal
counterflow which excites the Donnelly-Glaberson instability (DGI) \citep{gjo74,sbd83,tab03}, and the vortices evolve into an isotropic tangle of
reconnecting loops. In this regime, the friction force is 
of GM form, coupling the normal and superfluid components loosely.
Right after a glitch, the differential rotation ceases, so does the poloidal
counterflow, the vortex tangle decays, 
a rectilinear vortex array forms, and the mutual friction changes
to HV form, suddenly locking the normal and superfluid components together.
In previous simulations done for neutron star parameters
\citep{pmgo05a},
we make an order-of-magnitude estimate of the ratio $|{\bf F}_{\rm HV}|/{\bf F}_{\rm GM}|$ as
follows. Ignoring $B^\prime$, we find
$|{\bf F}_{\rm HV}|/{\bf F}_{\rm GM}| \sim [B \omega_s (v_{ns} - \nu_s/R_2)]/
[A^\prime \rho_n \rho_s v_{ns}^3 / \kappa \rho^2]$.
Taking $v_{ns} \sim R_2 \Omega_2$, with $R_2 \sim 10^{6}$ cm,
$\Omega_2 \sim 10^2$ Hz, and $A^\prime \sim 10^{-5}$,
we get $|{\bf F}_{\rm HV}|/{\bf F}_{\rm GM}| \sim 10^6$. From the numerical
simulations, we obtain a typical value $|{\bf F}_{\rm HV}|/{\bf F}_{\rm GM}| \sim 10^5$,
which is similar. Nevertheless, we note that
the microphysics of the GM force in superfluid turbulence has not
been worked out fully yet \citep{jm04}.

The results on superfluid turbulence summarised above
are also relevant to laboratory experiments
by \citet{tsatsa72,tsatsa73,tsatsa74,tsatsa75,tsatsa80},
the only systematic experimental
study of spherical Couette flow in superfluid helium
undertaken to date. 
\citet{tsatsa72} studied the deceleration of axisymmetric vessels
made of glass and plastic and filled with $^4$He
(in the temperature range $1.4\, {\rm K} \leq T \leq 2.0$ {\rm K}),
after an impulsive acceleration. They observed
``jerky" behavior, reminiscent of pulsar glitches, and developed an
empirical formula for the relaxation time as a function of the initial angular
velocity, the normal fluid fraction, and the radius of the vessel. The
results agree qualitatively with glitch data from the Crab and Vela.
These experiments lend support to the idea that the neutron superfluid
inside pulsars plays an important role in the glitch phenomenon.
More broadly, however, the Tsakadze experiments --- and by extension,
the theoretical results in this paper ---
are of general interest in understanding the physics
of superfluid turbulence in rotating systems \citep{orange_book}.
The spin-up problem in helium II, and superfluids
in general, is far more complicated than in a viscous fluid, because
the normal fluid component interacts nonlinearly with
the quantized Feynman-Onsager vortices in the superfluid component.
One interesting effect is that sudden, ``glitch-like", spin-up
events and other rotational irregularities are associated
with {\it patchy} mutual friction:
the DGI is excited in parts of the superfluid (e.g. near
the walls, on the rotation axis, and at the equator) but not
elsewhere \citep{pmgo06b}. This new phenomenon will be
investigated further in a forthcoming paper \citep{pmgo07c}.

\begin{acknowledgments}

We gratefully acknowledge
the computer time supplied by the 
Victorian Partnership for Advanced Computation (VPAC).
We also thank S. Balachandar, from the University of
Florida, for supplying us with his
original pseudo-spectral solver, designed for a single
Navier-Stokes fluid, from which our two-fluid, HVBK solver
was developed.
This research was supported by
a postgraduate scholarship from the University of Melbourne
and the Albert Shimmins Writing-up Award.
C. P. acknowledges the support of the Max-Planck Society
(Albert-Einstein Institut).
\end{acknowledgments}

\appendix
\section{Pseudospectral collocation grids}
\label{sec:appendixA}
The radial direction ($r$) is discretized using
Chebyshev polynomials and a collocation
scheme \citep{boyd02}. The Gauss-Lobatto collocation points in $r$ are defined
as \citep{canuto88}
\begin{equation}
\label{eq:glc} x_{i} = -\cos\left[ \frac{\pi (i-1)}{N_r - 1} \right], \hspace{1cm} \mbox{with} \, \, \, 1 \leq i \leq N_{r},
\end{equation}
where $N_r$ is the number of radial collocation
points. The computational points $r_i$ in physical space are related
to the Chebyshev grid $x_i$ through the mapping

\begin{equation}
\label{eq:xtor} r_i = x_i \left(\frac{R_2 - R_1}{2}\right) + \left(\frac{R_1 + R_2}{2}\right).
\end{equation}

The angular directions $\theta$ and $\phi$ are discretized using
a periodic grid over the intervals $0 \leq \theta \leq \pi$ and
$0 \leq \phi \leq 2 \pi$ respectively. In the azimuthal
direction, the collocation points are defined as
\begin{equation}
\label{eq:phigrid} \phi_{k} = \frac{2 \pi (k - 1)}{N_{\phi}}, \hspace{1cm} \mbox{with} \, \, \, 1 \leq k \leq N_{\phi},
\end{equation}
where $N_\phi$ is the number of grid points in $\phi$.
In the polar direction, the collocation points are defined as
\begin{equation}
\label{eq:thetagrid} \theta_{j} = \frac{\pi (j - 1/2)}{N_{\theta}}, \hspace{1cm} \mbox{with} \, \, \, 1 \leq j \leq N_{\theta},
\end{equation}
where $N_\theta$ is the number of grid points in $\theta$.

\section{Spectral expansions}
\label{appendix:expansions}
A scalar field $s$
(such as pressure) is expanded as 
\footnote{Note that, in (\ref{eq:glc})--(\ref{eq:phigrid}),  
the grid indices run from $1$ to $N_{r,\theta,\phi}$, whereas, in
(\ref{eq:expansion_scalar})--(\ref{eq:expansion_uphi}),
the wavenumber indices $l$, $j$, $k$ run over different ranges.
In the solver, the Navier--Stokes equations are discretized 
according to the prescription in
Appendix \ref{ap:solalgo}.}
\begin{equation}
\label{eq:expansion_scalar}
s  = \left\{ \begin{array}{lll}
     \displaystyle \sum_{l=0}^{N_r-1}  \sum_{j=0}^{N_\theta-1} \sum_{k=-N_\phi/2}^{N_\phi/2} \alpha_{ljk} T_l(r)\cos(j \theta)e^{ik\phi} & \, \, \mbox{for} \, \, k \, \, \mbox{even} &   \\
     \displaystyle \sum_{l=0}^{N_r-1}  \sum_{j=1}^{N_\theta} \sum_{k=-N_\phi/2}^{N_\phi/2} \alpha_{ljk} T_l(r)\sin(j \theta)e^{ik\phi} & \, \, \mbox{for} \, \, k \, \, \mbox{odd}  &
     \end{array}\right.
\end{equation}
The radial component of a vector is continuous across the poles, but the tangential
and azimuthal components change sign. Hence
the radial, tangential, and azimuthal components of the velocity field are expanded as
\begin{equation}
\label{eq:expansion_ur}
u_r  = \left\{ \begin{array}{lll}
     \displaystyle \sum_{l=0}^{N_r-1}  \sum_{j=0}^{N_\theta-1} \sum_{k=-N_\phi/2}^{N_\phi/2} \beta_{ljk} T_l(r)\cos(j \theta)e^{ik\phi} & \, \, \mbox{for} \, \,  k \, \, \mbox{even} &  \\
     \displaystyle \sum_{l=0}^{N_r-1}  \sum_{j=1}^{N_\theta} \sum_{k=-N_\phi/2}^{N_\phi/2} \beta_{ljk} T_l(r)\sin(j \theta)e^{ik\phi} & \, \, \mbox{for} \, \,  k \, \, \mbox{odd}  &
     \end{array}\right.
\end{equation}

\begin{equation}
\label{eq:expansion_uth}
u_\theta  = \left\{ \begin{array}{lll}
     \displaystyle \sum_{l=0}^{N_r-1}  \sum_{j=1}^{N_\theta} \sum_{k=-N_\phi/2}^{N_\phi/2} \gamma_{ljk} T_l(r)\sin(j \theta)e^{ik\phi} & \, \, \mbox{for} \, \, k \, \, \mbox{even} &  \\
     \displaystyle \sum_{l=0}^{N_r-1}  \sum_{j=0}^{N_\theta-1} \sum_{k=-N_\phi/2}^{N_\phi/2} \gamma_{ljk} T_l(r)\cos(j \theta)e^{ik\phi} & \, \, \mbox{for} \, \, k \, \, \mbox{odd}  &
     \end{array}\right.
\end{equation}
\begin{equation}
\label{eq:expansion_uphi}
u_\phi  = \left\{ \begin{array}{lll}
     \displaystyle \sum_{l=0}^{N_r-1}  \sum_{j=1}^{N_\theta} \sum_{k=-N_\phi/2}^{N_\phi/2} \delta_{ljk} T_l(r)\sin(j \theta)e^{ik\phi} & \, \, \mbox{for} \, \, k \, \, \mbox{even} &  \\
     \displaystyle \sum_{l=0}^{N_r-1}  \sum_{j=0}^{N_\theta-1} \sum_{k=-N_\phi/2}^{N_\phi/2} \delta_{ljk} T_l(r)\cos(j \theta)e^{ik\phi} & \, \, \mbox{for} \, \, k  \, \, \mbox{odd}  &
     \end{array}\right.
\end{equation}
where $T_l$ is the $l$-{\rm th} Chebyshev polynomial,
$j$ and $k$ are the $\theta$ and $\phi$ wavenumbers, and
$\alpha_{ljk}$, $\beta_{ljk}$, $\gamma_{ljk}$ and $\delta_{ljk}$ are
real coefficients.

\section{Solution Algorithm}\label{ap:solalgo}
\subsection{Numerical differentiation}
\label{cap2:numdif}

Spatial differentiation in $r$ and $\theta$ is carried out
in physical space. The first-order $r$ derivative is calculated by
performing the operation 
\begin{equation}
\frac{\partial f}{\partial r} =
\displaystyle \sum_{j=1}^{N_{r}} {\cal D}_{ij}^r f(x_j),
\, \, \, \, \, \, \, \, \, \, \mbox{with} \, \, \, \, \, \, 1 \leq i \leq N_{r},
\end{equation} 
in which $f(x_j)$ is a discrete vector array, and ${\cal D}^{(r)}_{ij}$ is the
differentiation operator for the variable $r$, defined below. A similar
formula applies for $\partial f/\partial \theta$.

For the Gauss-Lobatto distribution of radial points defined
in Section \ref{sec:grids}, the radial
derivative operator is defined as
\begin{equation}
\label{eq:chebop}
{\cal D}_{ij}^{r}  = \left\{\begin{array}{ll}
\displaystyle \displaystyle \frac{(-1)^{i+j}}{ \displaystyle  2 \sin \left[\frac{\pi}{2(N_r-1)(i+j-2)}\right] \displaystyle \sin \left[\frac{\pi (i-j)}{2(N_r-1)}\right]}  & \, \, \, \, \, \mbox{for} \, \, i \neq j  \\ \\
\displaystyle \frac{ \displaystyle \cos [\pi j/(N_r-1)]}{ \displaystyle 2 \sin^2[\pi j/(N_r-1)]}     &\, \,  \, \, \, \mbox{for} \, \, i<=j<N_r \\ \\
\displaystyle -\frac{2(N_r-1)^2+1}{6}     &\, \, \, \, \, \mbox{for} \, \,  i=j=1 \\ \\
\displaystyle \frac{2(N_r-1)^2+1}{6}     &\, \,  \, \, \, \mbox{for} \, \, i=j=N_r.
     \end{array}\right.
\end{equation}
The top-half entries
of the matrix ${\cal D}_{ij}^{r}$ ($1 \leq i,j \leq N_r/2$) are more accurately represented than the lower-half ones, for
a given machine precision $\epsilon$ \citep{ds95}.
A significant improvement in accuracy is obtained by using the property
${\cal D}_{ij}^{r} = -{\cal D}_{{N_r -i},{N_r - j}}^{r}$ [for $(N_r/2)+1 \leq i \leq N_r$],
and calculating only the more accurate top-half part of (\ref{eq:chebop}). This
reduces the overall round-off error from ${\cal O}(N_r^3 \epsilon)$ to ${\cal O}(N_r^2 \epsilon)$ \citep{ds95}.
Radial derivatives of order $n$ are computed by applying
(\ref{eq:chebop}) $n$ times.

For a periodic grid with $2 N_\theta$ points, the first-order $\theta$ differentiation operator can
be expressed as \citep{canuto88}
\begin{equation}
\label{eq:firstheta}
\displaystyle {\cal F}_{ij}^{\theta}  = \left\{\begin{array}{ll}
\displaystyle \frac{1}{2} \displaystyle (-1)^{i-j} \cot \displaystyle \frac{\pi(i-j)}{2 N_\theta} &\hspace{1cm} \, \, \, \, \mbox{for} \, \, i \neq j  \\ \\
0    &\hspace{1cm} \, \, \, \, \mbox{for} \, \, i=j,
     \end{array}\right.
\end{equation}
and the second-order differentiation operator as
\begin{equation}
\label{eq:secondtheta}
\displaystyle {\cal F}_{ij}^{\theta \theta}  = \left\{\begin{array}{ll}
\displaystyle \frac{1}{2} \displaystyle (-1)^{i-j+1} \sin^2 \displaystyle \frac{\pi(i-j)}{2 N_\theta}, & \, \, \,\, \, \,
\mbox{for} \, \, i \neq j  \\ \\
\displaystyle -\frac{2+N_{\theta}^2}{12},    & \, \, \, \, \, \, \mbox{for} \, \, i=j.
     \end{array}\right.
\end{equation}
The form of ${\cal D}_{ij}^\theta$ depends on the parity of the function $f(x_j)$
on which it acts, as explained in Section \ref{appendix:expansions}.
First-order cosine ($\cal C$) and sine ($\cal S$) operators can be constructed from
(\ref{eq:firstheta}) as
\begin{eqnarray}
\label{eq:cossin11}
 {\cal C}_{ij}^{\theta} = {\cal F}_{ij}^{\theta} + {\cal F}_{i,2N_\theta+1-j}^{\theta}  \hspace{1cm} \, \, \, \, \, \, \, \, \, \, \, \mbox{for} \, \, \, \, 1 \leq i,j \leq N_{\theta} \\
\label{eq:cossin12}
 {\cal S}_{ij}^{\theta} = {\cal F}_{ij}^{\theta} - {\cal F}_{i,2N_\theta+1-j}^{\theta} \hspace{1cm} \, \, \, \, \, \, \, \, \, \, \,  \mbox{for} \, \, \, \, 1 \leq i,j \leq N_{\theta},
\end{eqnarray}
and second-order cosine and sine operators can be defined as
\begin{eqnarray}
\label{eq:cossin21}
{\cal C}_{ij}^{\theta\theta} = {\cal F}_{ij}^{\theta \theta} + {\cal F}_{i,2N_\theta+1-j}^{\theta \theta} \hspace{1cm} \, \, \, \, \, \,  \, \, \, \, \, \mbox{for} \, \, \, \,  1 \leq i,j \leq N_{\theta} \\
\label{eq:cossin22}
{\cal S}_{ij}^{\theta\theta} = {\cal F}_{ij}^{\theta \theta} - {\cal F}_{i,2N_\theta+1-j}^{\theta \theta} \hspace{1cm} \, \, \, \, \, \,  \, \, \, \, \, \mbox{for} \, \, \, \, 1 \leq i,j \leq N_{\theta}.
\end{eqnarray}
The operators (\ref{eq:cossin11})--(\ref{eq:cossin22})
are applied in spectral space using a fast Fourier transform (FFT).
Since the expansions (\ref{eq:expansion_scalar})--(\ref{eq:expansion_uphi}) are different
for even or odd $k$, the $\theta$ differentiation  is performed
separately for even and odd $k$.
For even $k$, a $\cal C$ operator is applied;
for odd $k$, an $\cal S$ operator is applied .
The differentiation operator reverses the parity when applied
an odd number of times and leaves the parity unchanged when
applied an even number of times, which needs to be taken
into account when computing higher order derivatives.

For the $\phi$ derivatives, the periodicity of the grid makes it easier
to perform the differentiation in spectral space.
Differentiation in $\phi$ reduces to multiplying each Fourier coefficient
${\rm FFT}_\phi[f(x_k)] = \widehat{f}(x_k)$ by
$i=\sqrt{-1}$ times the corresponding wavenumber $k$, viz.
\begin{equation}
\label{eq:phidiff}
{\cal D}^{\phi} \widehat{f}(x_j) =  i k \widehat{f}(x_j)  \hspace{2cm}  \, \, \mbox{for} \, \, 1 \leq j\leq N_{r,\theta}.
\end{equation}
The function $\widehat{f}(x_j)$ is transformed back to physical space
using an inverse FFT. In order to compute $\phi$ derivatives of order $n$,
the operation (\ref{eq:phidiff}) is performed $n$ times.


\subsection{Temporal discretization}
\label{cap2:tempdisc}
The most common and efficient method to solve the Navier--Stokes equation in terms
of the primitive variables (i.e., the velocity  and pressure fields) is the fractional step approach \citep{chorin68, bcm01}.
This method proceeds in two time steps,
decoupling the calculation
of the velocity and pressure fields. In the first step, an intermediate velocity
field $\vv_{n,s}^{\star}$ is computed from the velocity field
at the $n$-th time level,
$\vv_{n,s}^{n}$, using the momentum equations
(\ref{eq:hvbk1}) and (\ref{eq:hvbk2}) and ignoring the pressure and
the incompressibility constraint (\ref{eq:incompress}). In the second step, the
pressure is calculated by solving a Poisson equation involving the intermediate
velocity field to obtain the final, divergence-free velocity field 
at the $n+1$-th time level, $\vv_{n,s}^{n+1}$. The second step can be seen as a projection of $\vv_{n,s}$
onto a space of divergence-free vectors.

Stable evolution
is achieved by treating all the viscous
terms implicitly, to avoid the
stringent upper bound on $\Delta t$ imposed by the CFL condition
 \citep{vgh84}. Nonlinear terms, on the other hand, are
treated explicitly \citep{o71b}.
Upon discretizing
equations (\ref{eq:hvbk1}) and (\ref{eq:hvbk2}) in time using
an explicit third-order Adams-Bashforth (AB3) method for the nonlinear
terms and an implicit second-order
Crank-Nicolson (CN2) for the viscous terms \citep{boyd02}, and advancing
the solution from the time level
$n$ to the time level $\star$, we obtain
\begin{eqnarray}
\label{eq:td1}
\frac{\vn^{\star} - \vn^{n}}{\Delta t} & =&
- \frac{1}{12} \left\{ 23 \left[ \vn \cdot \nabla \vn +
\f{\rho_s \nabla \omega_s}{\rho \Reys} -\f{\rho_s {\bf F}}{\rho}
- {D}_\alpha
\right]^{n}
-16 \left[ \vn \cdot \nabla \vn+\f{\rho_s \nabla \omega_s}{\rho \Reys}
\right. \right. \nonumber \\
& & \left. \left.
 -\f{\rho_s {\bf F}}{\rho} - {D}_\alpha \right]^{n-1}
+ 5 \left[ \vn \cdot \nabla \vn
+ \f{\rho_s \nabla \omega_s}{\rho \Reys} -\f{\rho_s {\bf F}}{\rho}
- \frac{{D}_\alpha}{\Reys} {\bf e}_\alpha  \right]^{n-2} \right\} 
\nonumber \\
& & + \f{{\bf e}_\alpha}{2 \Rey} \left[ \frac{1}{r^2 \sin^2 \theta}
\frac{\partial^2 v_{n\alpha}^\star}{\partial \phi^2}
+ \frac{1}{r^2 \sin^2 \theta}
\frac{\partial^2 v_{n\alpha}^{n}}{\partial \phi^2} 
+ \frac{1}{r^2}\frac{\partial}{\partial r}
\left(r^2 \frac{\partial v_{n\alpha}^\star}{\partial r}\right) 
\right. \nonumber \\
& & \left.
+ \frac{1}{r^2}\frac{\partial}{\partial r}
\left(r^2 \frac{\partial v_{n\alpha}^n}{\partial r}\right) \right], 
\end{eqnarray}
\begin{eqnarray}
\label{eq:td2}
\frac{\vs^{\star} - \vs^{n}}{\Delta t} & = &
 \frac{1}{12} \left\{ 23 \left[ \vs \cdot \nabla \vs +
+ \f{\rho_s \nabla \omega_s}{\rho \Reys} +\f{\rho_n {\bf F}}{\rho}+
\f{{\bf T}}{\Reys}
 \right]^{n}
-16 \left[ \vs \cdot \nabla \vs+\f{\rho_s \nabla \omega_s}{\rho \Reys}
\right. \right. \nonumber
\\
& & \left.
+\f{\rho_n {\bf F}}{\rho}+
\f{{\bf T}}{\Reys}
 \right]^{n-1}
\left.
+ 5 \left[ \vs \cdot \nabla \vs
+ \f{\rho_s \nabla \omega_s}{\rho \Reys} +\f{\rho_n {\bf F}}{\rho} +
\f{{\bf T}}{\Reys}
\right]^{n-2}
\right\}, 
\end{eqnarray}
where ${\bf e}_\alpha = ({\bf e}_r, {\bf e}_\theta,
{\bf e}_\phi)$ is the triad
of spherical polar unit vectors. The term ${D}_\alpha$
is defined as
\begin{eqnarray}
\label{eq:dterm}
{D}_r & = & \frac{1}{r^2 \sin \theta} \frac{\partial}{\partial \theta}
\left( \sin\theta \frac{\partial v_{nr}}{\partial \theta} \right)
-\frac{2 v_{nr}}{r^2} -\frac{2}{r^2} \frac{\partial v_{n\theta}}{\partial \theta}
- \frac{2 v_{n\theta} \cot \theta}{r^2}
- \frac{2}{r^2 \sin \theta} \frac{\partial v_{n \phi}}{\partial \phi}, \\
{D}_\theta & = & \frac{1}{r^2 \sin \theta} \frac{\partial}{\partial \theta}
\left( \sin\theta \frac{\partial v_{n \theta}}{\partial \theta} \right)
+ \frac{2}{r^2} \frac{\partial v_{nr}}{\partial \theta}
- \frac{v_{n \theta}}{r^2 \sin^2 \theta} -
\frac{2 \cos \theta}{r^2  \sin^2 \theta} \frac{\partial v_{n \phi}}{\partial \phi},
\\
{D}_\phi & =&  \frac{1}{r^2 \sin \theta} \frac{\partial}{\partial \theta}
\left( \sin\theta \frac{\partial v_{n \phi}}{\partial \theta} \right)
- \frac{v_{n \phi}}{r^2 \sin^2 \theta}
+ \frac{2}{r^2 \sin^2 \theta} \frac{\partial v_{nr}}{\partial \phi}
+ \frac{2 \cos \theta}{r^2  \sin^2 \theta} \frac{\partial v_{n \theta}}{\partial \phi}.
\end{eqnarray}
Note that the Laplacian operator
$\nabla^2 \vv$ splits into two parts: $\theta$ derivatives are computed explicitly
with an AB3 scheme, while $r$ and $\phi$
derivatives are computed implicitly with a CN2 scheme. This procedure
avoids numerical instabilities \citep{bagchi02} and allows us to rewrite equation (\ref{eq:td1}) at the time
level $\star$ as a Helmholtz equation for $\vv_n^\star$
(see Section \ref{appendix:sol_alg}).

The previous advection-diffusion step is followed by a pressure correction step
based on
\be
\label{eq:presscorr} \frac{\vv_{n,s}^{n+1} - \vv_{n,s}^{\star}}{\Delta t}
= -\nabla p_{n,s}^{n+1}.
\ee
By taking the divergence of this equation, and imposing the continuity
constraint
$\nabla \cdot \vv_{n,s}^{n+1} = 0$, we obtain a Poisson equation
for the pressure:
\be
\label{eq:presscorr2} \frac{\nabla \cdot \vv_{n,s}^{\star}}{\Delta t}
=  \nabla^2 p_{n,s}^{n+1}.
\ee
This equation is solved implicitly for $p_{n,s}^{n+1}$ at every time step,
in order to get divergence-free normal and superfluid velocity fields
accurate to second order in time.

A boundary condition is needed to solve equation (\ref{eq:presscorr2}). Since
physical boundary conditions do not apply to
$\vv_{n,s}^{\star}$ and $p_{n,s}^{n+1}$, a number of empirical choices have
been proposed \citep{km85,bcg89}. Unfortunately, all
the choices create
a spurious numerical boundary layer at the edge of the
computational domain.
The simplest and most common choice for the pressure is a homogeneous
boundary condition $(\partial p_{nr,sr}^{n+1}/\partial r)|_{r=R_1,R_2} = 0$ \citep{oid86}.
The radial component of the velocity field satisfies
no penetration, viz. ${\bf e}_r \cdot  \vv_{n,s}^{\star} =  {\bf e}_r \cdot \vv_{n,s}^{n+1} = 0$.
A boundary condition for the tangential components can be obtained by rearranging equation
(\ref{eq:presscorr}) to read $\vv_{n,s}^{\star}|_{r=R_1, R_2} = (\vv_{n,s}^{n+1}
+ \Delta t \nabla p_{n,s}^{n+1})|_{\partial \Gamma}$. However,  $\nabla p_{n,s}^{n+1}$
is unknown at this stage of the calculation: we wish
to solve for it in order to pressure-correct the fields. \citet{sh91}
argue that, since the time-stepping scheme is second-order accurate in time, a
second-order accurate (or better) estimate for the velocity fields
is sufficient to preserve global second-order accuracy. Expanding
$\nabla p_{n,s}^{n+1}$ in a Taylor series about $t=t^{n}$ and approximating
$\partial p_{n,s}^{n}/\partial t$ with a first-order backward difference
formula, we obtain $\nabla p_{n,s}^{n+1} = 2 \nabla p_{n,s}^{n} - \nabla p_{n,s}^{n-1} + {\cal O}(\Delta t^2)$
and the boundary condition for the $\theta$ and $\phi$ components of the $\star$
velocity fields then becomes
\be
\label{eq:tanveloc} \mbox{\boldmath$\tau$} \cdot \vv_{n,s}^{\star}
=  \mbox{\boldmath$\tau$} \cdot \left[ \vv_{n,s}^{n+1}
+ \Delta t (2 \nabla p^{n} - \nabla p^{n+1})\right] + {\cal O}(\Delta t^3),
\ee
with $\mbox{\boldmath$\tau$} = {\bf e}_\theta$ or ${\bf e}_\phi$. This
gives a small slip velocity at the boundary, at the time level $n+1$, of order
${\cal O} (\Delta t^3)$. The particular choices for $\vv^{n+1}_{n,s}$
are described in more detail below.

Once $p^{n+1}$ has been calculated, the updated divergence-free
velocity fields $\vv_{n,s}^{n+1}$ are calculated from (\ref{eq:presscorr}).
Note that, since there is no diffusive term in equation (\ref{eq:hvbk2}), the
explicit AB3 method is used to advance $\vs$
from the time level $n$ to the time level $\star$.

\subsection{Advancing the solution in time}
\label{appendix:sol_alg}
Starting from an initial choice of $\vn$ and $\vs$ that
satisfies the continuity equations (\ref{eq:incompress}), the 
time-stepping procedure consists of an advection step
(or, more precisely, an advection-diffusion step for the normal fluid),
in which the
linear terms are handled implicitly (Crank-Nicolson)
and the nonlinear terms explicitly (Adams-Bashforth),
followed by a pressure-correction step. Given the solution
at the $n$-th time level, and appropriate boundary conditions,
the algorithm proceeds as follows.
\begin{enumerate}
\item We calculate
the Fourier and Chebyshev grids, the differentiation matrices, the
matrix operations that only depend on the spatial grid needed by the
Helmholtz and Poisson solvers (see below), 
any exponential filters to be applied to $\vn$ and $\vs$,
and the pole filter
and anti-aliasing filter matrices (see Section \ref{subsec:polefilter}).

\item We calculate the nonlinear and linear parts of
(\ref{eq:td1}) and (\ref{eq:td2}). The time-stepping loop
starts. The AB3 method initializes
$\vv_{n,s}^{1}$ and $\vv_{n,s}^{2}$
using a lower-order, Euler method.

\item We perform the advection-diffusion step. All the velocity and
pressure variables are
Fourier expanded in $\phi$. 
A Helmholtz equation (equation \ref{eq:hnormal} below) is solved in order to get the intermediate
$\vv_{n}^{\star}$ at the time level $\star$. The AB3 algorithm
is used to step forward $\vs$ explicitly to the time level $\star$.
The boundary conditions are applied.

\item We apply filters to the expansion coefficients and a convective filter,
as described in Section \ref{subsec:polefilter}. We then
pressure-correct $\vn^\star$ and $\vs^\star$ in order
to get $\vn^{n+1}$ and $\vs^{n+1}$, satisfying the continuity equation
(\ref{eq:incompress}). Poisson's equation (\ref{eq:presscorr}) is
solved for $\vn^{\star}$ and $\vs^{\star}$ and the solution
is advanced to the time level $n+1$ using (\ref{eq:presscorr}).

\item We take the inverse Fourier transform of the velocity and pressure
fields and write out their values on the coordinate grid as a
restart file (if desired).

\end{enumerate}

We now describe in detail steps (c) and (d).
The $r$, $\theta$ and $\phi$ components of the
vector Laplacian $\nabla^2 \vv$ are coupled
and a completely implicit treatment of the
diffusive terms is cumbersome. To get around this
issue, in step (d), the linear diffusive terms in
equation (\ref{eq:hvbk1})
are treated semi-implicitly: only the linear terms in the Laplacian
are included in the implicit temporal discretization (using
the CN algorithm) \citep{bagchi02}, while the nonlinear
terms in (\ref{eq:hvbk1}) are treated in an explicit
way (using an AB3 algorithm).
Since $r \Delta \theta > \Delta r$, the
$\theta$ components of the Laplacian can be treated explicitly, without
affecting the stability of the time-stepping algorithm
(see Section \ref{cap2:tempdisc}). Moving the nonlinear
terms and $\theta$ derivatives of the Laplacian
to the right-hand side of equation (\ref{eq:td1}),
and the time derivatives to the left-hand side, and
Fourier transforming all the variables, we arrive at
a Helmholtz equation
for the normal
velocity field $\vn = (v_{nr},v_{n\theta},v_{n\phi})$ at the time step
$\star$ which, for the radial component, takes the form
\be
\label{eq:hnormal}
\sin^2 \theta \frac{\pa}{\pa r}
\left(r^2 \frac{\pa \widehat{v}_{nr}^\star}{\pa r} \right)
- \left(\frac{r^2 \sin^2 \theta \Rey}{\Delta t}
+ \frac{k_{\phi}}{2} \right) \widehat{v}_{nr}^\star  = \widehat{\left(RHS\right)}^n_{r},
\ee
where all Fourier-transformed quantities are indicated by a `$\widehat{\hspace{0.25cm}}$'.
Similar equations can be written for $\widehat{v}_{n\theta}^\star$
and $\widehat{v}_{n\phi}^\star$. The right-hand
side $\widehat{\left(RHS\right)}^{n}_r$
contains all the non-linear terms and $\theta$ derivatives
of the Laplacian evaluated at the time step $n$.
This equation is solved in spectral space for each wave number $k_{\phi}$,
as described below.

The absence of a viscous term
in equation (\ref{eq:hvbk2}) allows us to use an
explicit method for the evolution in time, as indicated
in equation (\ref{eq:td2}).
For $\vs = (v_{sr},v_{s\theta},v_{s\phi})$,
the solution advances in time from $n$ to $\star$ according
to an AB3 scheme,
\be
\label{eq:superstar}
\widehat{v}_{si}^\star = \widehat{v}_{si}^{\, n}
- \Delta t \left[ \frac{23}{12} \widehat{\left(NLS\right)}_i^n
- \frac{4}{3}\widehat{\left(NLS\right)}_i^{n-1} +
\frac{5}{12} \widehat{\left(NLS\right)}_i^{n-2} \right],
\ee
where $\widehat{\left(NLS\right)}_i$ contains all the nonlinear
terms in (\ref{eq:hvbk2}).

The velocity fields obtained from equations (\ref{eq:hnormal}) and
(\ref{eq:superstar}) are not divergence-free, and
therefore do not satisty the continuity equation. A correction
must be made by solving the Poisson equation (\ref{eq:presscorr2}) for
the pressure, which can be written as

\be
\label{eq:presscorr3} \frac{\pa }{\pa r}
\left( r^2 \frac{\pa \widehat{p}_{n,s}^{\, n+1}}{\pa r} \right)
+ \frac{1}{\sin \theta} \frac{\pa}{\pa \theta} \left( \sin \theta
\frac{\pa \widehat{p}_{n,s}^{\, n+1}}{\pa \theta} \right) -
\frac{k_{\phi} \widehat{p}_{n,s}^{\, n+1}}{\sin^2 \theta} =
r^2 \frac{\widehat{\nabla \cdot \vv_{n,s}^{\star}}}{\Delta t}.
\ee
The corrected velocity fields $\vn^{n+1}$ and $\vs^{n+1}$ are then
calculated using (\ref{eq:presscorr}), which gives
\begin{equation}
\label{eq:corregidos} \widehat{\vv}_{n,s}^{n+1}
=\widehat{\vv}_{n,s}^{\star} -\nabla \widehat{p}_{n,s}^{n+1} \Delta t.
\end{equation}

Equations (\ref{eq:hnormal}) and (\ref{eq:presscorr3}) are
solved using a matrix diagonalization method \citep{canuto88,trefethen}.
The Helmholtz and Poisson equations can be written in matrix form, viz.
\begin{eqnarray}
\left[A \right]\left[\widehat{v}\right]+\left[\widehat{v}\right][B]^{T}
-\alpha[\widehat{v}] = \widehat{\left[RHS\right]},
\label{eq:matrixdiag}
\end{eqnarray}
where $\left[\widehat{v}\right]$ is a $N_r \times N_\theta$ array containing
the velocity or pressure fields, and  $\left[A \right]$ is a $N_r \times N_r$ matrix that represents
the discrete operators on
the left-hand sides of (\ref{eq:hnormal}) and (\ref{eq:presscorr3}).
The matrix $\left[B\right]^T$ is the transpose of $\left[B\right]$, a
$N_\theta \times N_\theta$ matrix
which is zero for the Poisson equation (\ref{eq:presscorr3})
and contains the second and third terms in the Helmholtz
equation (\ref{eq:hnormal}). Finally, we have
$\alpha = 2 \Rey /\Delta t$ for (\ref{eq:hnormal})
and $\alpha =0$ for (\ref{eq:presscorr3}).

In order to solve equation (\ref{eq:matrixdiag}) by matrix
diagonalization \citep{canuto88}, we
decompose $\left[A \right]$ and $\left[B\right]$ in
eigenvectors and eigenvalues for the operators $\nabla^2_r$
and $\nabla^2_\theta$ as
\begin{eqnarray}
\left[A\right]=\left[M\right]\left[\lambda_r\right]\left[M\right]^{-1},
\quad \quad
\left[B\right]=\left[N\right]\left[\lambda_\theta\right]\left[N\right]^{-1},
\label{eq:decompose}
\end{eqnarray}
where $\left[M\right]$ and $\left[N\right]$ are matrices formed from the
eigenvectors of $\left[\nabla^2_r\right]$ and $\left[\nabla^2_\theta\right]$,
and
$\left[\lambda_r\right]$ and $\left[\lambda_\theta\right]$ are diagonal matrices
formed from the eigenvalues of $\left[\nabla^2_r\right]$ and $\left[\nabla^2_\theta\right]$, respectively.
Substituting equation (\ref{eq:decompose}) in equation (\ref{eq:matrixdiag}) leads to
\begin{equation}
\label{eq:decompose2}
\left[\lambda_r\right]  \left[u\right] + \left[u\right] \left[\lambda_\theta\right]
- \alpha \left[u \right] = \left[s\right],
\end{equation}
with $\left[u\right] = \left[M\right]^{-1} \left[\widehat{v}\right] \left[N\right]$
and $\left[s\right] = \left[M\right]^{-1} \widehat{\left[RHS\right]} \left[N\right]$. Since
$\left[\lambda_r\right]$ and $\left[\lambda_\theta\right]$ are diagonal matrices, this
expression can be simplified to $[u] = [\lambda_r+\lambda_\theta-\alpha]^{-1} [s]$
and written explicitly as
\be
\label{eq:decompose3}
\left[u\right]_{ij} = \frac{\left[s\right]_{ij}}{\left[\lambda_r\right]_{ii}+\left[\lambda_\theta\right]_{jj}
-\alpha}, \, \, \hspace{3cm} \mbox{for} \, \, 1 \leq i,j \leq N_{r,\theta}.
\ee
Finally, $\left[\widehat{v}\right]$ can be obtained from (\ref{eq:decompose3}) using
$\left[\widehat{v}\right] = [M][u][N]^{-1}$.

To impose boundary conditions in the $r$ and $\theta$ directions, the
last and first rows of the
operator matrices $\left[A\right]$ and $\left[B\right]$ must
be modified to include the known
boundary values on the right-hand side of the equation \citep{trefethen}.
Dirichlet boundary conditions are applied to the
Helmholtz equation for $\vn$ (see Section \ref{ssubsec:rsbc}).
Neumann boundary conditions are applied to
the Poisson equation for $p_{n,s}$ (see Section \ref{cap2:tempdisc}).

\bibliographystyle{jfm}
\bibliography{pmgo08_jfm}

\end{document}